\documentclass[useAMS,usenatbib]{style/mnras} \usepackage[utf8]{inputenc}

\usepackage{times}
\usepackage{setspace}
\usepackage{graphics}
\usepackage{graphicx}
\usepackage{amssymb}
\usepackage{longtable}
\usepackage{amsmath}
\usepackage{amsbsy,fixmath}
\usepackage{epstopdf}
\usepackage{color}
\usepackage[T1]{fontenc}
\usepackage{aecompl}
\usepackage[none]{hyphenat}
\usepackage[normalem]{ulem}
\usepackage{physics}
\usepackage{mathrsfs}
\bibpunct[,]{(}{)}{;}{a}{}{,}
\usepackage{colortbl}
\usepackage{xcolor}
\usepackage{bm}
\usepackage{float}
\usepackage{multicol}
\usepackage{tabularx}
\usepackage{csquotes}
\usepackage{multirow}
\usepackage{rotating}

\let\sun=\odot 
\newcommand{\gadget}{{\sc Gadget}}

\newcommand{\mbh}{\ensuremath{\mathrm{MBH}}}

\newcommand{\mbhb}{\ensuremath{\mathrm{MBHB}}}

\newcommand{\Msun}{\ensuremath{M_{\odot}}}
\newcommand{\msun}{\ensuremath{M_{\odot}}}

 \newcommand{\lowa}{\texttt{50k}}
\newcommand{\lowb}{\texttt{200k}} \newcommand{\higha}{\texttt{500k}}
\newcommand{\highb}{\texttt{1m}} \newcommand{\F}[1]{\ensuremath{F=#1}}
\newcommand{\md}{\textrm{mini-disc}} \newcommand{\mds}{\textrm{mini-discs}}
\newcommand{\runa}{\texttt{RunA}} \newcommand{\runb}{\texttt{RunB}}
\newcommand{\runc}{\texttt{RunC}}
\newcommand{\paperbhb}{\citet{Goicovic2018}}
\title[Clumpy accretion onto MBHBs: gas dynamics]{Accretion of clumpy cold gas onto massive black holes
binaries:\\ the challenging formation of extended circumbinary structures}

\author[Maureira-Fredes et al.]{
    Cristián Maureira-Fredes$^{1}$\thanks{E-mail: cmaureirafredes@gmail.com},
    Felipe G. Goicovic$^{2,3}$,
    Pau Amaro-Seoane$^{4,5,6,7}$\newauthor and
    Alberto Sesana$^{8}$\\
    $^{1}$Max Planck Institute for Gravitational Physics (Albert Einstein Institute),
    Am M\"ulenberg 1, 14476, Potsdam-Golm, Germany\\
    $^{2}$Heidelberg Institute for Theoretical Studies
    (HITS), Schloss-Wolfsbrunnenweg 35, D-69118 Heidelberg, Germany\\
    $^{3}$Instituto de Astrof\'isica,
    Pontificia Universidad Cat\'olica de Chile, Av. Vicu\~na Mackenna 4860, 7820436
    Macul, Santiago, Chile\\
    $^{4}$Institute of Space Sciences (ICE, CSIC) \& Institut d'Estudis Espacials
    de Catalunya (IEEC)\\ at Campus UAB, Carrer de Can Magrans s/n 08193 Barcelona,
    Spain\\
    $^{5}$Institute of Applied Mathematics,\ Academy of Mathematics and
    Systems Science,\ CAS, Beijing 100190, China\\
    $^{6}$Kavli Institute for
    Astronomy and Astrophysics,\ Beijing 100871, China\\
    $^{7}$Zentrum f\"ur Astronomie und Astrophysik, TU Berlin,
    \ Hardenbergstra{\ss}e 36, 10623 Berlin, Germany\\
    $^{8}$School of Physics and Astronomy, University of Birmingham,
    Edgbaston, Birmingham B15 2TT, United Kingdom\\ }

\begin{document} \date{\today}
\pagerange{\pageref{firstpage}--\pageref{lastpage}} \pubyear{2017}
\maketitle
\label{firstpage}

\begin{abstract}
{Massive black hole binaries (\mbhb s) represent an unavoidable outcome of
hierarchical galaxy formation, but their dynamical evolution at sub-parsec
scales is poorly understood, due to a combination of uncertainties in
theoretical models and lack of firm observational evidence.  In gas rich
environments, it has been shown that the presence of a putative extended,
steady circumbinary gaseous disc plays an important role in the {\mbhb}
evolution, facilitating its coalescence. How gas on galactic scales is
transported to the nuclear region to form and maintain such a stable structure
is, however, unclear. One possibility is that, following the violent merger
dynamics, turbulent gas condenses in cold clumps and filaments that are
randomly scattered towards the nucleus. In this scenario, gas is naturally
transported on parsec scales and interacts with the {\mbhb} in discrete
incoherent pockets. The aim of this work is to investigate the gaseous
structures arising from this interaction.  We employ a suite of
smoothed-particle-hydrodynamic simulations to study the formation and evolution
of gaseous structures around a {\mbhb} constantly perturbed by the incoherent
infall of molecular clouds.  We investigate the influence of the infall rate
and angular momentum distribution of the clouds on the geometry and stability
of the arising structures. We find that the continuous supply of incoherent
clouds is a double-edge sword, resulting in the intermittent formation and
disruption of circumbinary structures. Anisotropic cloud distributions
featuring an excess of co-rotating events tend to generate more prominent
co-rotating circumbinary discs. Similar structures are also seen when mostly
counter-rotating clouds are fed to the binary, even though they tend to be more
compact and less stable. In general, our simulations do not show the formation
of extended smooth and stable circumbinary discs, typically assumed in
analytical and numerical investigations of the the long term evolution of \mbhb
s.}
\end{abstract}


\section{Introduction}

Thanks to advances in high angular resolution instrumentation, our
understanding of the central regions in galaxies has gone through a major
breakthrough.  Space-borne telescopes such as the Hubble Space Telescope and
adaptive optics from the ground allow us to investigate the kinematics and
distribution of gas and stars at sub-parsec scales for external galaxies
\citep[see e.g.][]{Kormendy03,KormendyHo2013} and to milli-pc for the Milky Way
(see e.g.
\citealt{SchoedelEtAl03,SchoedelEtAl2014b,Gallego-CanoEtAl2017,SchoedelEtAl2017}
and, in particular, the review of \citealt{GenzelEtAl10}).  A capital
conclusion is that massive dark compact objects, very likely massive black
holes ({\mbh}), with a mass typically ranging between $~10^{6-9}M_{\sun}$ are
lurking at the innermost centre of most large galaxies in the observable range.
Moreover, the formation and evolution of these objects and their host galaxies
seem to be linked
\citep{MagorrianEtAl1998,FerrareseMerrit2000,GebhardtEtAl2000,HaeringRix2004}.

On the other hand, from a theoretical standpoint hierarchical models best
explain the formation of structures in the Universe, down to the size of a
galaxy.  This implies that galaxies, during their lifetime, have suffered at
least one merger, if not several \citep[see e.g.][ for hierarchical merger
studies in $\Lambda$CDM Cosmology]{VolonteriEtAl03}. {If these galaxies harbour
a {\mbh} in their centre, during the collision they will sink to the centre of
the merged galaxy, form a binary and shrink its semi-major axis and become
``harder'' thanks to the interaction with stars coming from the surrounding
stellar system in which they are embedded
\citep[seee.g.][]{BegelmanEtAl1980,1996NewA....1...35Q,2007ApJ...660..546S,ColpiDotti2011}.
A star will strongly interact with the binary of \mbh s by removing energy and
angular momentum out of it, so that it will be re-ejected into the stellar
system with a higher kinetic energy, and the semi-major axis of the binary will
shrink a bit more.  Nevertheless, the loss-cone \citep[see
e.g.][]{FR76,Amaro-SeoaneEtAl2004}, which provides the {\mbh} binary (\mbhb)
with stars to interact with, is soon empty, halting the evolution of the
binary. The pace of subsequent shrink of the {\mbhb} is dictated by the rate at
which new stars are scattered into the loss cone.  For spherical stellar
distributions, the relevant scale is set by the two-body scattering timescale,
that can exceed the Hubble time \citep{MilosavljevicMerrit2001}. When this
happens, the binary stalls: the MBHs are still bound but they will not coalesce
in a Hubble time.  They are emitting gravitational waves, but the energy loss
is too weak.

This situation is known as ``the last parsec problem'', because the separation
between the {\mbh} s is typically of the order of $\sim 1$ pc
\citep{BegelmanEtAl1980,MM03,MM05}.  Whether or not these binaries will merge
in a Hubble time is a question that depends on various issues and is currently
debated.  However, the emerging consensus is that such binaries will in most of
the cases overcome this hang-up, due to efficient loss cone re-population in
the triaxial, dynamically un-relaxed stellar distribution produced by galaxy
mergers
\citep{2006ApJ...642L..21B,2011ApJ...732L..26P,2011ApJ...732...89K,2015ApJ...810...49V,2015MNRAS.454L..66S}.
{\mbhb} s will therefore get to the phase in their evolution in which they
efficiently emit gravitational waves to be detected with a GW observatory such
as the Laser Interferometer Space Antenna \citep[LISA,
see][]{Amaro-SeoaneEtAl2017,Amaro-SeoaneEtAl2013,Amaro-SeoaneEtAl2012}.

In gas rich galaxies, which become dominant at low galaxy masses and/or high
redshifts, a key factor to surmount this last stretch is the role of the gas,
which may be crucial in the evolution of the binary
\citep{EscalaEtAl04,EscalaEtAl05}.  For instance, the work of
\cite{CuadraEtAl09} found that gaseous discs should commonly help in the merger
of {\mbh} s with masses of interest for \emph{LISA}, whilst this mechanism
fails for masses larger than $\sim 10^7\,M_{\odot}$.

The evolution of {\mbhb} s interacting with a circumbinary gaseous structure
has been investigated by a number of authors. In most studies, gas is assumed
to be distributed in a steady, extended circumbinary disc that is either
co-rotating \citep[see e.g.][]{ArmNat05, CuadraEtAl09, Haiman2009,
Kocsis2012,Pau2013,DOrazio2013,Tang2017} or a counter-rotating \citep[see
e.g.][]{Roedig2014,Nixon2015,Amaro-SeoaneEtAl2016a} with respect to the
rotation of the {\mbhb}. The binary-disc system is assumed to evolve in
isolation, and almost no attempt has been made to connect the system with the
larger scale galactic environment that is providing the mass supply to the
gaseous structure. The violent interaction of two merging, gas rich galaxies is
expected to produce a turbulent environment in which cold clumps and filaments
of gas continuously interact exchanging angular momentum and eventually
infalling toward the centre of the merger remnant.  The initial orbit of the
galaxy merger provides a large source of angular momentum, the geometry and
angular momentum distribution of the infalling matter is therefore set by the
competing effect of ordered dynamics due to the large scale rotation of the
merger remnant and chaotic motions driven by turbulence and clumps and filament
scattering \citep{Sesana2014}.

It has been proposed that accretion of gas on to the binary can happen either
(i) in a coherent way, so that the angular momentum direction is basically kept
constant through all the episodes of accretion \citep{DottiEtAl2010}, (ii)
stochastically, meaning that the gas accretes on to the black holes in randomly
oriented planes \citep{KingEtAl2005,KingPringle2006}, for which we have
observational evidences, such as in the nucleus of the Abell 2597 galaxy
cluster \citep{TremblayEtAl2016} {or in the AGN PKS B1718-649
\citep{2018arXiv180103514M}}. Cold chaotic accretion has been linked to several
physical mechanisms acting on the interstellar medium, such as turbulence,
rotation, AGN/Supernovae feedback, among others
\citep{Hobbs11,Gaspari2013,Gaspari2015,Gaspari2017b}. Also, it has been pointed
that (iii) actually accretion might be a mixture of both, coherent and
stochastic accretion, with different degrees of anisotropy, as investigated by
\cite{DottiEtAl2013} to address possible anisotropies present  in the gas in
the nuclear regions of active galaxies.

Accretion on to a {\mbhb} of single clouds has been investigated numerically by
\cite{Dunhill2014}, \cite{GoicovicEtAl2016} and \cite{Goicovic2017}, taking
into account different orbital configurations and cooling rates.  {The
authors found that the interaction prompts a transient phase of high accretion
onto the {\mbhb}, while part of the leftover gas settles into a circumbinary
disc of various masses and sizes depending on the initial orbit of the cloud.
Whether such gaseous structure can be kept stable and grow in time under the
influence of a series of episodic accretion events is unclear, as the
successive infall of various clouds on to a {\mbhb} has not been addressed yet.
In this article, we present for the first time to our knowledge dedicated,
detailed numerical simulations of repeated gaseous infall episodes and
accretion on to a {\mbhb} in a post-merger galactic nucleus to assess the
architecture of the gas forming around the binary. We consider stochastic
feeding of the binary, assuming different degrees of anisotropy in the
distribution of the infalling clouds, as well as different feeding rates.}

The paper is organised as follows. In \S\ref{sec:methods} we describe the main
technical features of the simulation, focusing on physical ingredients and
numerical details that have been developed specifically for this suite of
simulations. In \S\ref{sec:experiments} we define the initial conditions of
each individual run, providing all the cloud parameter details necessary to
reproduce our runs. We test the stability and convergence of our numerical
scheme against critical parameters such as the assumed sink radius and the
number of particles used to simulate each cloud in \S\ref{sec:tests}. The
results of all simulations, including an extensive description of the transient
and long term circumbinary structures, and individual {\mds} are presented in
\S\ref{sec:results} and the significance of our main finding and future outlook
are discussed in \S\ref{sec:disc}. The impact on the evolution of the binary
itself is presented in a companion paper, \paperbhb.

Additional data and multimedia material of {\paperbhb} and the current publication
can be found in the website of the project~\footnote{\url{http://multipleclouds.xyz/}}.

\section{Methods}

\label{sec:methods} We study the evolution of the
{\mbh}-clouds system using {\gadget}-3, a Smooth Particle Hydrodynamics (SPH)
and $N-$body code, which is a modified version of
{\gadget}-2~\footnote{\url{http://wwwmpa.mpa-garching.mpg.de/galform/gadget/}}.
Every particle is represented as a point-mass, characterised by its 3-D
position and velocity, and the code solves for the hydro-dynamical and
gravitational interaction of the elements, which are organised in a tree
structure~\citep{BarnesHut1986}.

Specific to our implementation is the modelling of the {\mbhb} as two
equal-mass sink particles, initialised in a Keplerian circular orbit with
centre of mass at rest in the origin of our Cartesian reference frame. We
established the code units such as the initial semi-major axis, orbital period
and mass of the black hole binary are equal to one, i.e.
$a_0=P_{0}=M_0=1$\footnote{Throughout the paper the subscript $_0$ will refer
to initial system parameters.}, which can be re-scaled to an astrophysical
system as detailed in \cite{GoicovicEtAl2016}. Along the paper, we will
present results for a physical {\mbhb} with initial semi-major axis of
$a=0.2 \rm{pc}$, a total initial mass of $M_{\rm bin}=10^{6} \Msun$, and an orbital period of
$P=8400$ yr. We will refer to this astrophysical rescaling as our {\it fiducial
system}.

Each infalling molecular cloud is modelled as a spherical distribution of
equal-mass SPH particles. Each cloud has a mass of $M_{c} = 0.01\ M_0$ and, to
verify the convergence of our simulations, is modelled with increasing
resolution, using four different number of particles \begin{itemize} \item
$N_{c} = 5\times10^4$ ({\lowa}, default simulation value), \item $N_{c}
= 2\times10^5$ ({\lowb}), \item $N_{c} = 5\times10^5$ ({\higha}), \item $N_{c}
= 1\times10^6$ ({\highb}).  \end{itemize}

The main configuration of our simulation set-up is described in
\cite{GoicovicEtAl2016} and visualised in Figure~\ref{fig:cloud_configuration}.
It is an adaptation of the scheme presented in \cite{BonnellRice2008}, in which
we:

\begin{itemize}
    \item replace the central black hole by a {\mbhb};
    \item set the initial separation between the cloud and the {\mbhb} to 15
          (in code units, corresponding to 3 pc rescaled to our fiducial system);
    \item adopt an initial velocity modulus of the cloud of
          $v_c =0.5\ v_0 = 0.5\ \sqrt{GM_0/a_0}$;
    \item gauge the encounter impact parameter by varying the $\theta_{v_c}$ angle
          formed by the cloud velocity vector with the direction connecting the
          centre of masses of the cloud and of the {\mbhb}.
\end{itemize}

\begin{figure}
    \resizebox{\hsize}{!}
              {\includegraphics[scale=1,clip]{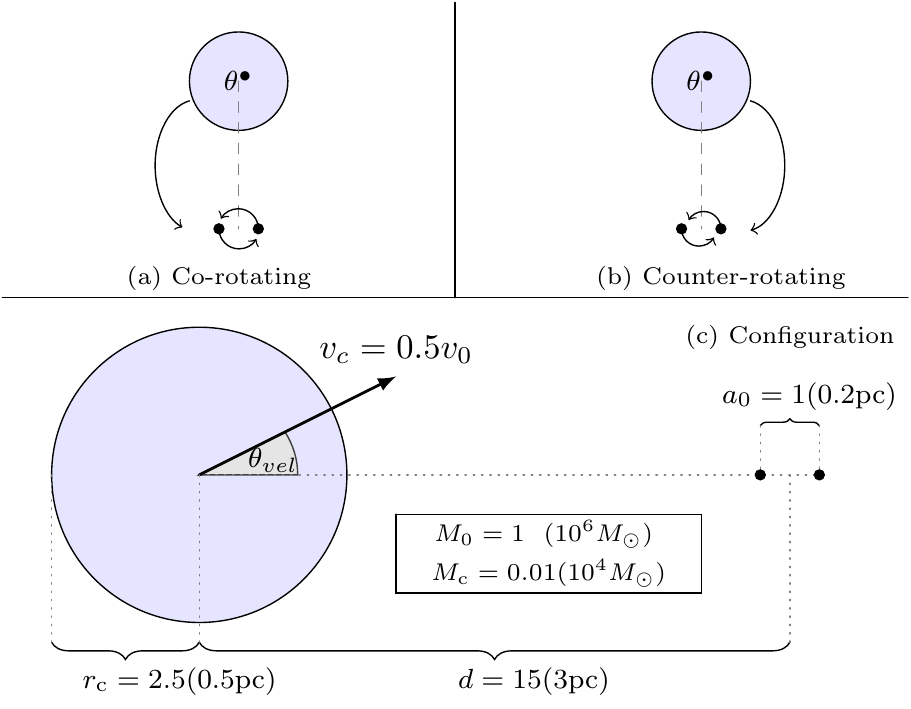}}
    \caption{Schematic representation of selected configurations of the simulations:
    (a) Co-rotating case, the gas cloud infalls following the {\mbhb} orbit;
    (b) Counter-rotating case, the gas cloud infalls oppositely to the {\mbhb}
    orbit. Panel (c) sketches how the cloud impact parameter is tuned by changing
    the angle  $\theta_{\rm vel}$ (see description in main text).
    The parameters of the system (initial separation between the cloud and {\mbhb},
    initial cloud velocity, masses and sizes of both the cloud and the {\mbhb}) are
    listed in the figure.}
    \label{fig:cloud_configuration}
\end{figure}

In the top panels of figure~\ref{fig:cloud_configuration}, we sketch the
set up of a co-rotating and a counter-rotating encounter. In the first case,
the angular momentum vector of the cloud orbit, $\vec{L}_c$, is aligned with
the orbital angular momentum vector of the \mbhb, $\vec{L}_{\rm bin}$; in
the latter, the two angular momenta have opposite direction. In general, the
two angular momenta can have arbitrary relative orientation. In our set-up, we
fix the reference frame so that the {\mbhb} is initially at rest in the origin
and $\vec{L}_{\rm bin}$ is along the positive $z$ axis. We then generally refer to
a cloud as ``co-rotating''(``counter-rotating'') if the $z$ component of its
$\vec{L}_c$ is aligned(counter-aligned) to $\vec{L}_{\rm bin}$.

When added to the simulation, the cloud is composed by spherically distributed
particles with uniform density, sustained against gravitational collapse by
a turbulent internal velocity field. We set up this turbulence by drawing from
a Gaussian random distribution with a power spectrum $P_v(k)\propto k^{-4}$,
where $k$ is the wave-number of the velocity perturbation in Fourier space.
This distribution is chosen to match the observed velocity of molecular clouds
\citep{Larson1981}.

\subsection{Additional physics}

\subsubsection{Accretion}

Each {\mbh} in the simulation is represented by a sink particle, which means
that it will accrete any particle entering a pre-defined sink radius, but
otherwise it interacts only gravitationally with other SPH particles. We do not
use the built-in stochastic accretion model that is included in {\gadget},
which is not appropriate for the problem in hand, where a reliable accretion
procedure for individual gas particles, beyond a stochastic selection scheme,
is required. Entering the sink-radius must be a necessary but not sufficient
condition for accretion. In fact physically unbound particles can simply fly-by
across the sink radius. Therefore, to avoid spurious accretion, we verify that
each candidate particle entering the sink radius is bound to the associated
{\mbh}. In practice, a gas particle will be accreted if the following
conditions are satisfied:

\begin{align}
    r_{\rm gas} &\leq r_{\rm sink} \\
    E_{\rm bind} &< 0
\end{align}

where $r_{\rm gas}$ is the relative distance between the {\mbh} and gas
particle, and we take a fiducial $r_{\rm sink} = 0.1$ which is a 10\% of the
initial separation of the {\mbhb}. A convergence study of the results against
the chosen value of $r_{\rm sink}$ is presented in \S\ref{sec:accretion}. In
order to be able to verify angular momentum conservation, it is important to
keep track of the accreted gas particles. The positions and velocity of those
particles at the moment of accretion are locally stored, to be used in the
computation of the angular momentum of the system when required.

\subsubsection{Thermodynamics}

The thermodynamics of the gas is modelled using a barotropic equation of state
\citep{Bonnell1994,EscalaEtAl2005,DottiEtAl2006},

\begin{align}
    P = K \rho^{\gamma},
\end{align}

where the constant $\gamma$ depends on the gas
density. The value of $K$ represents the entropic function of each gas particle
in the code \citep{SpringelHernquist2002}, and because it depends explicitly of
$\gamma$, it must be calculated so that the pressure behaves as a continuous
function of density. We choose an effective equation of state that represents
the behaviour of a collapsing protostellar cloud, whereby the low density gas
evolve isothermally up to some critical density ($\rho_c$), at which point it
becomes adiabatic \citep{Bate1995}. This can be represented as:

\begin{align}
    \gamma &= 1.0\quad\mbox{for}\quad \rho \leq \rho_{c};\\
    \gamma &= 1.4\quad\mbox{for}\quad \rho > \rho_{c}
\end{align}

As explained in \cite{GoicovicEtAl2016}, the addition of these two regimes breaks the
scale-free nature of our simulations, although the results can be scaled within
a certain range of critical densities, as explained in \cite{GoicovicEtAl2016}.

For the simulations presented in this paper we choose
$\rho_c=10^{-2} M_0 a_0^{-3}$ in code units, corresponding to
$10^{-16}$g cm$^{-3} \approx 1.5\times 10^6\msun$ pc$^{-3}$ when scaled
to our fiducial system. Note that this is two orders of magnitude
smaller than the value adopted in
\cite{GoicovicEtAl2016}. This modification was introduced to save computational
cost, as some of the configurations produce a large number of gas clumps.
Evolving these high density regions requires very small time-steps. Hence, by
effectively stopping the collapse at much lower densities, this modification
prevents simulations from stalling due to excessively small time-steps.

With this approach we are mimicking the evolution of a multi-phase gas without
implementing sophisticated cooling mechanisms or radiative transfer schemes.
The isothermal treatment of the gas is qualitatively representative of
optically thin media, where cooling is very efficient. Since our goal is to
study the dynamics of the gas and not the detailed thermodynamic evolution of
the dense regions, this treatment suites our purposes. Because the gas dynamics
is mostly dominated by the gravitational potential generated by the binary, we
do not expect the thermodynamics to have a major impact in our results.
Nevertheless, future studies should include a proper thermodynamic treatment of
the gas.

\subsubsection{External potential}

{\mbhb}s live in the dense environment of galactic nuclei, sitting at the
bottom of the galactic potential well. Although this potential does not
significantly affect the cloud-{\mbhb} individual interaction, it is important
to take it into account when performing an extensive experiment including the
interaction of multiple clouds. In fact, the external potential is relevant in
two ways:

\begin{itemize}
    \item it changes the dynamics of the gas flung away by the binary, keeping it
          bound to the system and allowing it to come back for further interactions;
    \item it acts as a restoring force, keeping the binary close to the origin of
          the reference frame (i.e. to the bottom of the potential well).
\end{itemize}

In practice the inclusion of the potential does not greatly affect the dynamics
of the close {\mbhb}-cloud encounters -- which remains dominated by the gravity
exerted by the {\mbhb} -- and it ensures angular momentum and energy
conservation due to its spherically symmetric nature. Most importantly, it
avoids drifting of the system away from the coordinate origin, which is
problematic when clouds are added to the system at different times. Instead,
the {\mbhb} experiences a gentle wandering with no secular
effects~\ref{fig:accr} and, together with its surrounded gas structures, is
kept close to the reference frame origin.

To determine the potential, we assume that matter is distributed around the
origin following an Hernquist density profile \citep{Hernquist1990}:

\begin{equation}
    \rho(r) = \frac{M_*}{2\pi}\frac{a_*}{r}\frac{1}{(r+a_*)^3},
\end{equation}

which implies a cumulative mass distribution given by

\begin{equation}
    m_*(r) = M_*\frac{r^2}{(r+a_*)^2},
\end{equation}

where $M_*$ and $a_*$ are the total mass and scale radius of the distribution.
Based on our fiducial system where $M_{\rm bin}=10^{6} \Msun$, we derived $M_*$ by
assuming the hole to bulge mass relation of \cite{MagorrianEtAl1998} and we
computed the scale radius using the radius to stellar mass relation of
\cite{Dabringhausen2008}. In code units, this gives

\begin{itemize}
    \item $M_*=4.78\times 10^2M_0$,
    \item $a_*=3.24\times 10^2a_0$,
\end{itemize}

which implies $m_*(<a_0) \approx 5\times10^{-3} M_0$, thus a negligible effect on the
Keplerian nature of the {\mbhb}.

\subsection{Other technical adjustments and numerical calibration}

\subsubsection{Dynamics of the sink particles}
\label{sec:sink_dyn}

The simple inclusion of dynamical 'sinks' several order of magnitude more massive
than the other SPH particles, introduced a number of issues with the SPH scheme.
This was already noticed in \cite{CuadraEtAl09},
who proposed to extract the sink particles from the tree for a better integration
of their trajectories. We adopted the same strategy here, integrating the
{\mbhb} orbit with a fixed time step $\Delta T = 0.02 P_{0}$, thus allowing the
binary positions and velocities to be updated more often then typical SPH
particles. We verified that this made the evolution more reliable, ensuring
that no crucial interactions between particles and the {\mbh}s were missed
along the integration.

Still, close inspection of the {\mbhb} evolution showed unphysical jumps in the
angular momentum of the system. We verified that this was related to the
frequency of update of the SPH tree. In SPH simulations one can choose how
often the particle tree (that defines how particles are grouped in computing
mutual forces) is generated and updated, which can be controlled by a adjustable
parameter in {\gadget}.
So long as the system does not experience dramatic
changes, simulations run smoothly with sparse tree updates. However, we are
dealing here with multiple clouds infalling onto a \mbhb from different
directions, triggering violent episodes of accretion, which is clearly not the
standard system handled by SPH codes of this type. We found that this required
reconstructing the tree 100 times more often than in the default {\gadget}
configuration.

\subsubsection{Injection of clouds}

At the beginning of each simulation, only one cloud is present in the system
besides the {\mbhb}. All the following clouds interact with the binary at later
times and therefore need to be included into the system `on the fly'. Once the
injection time for the new cloud is reached, the simulation is stopped and the
new cloud is added. Each cloud is characterised by a specific set of ids, so
that particles can always be tracked back to their original cloud. This is
useful to track the relative importance of each cloud in the accretion process,
or in the formation of specific circumbinary structures. Once the cloud is
added, the simulations is resumed and the integrator can adapt to the new
particles, forming the tree again, and handling this new scenario. The
procedure is repeated for each cloud. Note that due to the inclusion of new
clouds, the total angular momentum is not conserved. The angular momentum of
each injected cloud is, however, known, and it is therefore easy to track
angular momentum conservation along the integration of the system.

With the exception of the ones accreted by the {\mbhb}, we do not remove any
particle from the simulation. In fact, because of the way the tree is
constructed, particles that are flung far away from the binary are grouped in
large structures and integrated rarely, representing a negligible contribution
to the computational burden. Moreover, the addition of the external potential
keeps the structure compact, minimising the number of particles escaping at
distances larger than $100a_0$.

\subsubsection{Softening and sink radius}

Finally it is important the selection of appropriate softening parameter. We
choose for the sink particles a value of $\epsilon_{\textrm{BH}}=0.01 a_{0}
= 0.002$pc, and for SPH particles $\epsilon_{\textrm{gas}}=0.001 a_{0}
= 0.0002$pc. These values are small enough to ensure we are not bypassing
gravitational interactions, and are an order or magnitude smaller than the
maximum value recommended by \cite{BonnellRice2008}. As mentioned above, we fix
the sink radius at $r_{\rm sink}=0.1a_0$. We performed (see
\S\ref{sec:sink_dyn}) a series of tests ensuring that neither the dynamics of
the {\mbhb} nor the accretion of SPH particles is sensitive to the specific
choice of $r_{\rm sink}$.

\section{Initial conditions and run description}
\label{sec:experiments}

\begin{figure}
    \begin{tabular}{c}
        \includegraphics[width=0.45\textwidth]{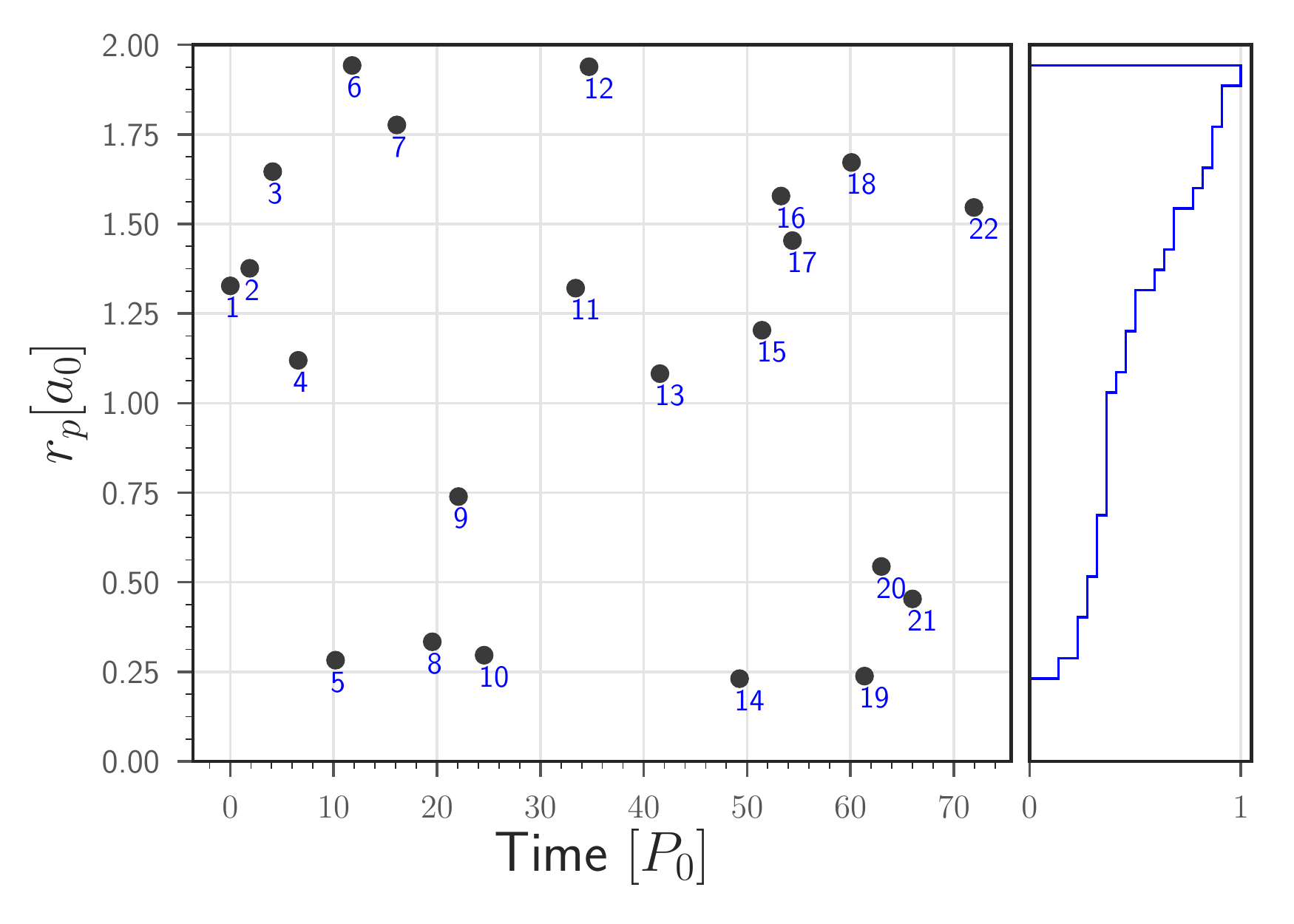}\\
        \includegraphics[width=0.45\textwidth]{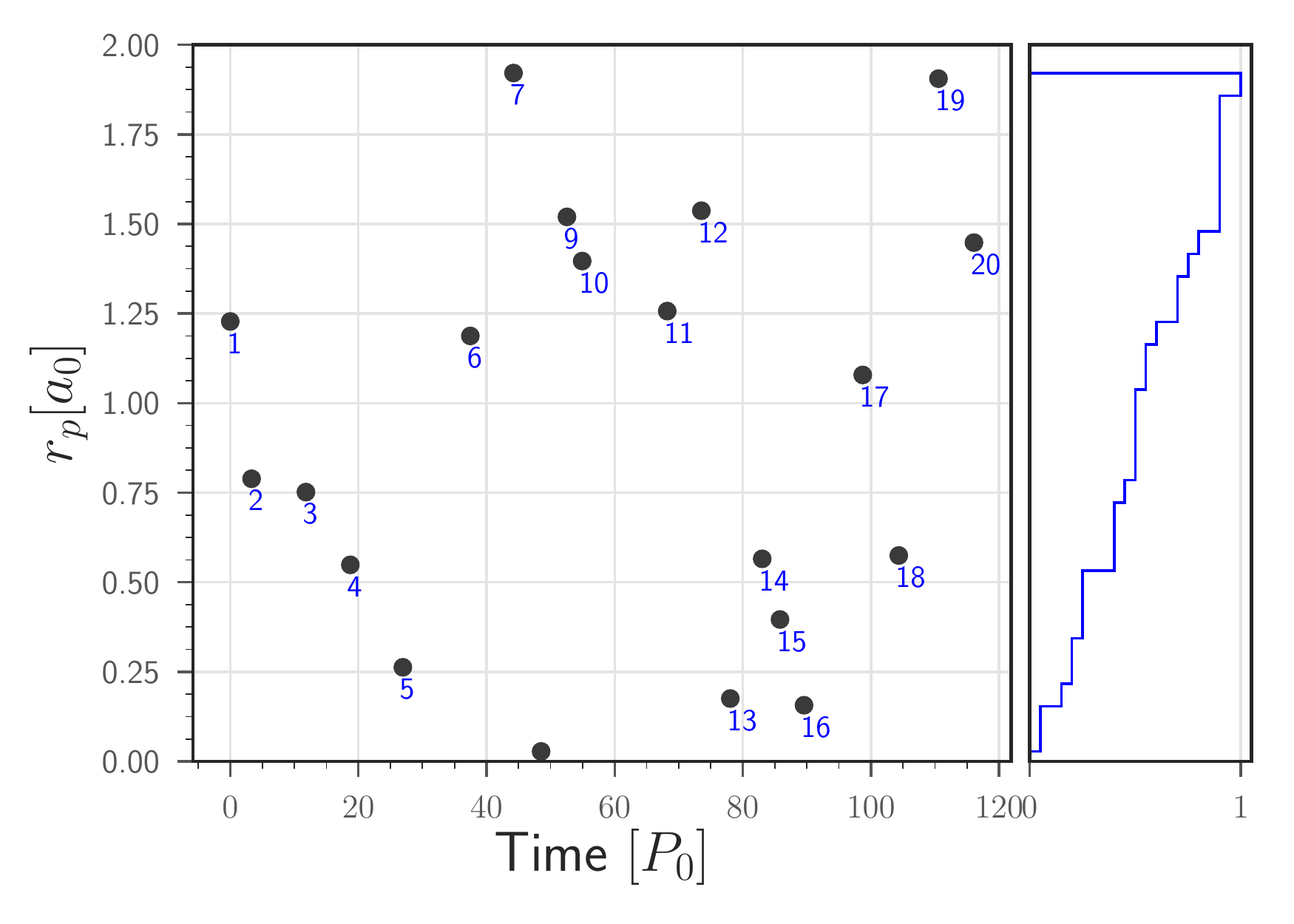}
    \end{tabular}
    \caption{Pericentre distance $r_p$ and injection time of the
    clouds in {\runa} (top panel) and {\runb} (bottom panel). Each individual cloud
    is represented as a black dot (numbered in ascending order). The small panels
    to the right show the $r_p$ cumulative distribution of the injected clouds.}
    \label{fig:dist_pericentre}
\end{figure}

\begin{figure*}
    \centering
    \begin{tabular}{c}
        \includegraphics[width=0.95\textwidth]{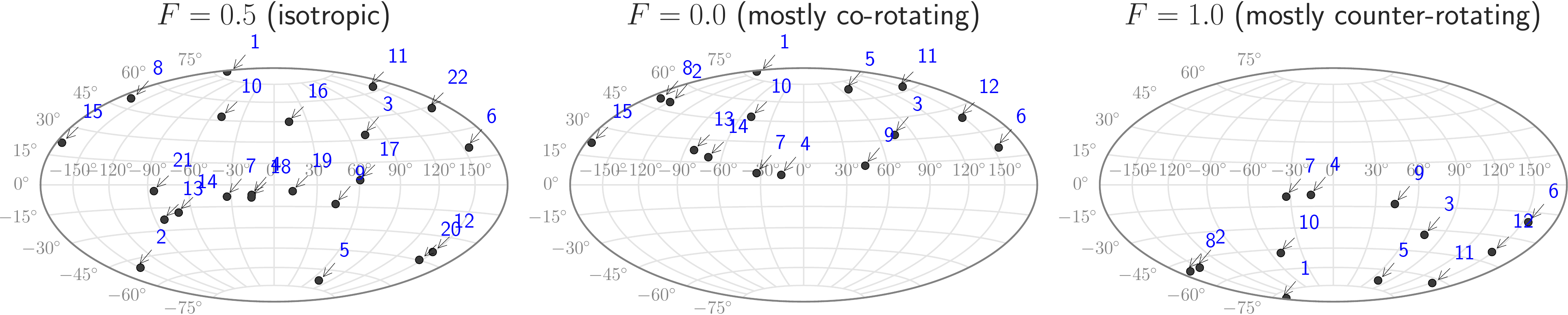}\\
        \includegraphics[width=0.95\textwidth]{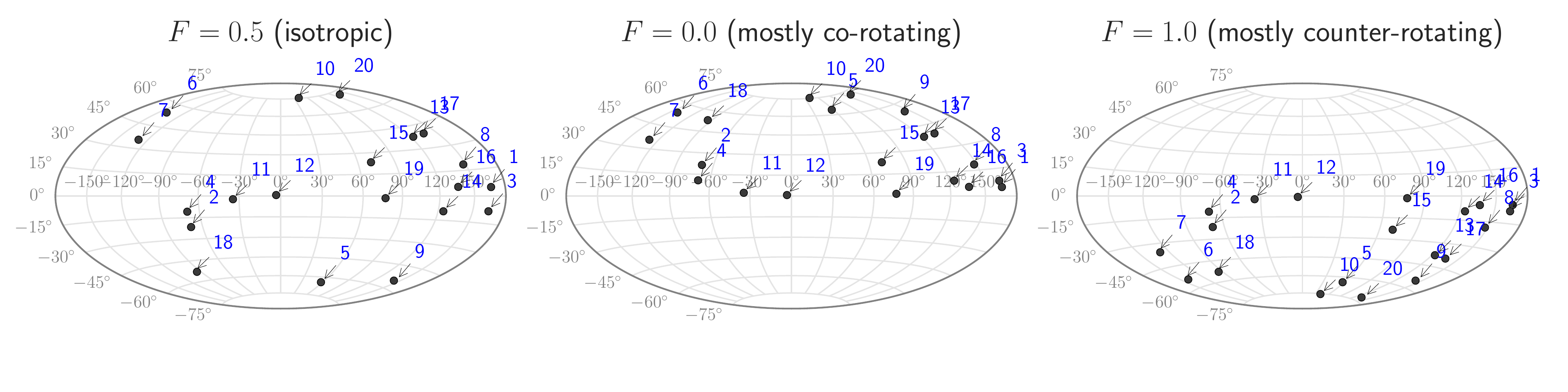}
    \end{tabular}
    \caption{Orientation of the angular momentum of each cloud on the
    sphere for the three $F$ distributions. In this representation $\vec{L}_{\rm bin}$
    points to the north pole. Each panel reports only the clouds that were
    integrated in that specific run, numbered by injection order. The top row is
    for {\runa} and the bottom row for {\runb}.}
    \label{fig:dist_orientation}
\end{figure*}

\begin{table}
    \centering
    \small
    \begin{tabular}{lrrrr}
        \hline
        & \multicolumn{2}{c}{\texttt{RunA}} & \multicolumn{2}{c}{\texttt{RunB}} \\
        \hline
        \textbf{Cloud} & \textbf{$t_{i}[P_0]$} & \textbf{$\Delta t_{i}[P_0]$}
        & \textbf{$t_{i}[P_0]$} & \textbf{$\Delta t_{i}[P_0]$} \\ \hline
            1   &   0.0   &  1.87  & 0.00   &  3.33    \\
            2   &   1.88  &  2.22  & 3.33   &  8.48    \\
            3   &   4.10  &  2.46  & 11.81  &  6.93    \\
            4   &   6.57  &  3.61  & 18.74  &  8.20    \\
            5   &  10.18  &  1.60  & 26.94  & 10.53    \\
            6   &  11.78  &  4.32  & 37.47  &  6.74    \\
            7   &  16.11  &  3.42  & 44.21  &  4.31    \\
            8   &  19.54  &  2.53  & 48.52  &  4.03    \\
            9   &  22.08  &  2.47  & 52.55  &  2.38    \\
            10  &  24.55  &  8.85  & 54.94  & 13.28    \\
            11  &  33.41  &  1.29  & 68.21  &  5.33    \\
            12  &  34.71  &  6.86  & 73.54  &  4.52    \\ \cline{2-3} 
            F1p0
            13  &  41.57  &  7.70  & 78.06  &  4.98    \\
            14  &  49.27  &  2.16  & 83.04  &  2.78    \\
            15  &  51.44  &  1.84  & 85.82  &  3.75    \\ \cline{2-3} 
            F0p0
            16  &  53.28  &  1.10  & 89.57  &  9.16    \\
            17  &  54.39  &  5.71  & 98.73  &  5.63    \\
            18  &  60.10  &  1.25  & 104.36 &  6.20    \\
            19  &  61.36  &  1.63  & 110.55 &  5.53    \\
            20  &  62.99  &  3.01  & 116.09 &  3.12    \\
            \cline{4-5} 
            21  &  66.01  &  5.94  & --- & ---    \\
            22  &  71.95  &  3.49  & --- & ---    \\ \cline{2-3} 
        \hline
    \end{tabular}
    \caption{Cloud injection times in {\runa} and
        {\runb}. For each cloud we report the time of injection (second and
        fourth columns) and the time to the next cloud injection (third and
        fifth columns). The two horizontal lines in column 2 and 3 identify the
        last cloud injection for {\runa} \F{1.0} (after cloud 12) and {\runa}
        \F{0.0} (after cloud 15), while {\runa} \F{0.5} reached cloud 22.
        Conversely all {\runb} configurations reached cloud 20.}
    \label{tab:times}
\end{table}

In the previous Section, we defined the main physical ingredients and
technical features of our simulations, we now proceed in detailing the initial
conditions of our set of runs. Our goal is to simulate a series of clouds
interacting with a central {\mbhb} sitting at the bottom of a fixed potential
well. The {\mbhb} is initially in the coordinate frame origin and has
a separation $a_0=1$ in code units. Each cloud is injected at a distance
$d=15$ and needs the specification of a time of injection, impact parameter
and direction of the orbit.

We construct two series of 30 events, drawing the time between each event from
a cumulative Gamma distribution with $k = 2$ and $\theta = 2.5 P_{0}$. We
made several draws from the Gamma distribution and picked two markedly
distinct sets. In the first set, hereinafter {\runa}, we perform an
``aggressive'' feed to the {\mbhb}, with an average $\Delta t\approx3P_0$
between events. On the other hand, in the second set, hereinafter {\runb},
clouds are fed to the {\mbhb} with an average $\Delta t\approx6P_0$ , allowing
the system more time to relax in between each infalling cloud. Note that when
scaled to our fiducial system, the above infall rates correspond to
0.4$\Msun$yr$^{-1}$ and 0.2$\Msun$yr$^{-1}$ entering the inner parsec
respectively, comparable to what is typically found in high resolution
simulations of gas-rich high redshift galaxies
\citep[see, e.g.][]{2017ApJ...836..216P},
and is in broad
agreement with observations of post-merger galaxies \citep[see
e.g.][]{SandersMirabel1996,NaabBurkert2001}.

Cloud impact parameters are drawn so that the periapsis passage is uniformly
distributed in the range $r_p\in[0,2a_0]$, if the {\mbhb} was replaced by
a single {\mbh} sitting at the origin of the coordinate system and the
potential well was ignored. A uniform periapsis distribution corresponds to
a standard impact parameter distribution $p(b)\propto b$ at infinity, when the
trajectory is dominated by gravitational focusing of the central object, as it
is the case in our simulations. The injection time and $r_p$ value of each
cloud for both {\runa} and {\runb} are shown in
Figure~\ref{fig:dist_pericentre}.

After specifying the time of injection and impact parameter, we define the
orbit of the incoming cloud by assigning a direction to its orbital angular
momentum, $\vec{L}_c$, with respect to $\vec{L}_{\rm bin}$. We explore three different
sets of initial conditions, defined by the fraction $F$ of counter-rotating
clouds interacting with the {\mbhb} \citep{DottiEtAl2013}: \begin{enumerate}
\item  \F{0.5}: $\vec{L}_c$ are randomly distributed on the sphere. In this
case, on average, 50\% of the clouds will be co-rotating and 50\% will be
counter-rotating with respect to the \mbhb. This is known as `chaotic'
accretion scenario and is visualised in the left panels of
Figure~\ref{fig:dist_orientation}, \item \F{0.0}: all clouds are co-rotating
with the \mbhb, i.e. they all have $L_{c,z}$ aligned to $\vec{L}_{\rm bin}$, as shown
in the central panels of Figure~\ref{fig:dist_orientation}, \item  \F{1.0}: all
clouds are counter-rotating with the \mbhb, i.e. they all have $L_{c,z}$
counter-aligned to $\vec{L}_{\rm bin}$, as shown in the right panels of
Figure~\ref{fig:dist_orientation}.  \end{enumerate}

We first generate 30 random clouds (\F{0.5} case) and obtain the \F{0.0} and
\F{1.0} cases by simply `mirroring' $\vec{L}_c$ with respect to the equatorial
plane, as shown in Figure~\ref{fig:dist_orientation}. The mirroring procedure
is crucial to single out the effect of co- and counter-rotation both on the
formation of gaseous structures and on the evolution of the binary, because it
allows us to consider systems that, besides the flipping of $\vec{L}_c$, are
otherwise identical. Note that once $\vec{L}_c$ and $r_p$ are specified, one
still has the freedom to rotate the orbit of the cloud within its orbital
plane. To define the orientation of the cloud orbit, we consider the
intersection of its orbital plane with the $x,y$ plane defined by our
coordinate system, and we place $r_p$ at an angle $\Theta$ randomly drawn in
the range $[0,2\pi]$. All the mathematical details of the generation of the
initial conditions are given in Appendix~\ref{Ap:ics}.

Although we generate 30 clouds for each set of initial conditions, we only show
initial conditions for 22 clouds for {\runa}, and 20 clouds for {\runb} in
Figure~\ref{fig:dist_pericentre} and \ref{fig:dist_orientation}. Due to time
constraints and necessary maintenance of the computer clusters employed
for the calculation, we were in fact only able to integrate {\runa} \F{0.0} up
to cloud 15, {\runa} \F{0.5} up to cloud 22 and  {\runa} \F{0.0} up
to cloud 12. All {\runb} were integrated up to cloud 20.
The full information about the initial condition of
each cloud, including initial positions and velocities, are given in
Appendix~\ref{Ap:ics}.

In summary, we generated two sets of runs, {\runa} and {\runb} defined by
different cloud impact parameters and injection times. For each of the runs we
considered three angular momenta distributions, \F{0.0}, \F{0.5} and \F{1.0},
for a total of six different sets of initial conditions. Each of the sets is
integrated at four single cloud resolutions: {\lowa}, {\lowb}, {\higha} and
{\highb}. In the following, we will concentrate on the results of the {\lowa}
simulations, which reached the larger number of clouds in the system. Runs at
higher resolutions are obviously slower; for example, only 4-5 clouds are
generally injected in the {\highb} case. Higher resolution runs are used as
benchmark for comparison and to assess convergence of the simulations.

To study the relaxation of the system after the infall of several clouds, we
also `forked' each of the {\lowa} runs after the injection of 5 and 10 clouds.
In practice, we ran in parallel two additional
sets of simulations in which the system was allowed to evolve unperturbed after
5 and 10 clouds interacted with the binary, to study the long term properties
of the relaxed system.

Commonly for this type of numerical investigations, a large computational
infrastructure was needed to handle the required set of runs and tests.

\section{Resolution and convergence tests}
\label{sec:tests}

The simulations presented in this
work feature complex dynamics of multiple clouds interacting with a {\mbhb}, it
is therefore important to test the incidence of our main numerical assumptions
on the evolution of the system, to keep the impact of spurious numerics under
control. The tree reconstruction frequency has been tuned to optimise angular
momentum conservation, as described in Section \ref{sec:sink_dyn}, and
softening has been chosen to guarantee a proper resolution of the gravitational
interaction between particles. The other numerical `degrees of freedom' are the
choice of $r_{\rm sink}$ and of the number of particles used to simulate each
cloud, $N_c$ (i.e. the `resolution of the simulation'). In this section we
check the robustness of our set-up against our choice of these parameters.

\subsection{$r_{\rm sink}$ value and accretion convergence}
\label{sec:accretion}

\begin{figure}
    \resizebox{\hsize}{!}
            {\includegraphics[clip]{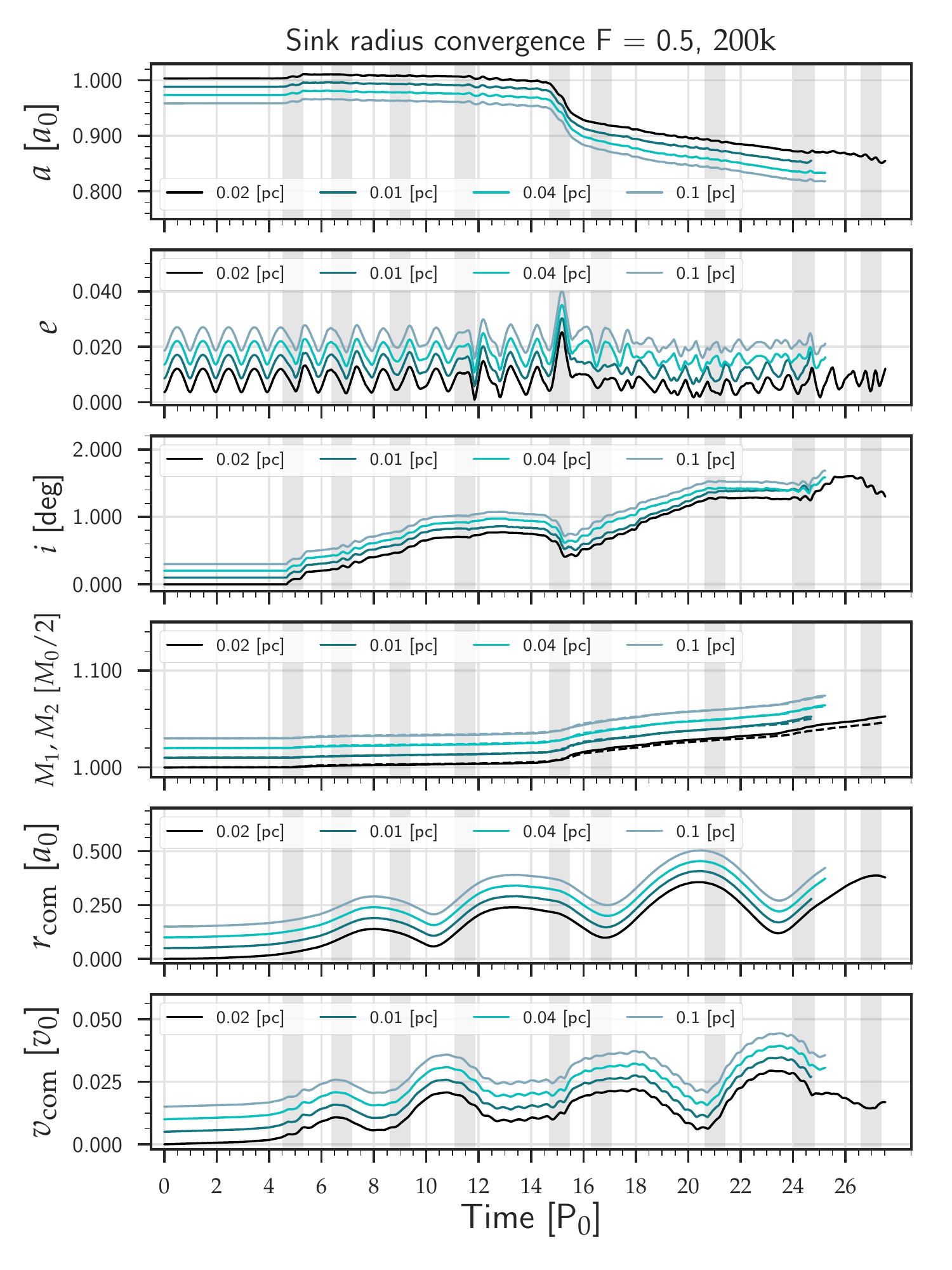}}
    \caption{Time evolution of the relevant {\mbhb} parameters in {\runa}
    \F{0.0}, with {\lowb}, for different values of $r_{\rm sink}$, displayed in
    simulation units. In each plot, the black line is for $r_{\rm sink}=0.1a_{0}$
    (the adopted default values), while different shades of blue are for
    $r_{\rm sink}=0.05a_{0}, 0.2a_{0}, 0.5a_{0}$.
    Note that lines have been progressively shifted upwards for clarity, since they
    would otherwise almost perfectly overlap. Grey vertical stripes indicate the
    `arrival time' of each new cloud, i.e. the time of first periastron passage in
    its orbit around the {\mbhb}. In the mass panel (fourth from the top), solid
    and dashed lines represent $M_1$ and $M_2$ respectively, normalised to their
    respective initial values.}
    \label{fig:accr}
\end{figure}

Ideally, a particle will be accreted when it approaches the {\mbh} at about
$r=6GM/c^2=3R_{\rm S}$, where $R_{\rm S}$ is the Schwarzschild radius. If we
consider our fiducial system, this distance is $\approx 5\times 10^{11}$ cm,
equivalent to $2\times10^{-6}$ in code units. It is clear that a realistic
condition for particle accretion is beyond any feasible resolution in our
numerical scheme. Thus a fictitious sink radius is introduced by hand, as
explained in Section \ref{sec:sink_dyn}. The numerical value of $r_{\rm sink}$
is set \emph{ad-hoc} for numerical convenience. To test its impact on the dynamics of
the system we ran four otherwise identical simulations with $r_{\rm sink}=0.1$
(standard model), $0.05$, $0.2$ and $0.5$ in code units thus
spanning an order of magnitude. For these test simulations we considered
{\runa} \F{0.0}, with {\lowb} resolution.

Figure~\ref{fig:accr} shows the evolution of the key parameters describing the
evolution of the {\mbhb} in the four runs, evolved for about 25 initial binary
orbital periods, sufficient to follow the strong dynamical interaction with
eight subsequent clouds. Results match so well across the runs that we had to
offset the lines, otherwise they would overlap almost perfectly. The value of
$r_{\rm sink}$ does not appreciably impact any of the {\mbhb} parameters, not
even the eccentricity evolution, which depends on a fine balance between energy
and angular momentum exchanges, and is therefore sensitive to minor
fluctuations in the dynamics. Note that the pool of interacting clouds span
a large dynamical range, including clouds with $r_p<a_0$ (clouds 5 and 8, see
figure~\ref{fig:dist_orientation}) whose dynamics might in principle be
severely affected by an improper treatment of  $r_{\rm sink}$.

Critically, the evolution of the two {\mbh} masses is independent on
$r_{\rm sink}$, which indicates that gas accretion is not affected by its
unphysically large value. This is because of the conditions spelled in
\S\ref{sec:sink_dyn},
whereby particles are required to be {\it bound} to the {\mbh} for being
accreted. In practice only particles that settle into orbits enclosed in the
{\mbh} Roche Lobe can be accreted. These particles form eccentric (either
transient or persistent) mini-disks that are continuously perturbed
by infalling material and are swiftly drained into the sink.
Therefore, setting a smaller sink
radius only causes a small delay in the time at which a particle is
recorded as accreted. Note that this does not mean that all particles crossing
$r_{\rm sink}$ will be accreted in reality, as we will discuss in the next
section.

\subsection{Robustness against $N_c$}
\label{sec:resolution}

\begin{figure}
    \resizebox{\hsize}{!}
    {\includegraphics[scale=1,clip]{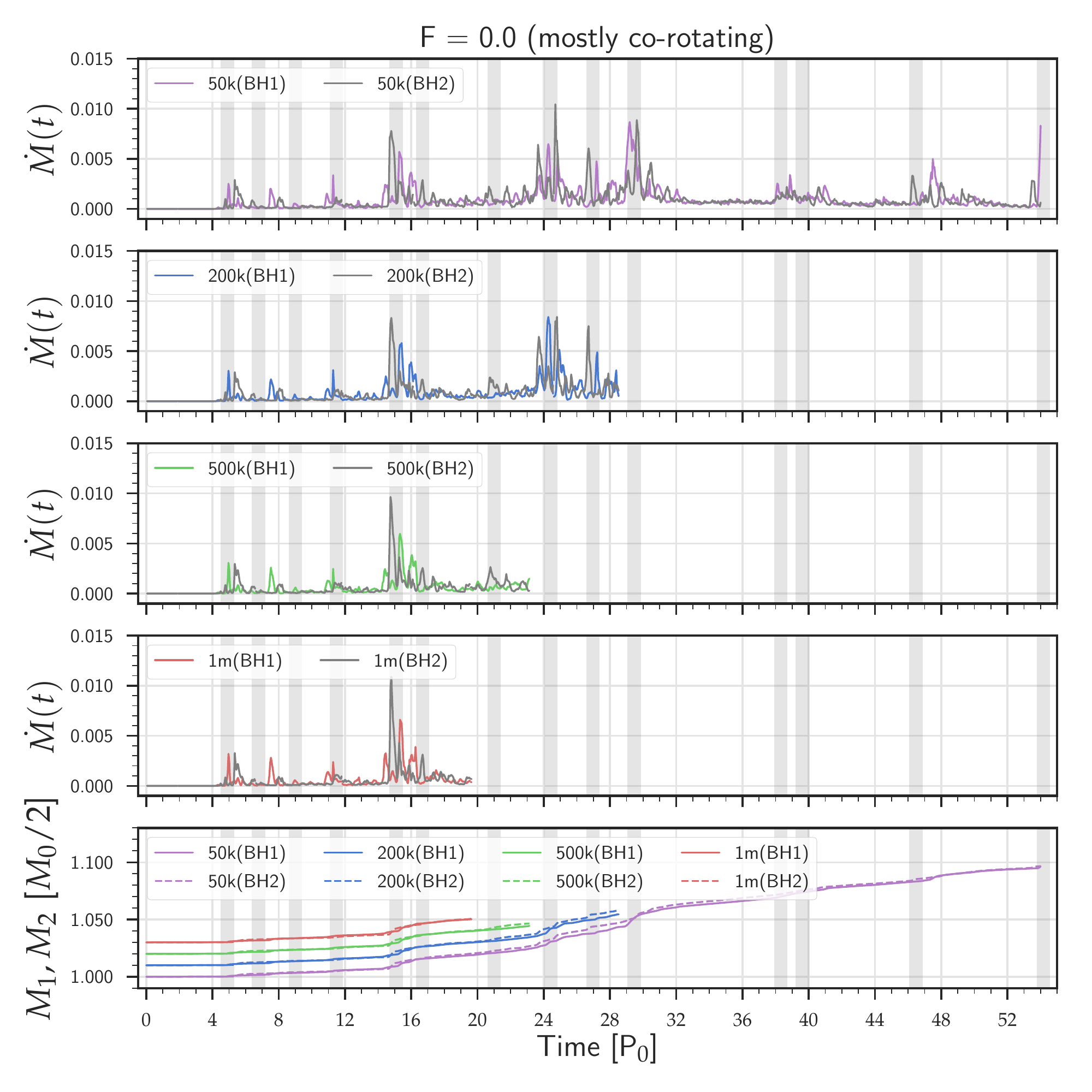}}
    \caption{Accretion rate for {\runa} \F{0.0} for all the investigated resolutions: {\lowa},
    {\lowb}, {\higha} and {\highb}, from top to bottom. The two lines on each panel
    represent accretion onto each {\mbh}. Orange vertical lines mark the moment
    each cloud is injected into the system, and grey vertical stripes the time of
    the first close interaction with the {\mbhb}. Rates are displayed in simulation
    units $[M_0/P_0]$. The bottom panel shows the mass evolution of each MBH
    at all resolutions. Lines have been progressively shifted upwards for clarity.}
    \label{fig:mdot}
\end{figure}

\begin{figure*}
    \resizebox{\hsize}{!}
    { \includegraphics[scale=1,clip]{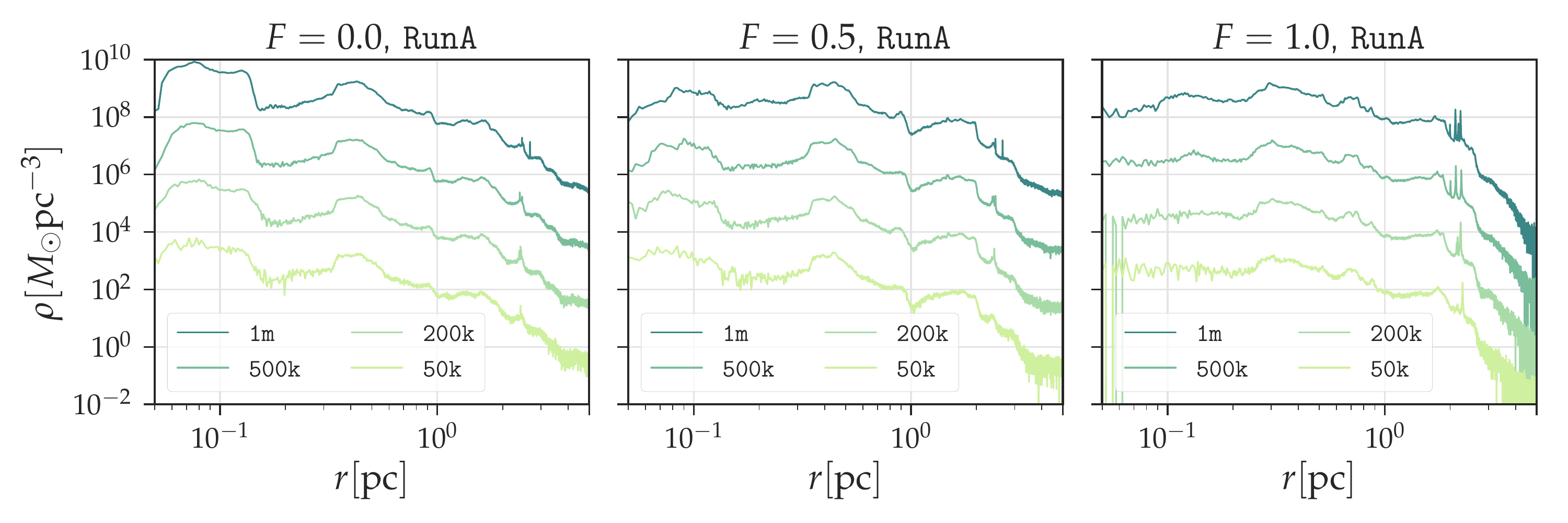}}
    \caption{Angle-averaged density profile at $T=10.2 [P_0]$. In each panel,
    from bottom to top, lines corresponds to resolution {\lowa}, {\lowb}, {\higha}
    and {\highb}, and have been shifted upwards to ease comparison. The three
    different $F-$distributions are represented from left to right, as indicated in
    figure. The $y-$axis normalisation of the {\lowa} resolution is fixed to the
    scale of our fiducial system, $M_{\rm bin}=10^6\msun$.} \label{fig:vol_density_5}
\end{figure*}

\begin{figure*}
    \resizebox{\hsize}{!}
        {\includegraphics[scale=1,clip]{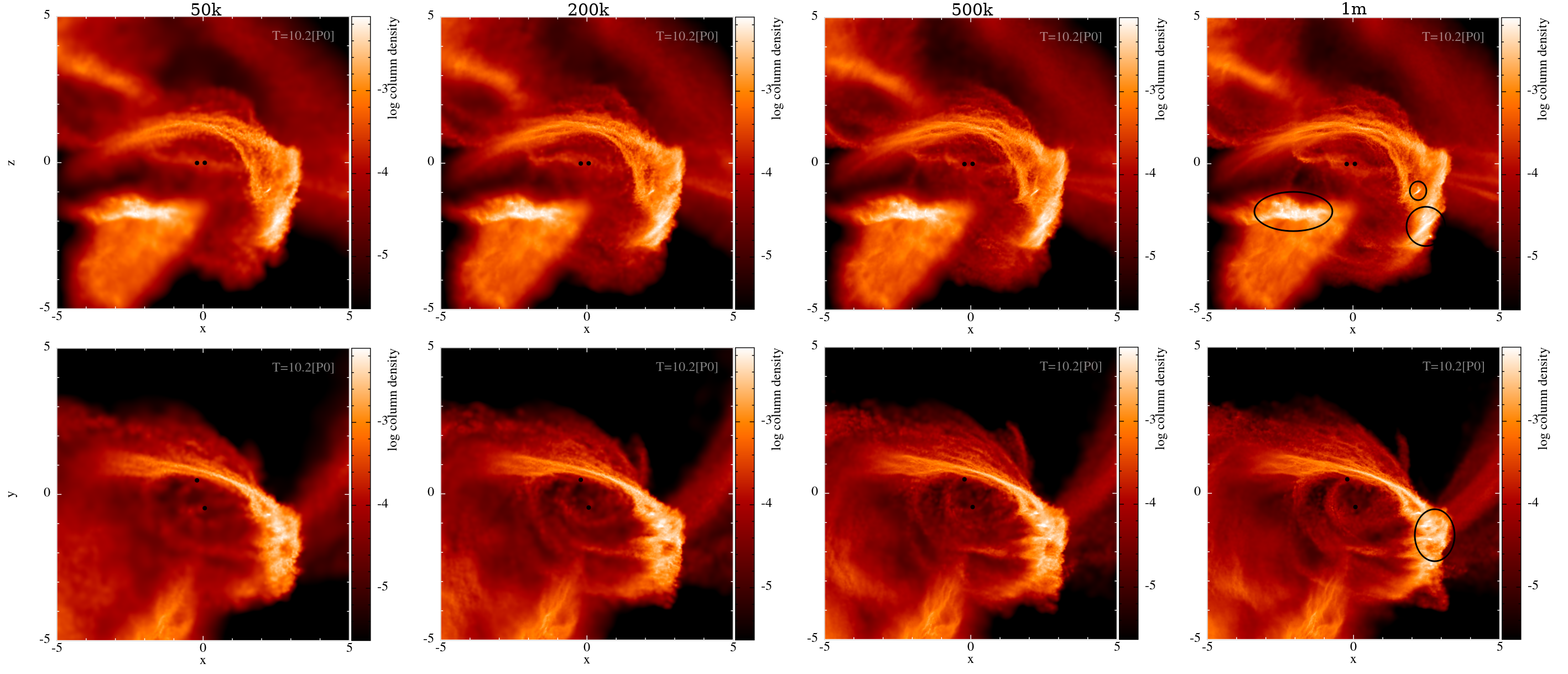}}
    \caption{Column density rendering of {\runa}, \F{1.0} at $T=10.2
        [P_0]$. Each column represents a cloud resolution, {\lowa}, {\lowb},
        {\higha} and {\highb} from left to right. In each column, the top and
        the bottom panels represent views in the $x-z$ and $x-y$ plane
        respectively.}
    \label{fig:vol_density_render}
\end{figure*}

It is obvious that the level of dynamical detail that traceable in a numerical
simulation critically depends on its resolution, and the SPH technique is no
exception to this rule. Simulations shall be performed with a number of
particles sufficient to resolve the physical features of interest in an
$N-$independent fashion. Of particular interest for this work are accretion
onto the {\mbhb} and the distribution of non accreted gas around the binary. To
test that our simulations are `well behaved' we ran each of them at the four
particle resolutions {\lowa}, {\lowb}, {\higha} and {\highb} particles, thus
spanning a range of 20.

\begin{figure*}
    \centering
    \begin{tabular}{c}
        \includegraphics[width=0.8\textwidth]{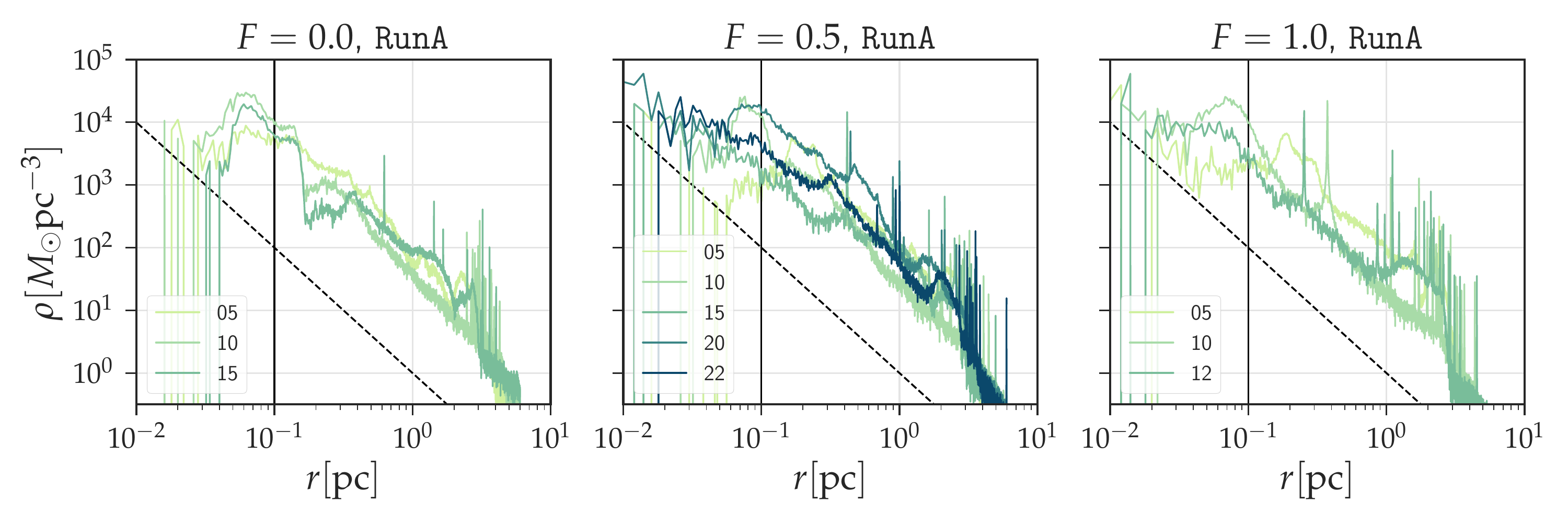}\\
        \includegraphics[width=0.8\textwidth]{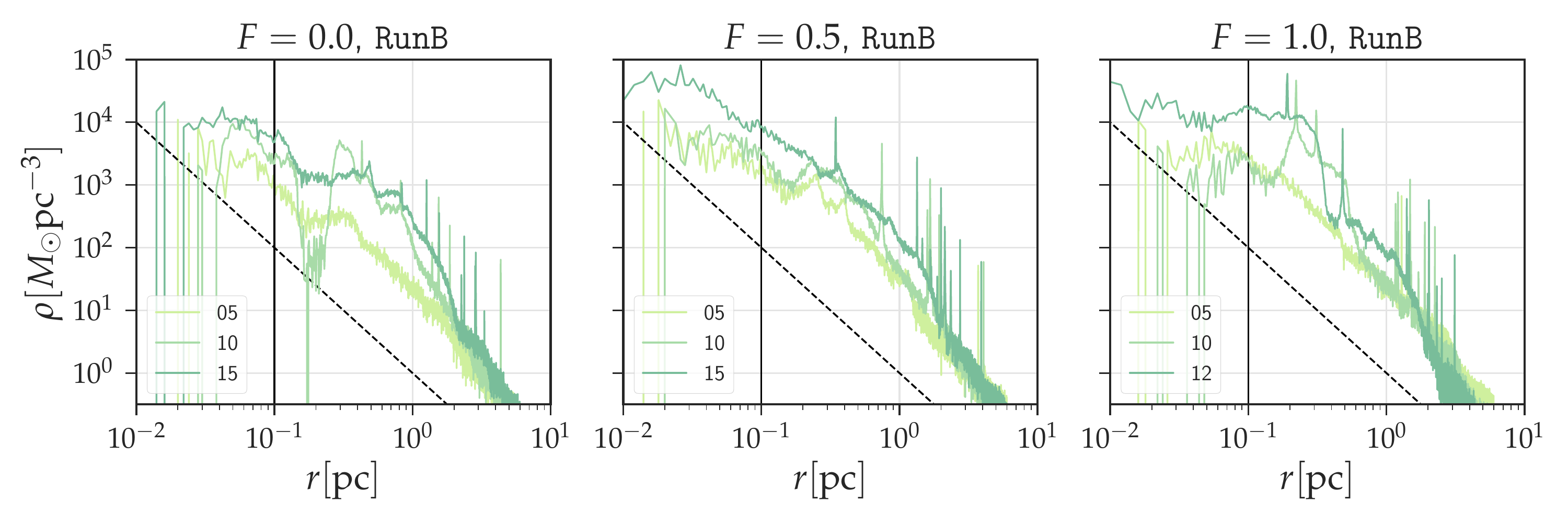}
    \end{tabular}
    \caption{Angle-averaged mass density profile for all {\lowa} runs
    (as indicated in each panel) at different snapshots, in physical units, scaled
    to our fiducial system. Curves are labelled by the number of clouds present in
    the system, and represent the density profile at the last recorded snapshot
    before the subsequent cloud is injected in the system (e.g.: 05 represents the
    status of the system at the last snapshot before adding the 6th cloud). In each
    panel, the dashed line refers to a $\rho\propto r^{-2}$ profile, for
    comparison.
    The solid vertical black
    line shows the initial position of each of the {\mbhb} components.}
    \label{fig:vol_density}
\end{figure*}

In Figure~\ref{fig:mdot}, we show the accretion rate on each individual {\mbh}
in the case {\runa} \F{0.0}, for all values of $N_c$. Accretion is highly
variable, showing prominent intermittent spikes in correspondence to the
arrival times of clouds with small impact parameters. Note that, at peak,
$dM/dt> 0.005 P_0^{-1}$ on each individual {\mbh}. Converted to our fiducial
system this is about $0.6\msun {\rm yr}^{-1}\approx 60\dot{M}_{\rm Edd}$. The
actual fate of the gas during these high accretion episodes is unclear. Photon
trapping might allow gas to be accreted at super-Eddington rates
\citep[e.g.][]{1988ApJ...332..646A,2005ApJ...628..368O}, or alternatively,
radiation pressure might cause the expulsion of the majority of the gas in
powerful winds, as observed in
\citep[e.g.][]{2010ApJ...719..700T,2015Natur.519..436T}. As discussed in
\cite{Goicovic2017}, the fate of the gas in itself has only a minor impact on
the dynamical evolution of the system, which is driven more by gas {\it
capture} from the {\mbhb}, rather than gas {\it accretion}. Winds can, however,
strongly interact with the surrounding infalling clouds, affecting their
dynamics. We caution that this effect is not captured in our simulations.
Nevertheless, Figure~\ref{fig:mdot} shows that, despite minor differences in
the definition of the accretion peaks, neither the accretion rate nor the total
mass growth of the {\mbh}s have an appreciable dependence on the amount of
particles used in the simulation.

To test the dependence of the gas distribution on $N_c$, we show in
Figure~\ref{fig:vol_density_5} the angle-averaged gas density as a function of
radius for all {\runa} at $T=10.2P_0$, after the disruption of the third cloud.
The density is displayed in physical units, by scaling the results to our
fiducial system ($M_{\rm bin}=10^6\msun$). Density profiles are equivalent at all
resolutions, even though they are noisier in the {\lowa} and {\lowb} runs, due
to the smaller $N_c$. Increasing the number of particles allows to better
capture fine details in the gas distribution. For example in the \F{0.0} case
(left panel), the sharp density drop around 0.15pc is better resolved in the
{\highb} run. Likewise, in the \F{1.0} case (right panel) the larger amount of
particles in run {\higha} and {\highb} allows the resolution of two density
peaks around 2.2pc, which are blended into a single peak in runs {\lowa} and
{\lowb}. Nevertheless, differences are minor, and the overall structure of the
gas distribution is preserved across resolutions.

A visualisation of the 3-D particle distribution is shown in
Figure~\ref{fig:vol_density_render}, where snapshots of the \F{1.0} case at
$T=10.2P_0$ are shown in the $x-y$ and $x-z$ plane. The overall gas
distribution is exquisitely consistent at all resolutions, even though
structures appear slightly blurrier going from the right to the left. The
aforementioned difference in the density peaks at 2.2pc, is due to the regions
highlighted with black circles on the rightmost column. Looking at this
critical areas, and moving left in the panel sequence (going down in
resolution), the dense areas become less defined eventually blending the fine
structures in larger clumps. These differences, however, do not affect the
overall dynamical evolution of the systems and are relevant only in the
determination of the statistics of dense clumps prone to star formation. Since
this specific investigation is beyond the scope of the current work, we deem
the {\lowa} runs sufficiently accurate for our purposes.

\section{Results}
\label{sec:results}

In this section we present the main results of the simulations concerning the
formation and evolution of gaseous structures around the {\mbhb}
\footnote{Animations of all the simulations can be seen in
\url{http://multipleclouds.xyz/movies/}}.
In a companion paper (\paperbhb) we focus on the evolution of the
{\mbhb}. We will describe the general outcome of both {\runa} and {\runb}, in
order to make comparisons between the two. When a specific aspect is
investigated in more depth, we consider {\runa} as default case. We also ran an
extra suite of simulations, that we call {\runc}, mixing the cloud
configuration of {\runa} and the time distribution of {\runb}. Some relevant
results for this extra suite of runs is presented in
Appendix~\ref{appendix:runc}. In the figures of this Section, relevant
quantities are displayed in physical units, scaled to our fiducial system (i.e.
a {\mbhb} with initial mass $M_{\rm bin}=10^6\msun$ and initial separation $a=0.2$pc),
whereas run snapshots are shown in simulation units ($M_{\rm bin}=a=1$).

\subsection{Overall evolution of the gas distribution}

\begin{figure*}
    \centering
    \begin{tabular}{ccc}
        \includegraphics[width=0.3\textwidth]{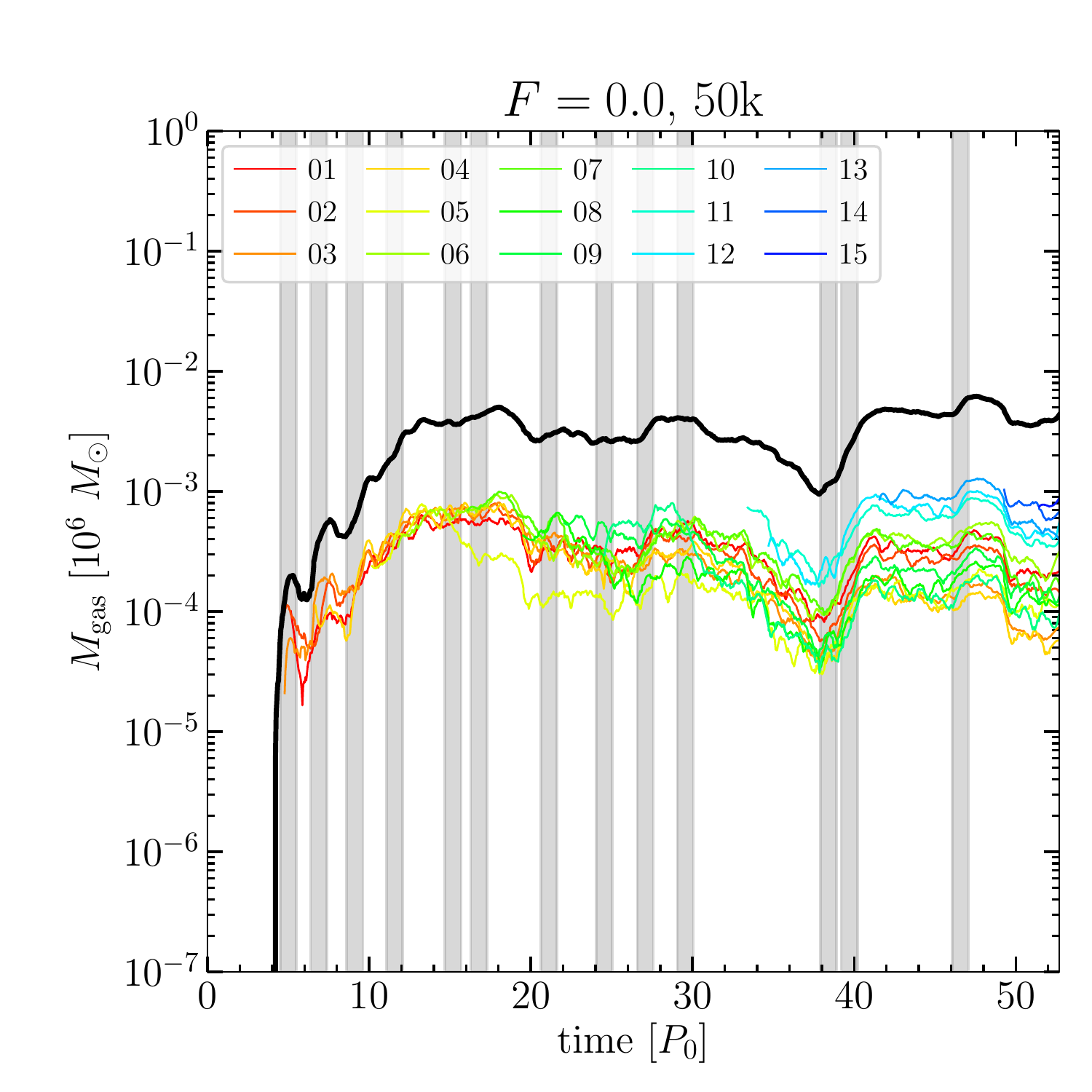}
      & \includegraphics[width=0.3\textwidth]{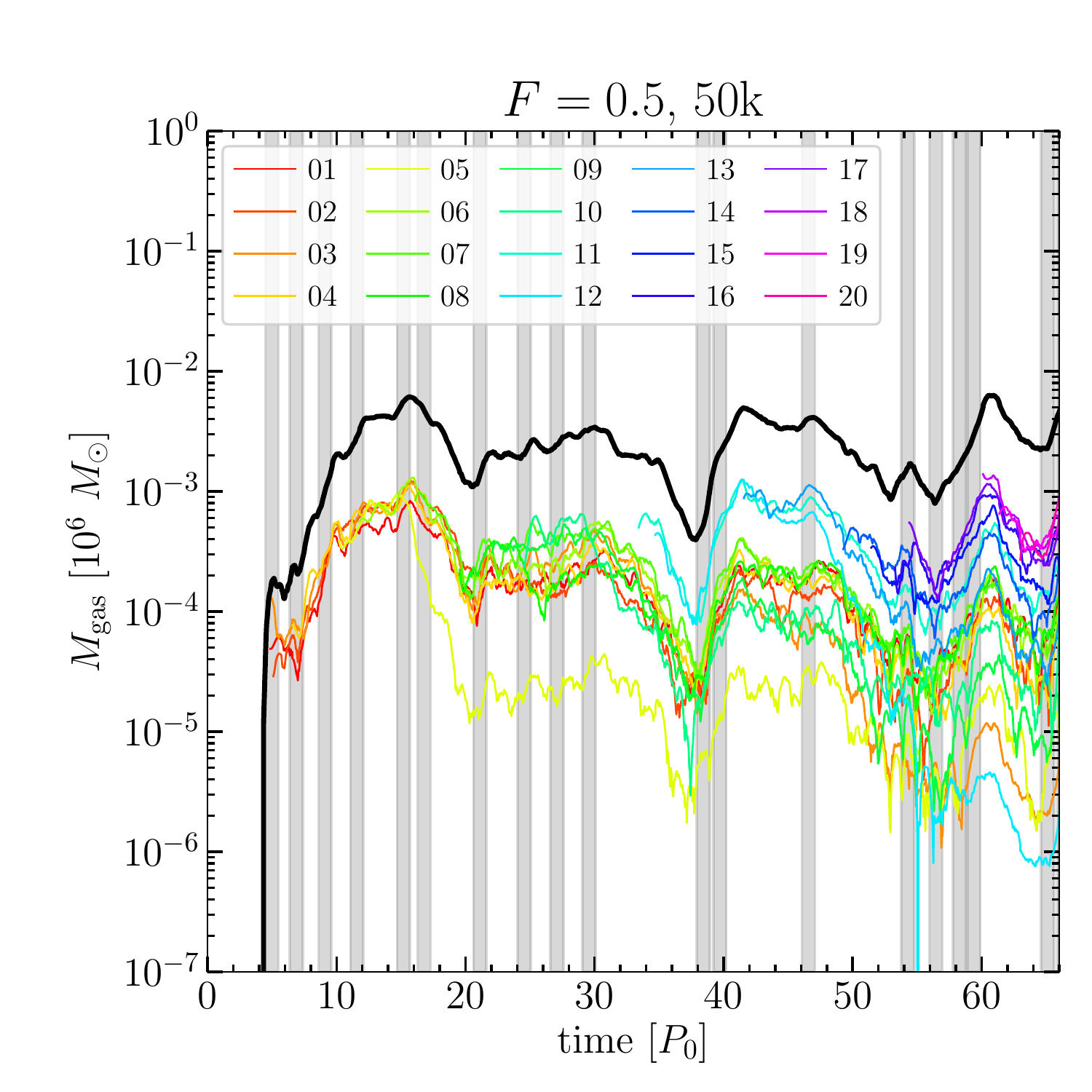}
      & \includegraphics[width=0.3\textwidth]{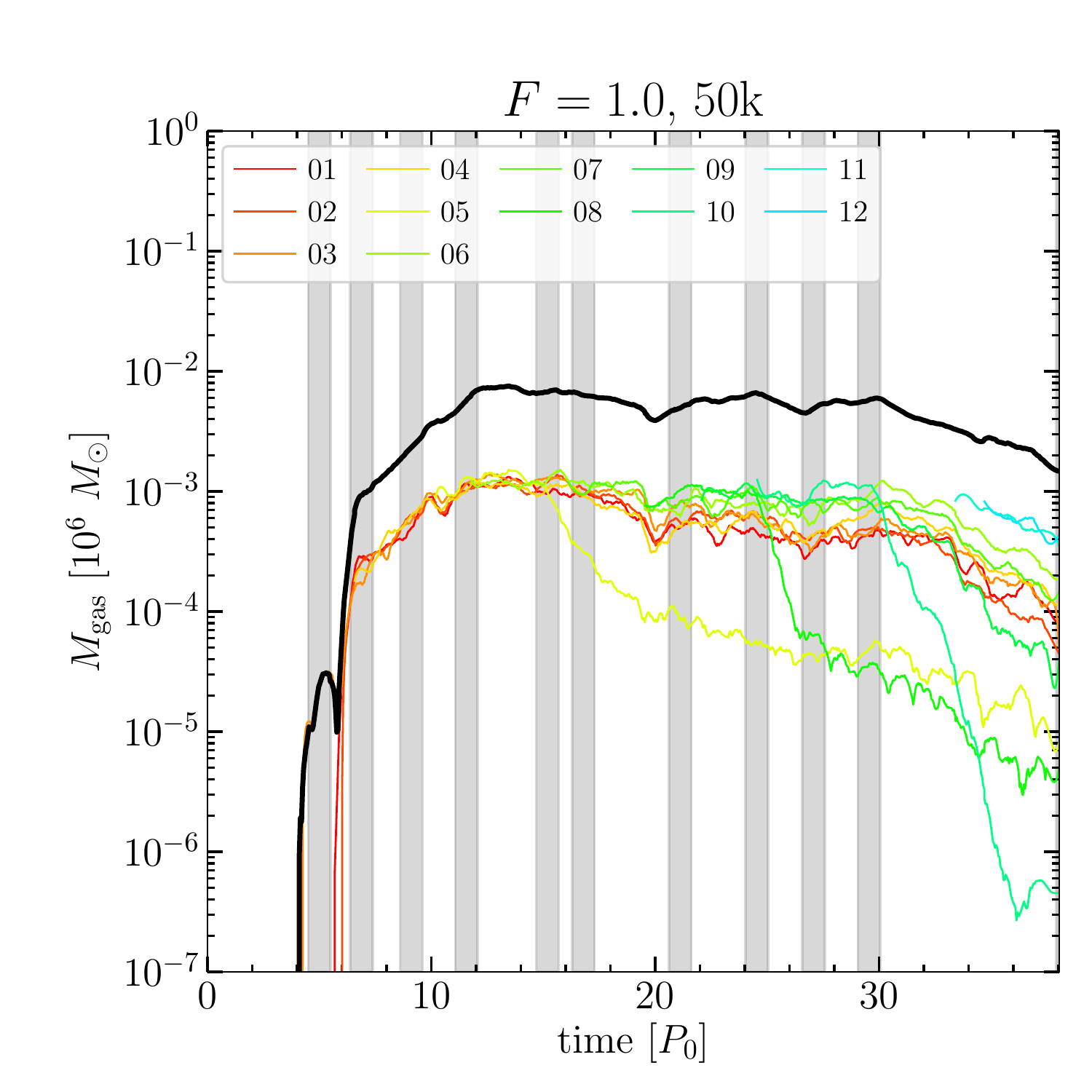}\\
        \includegraphics[width=0.3\textwidth]{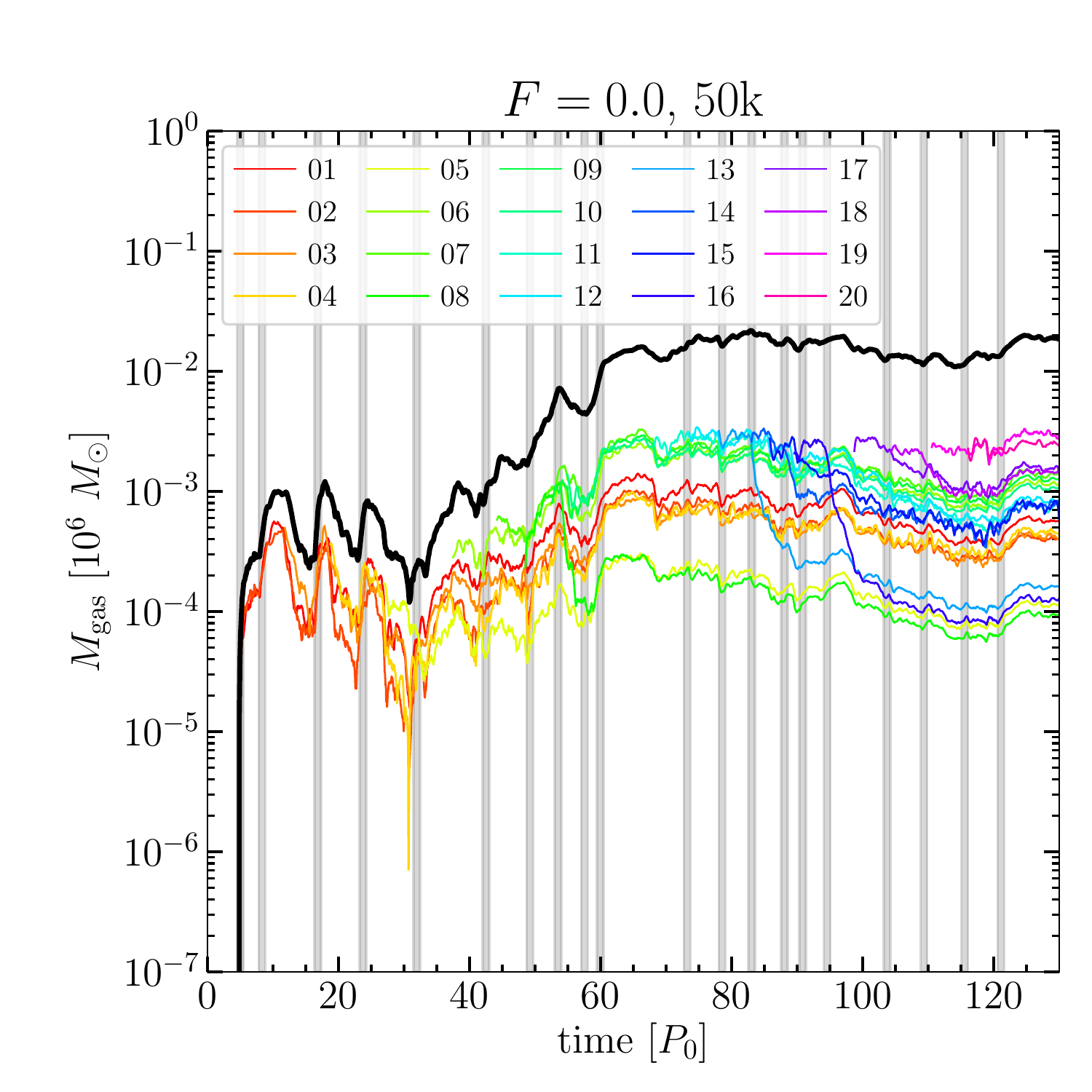}
      & \includegraphics[width=0.3\textwidth]{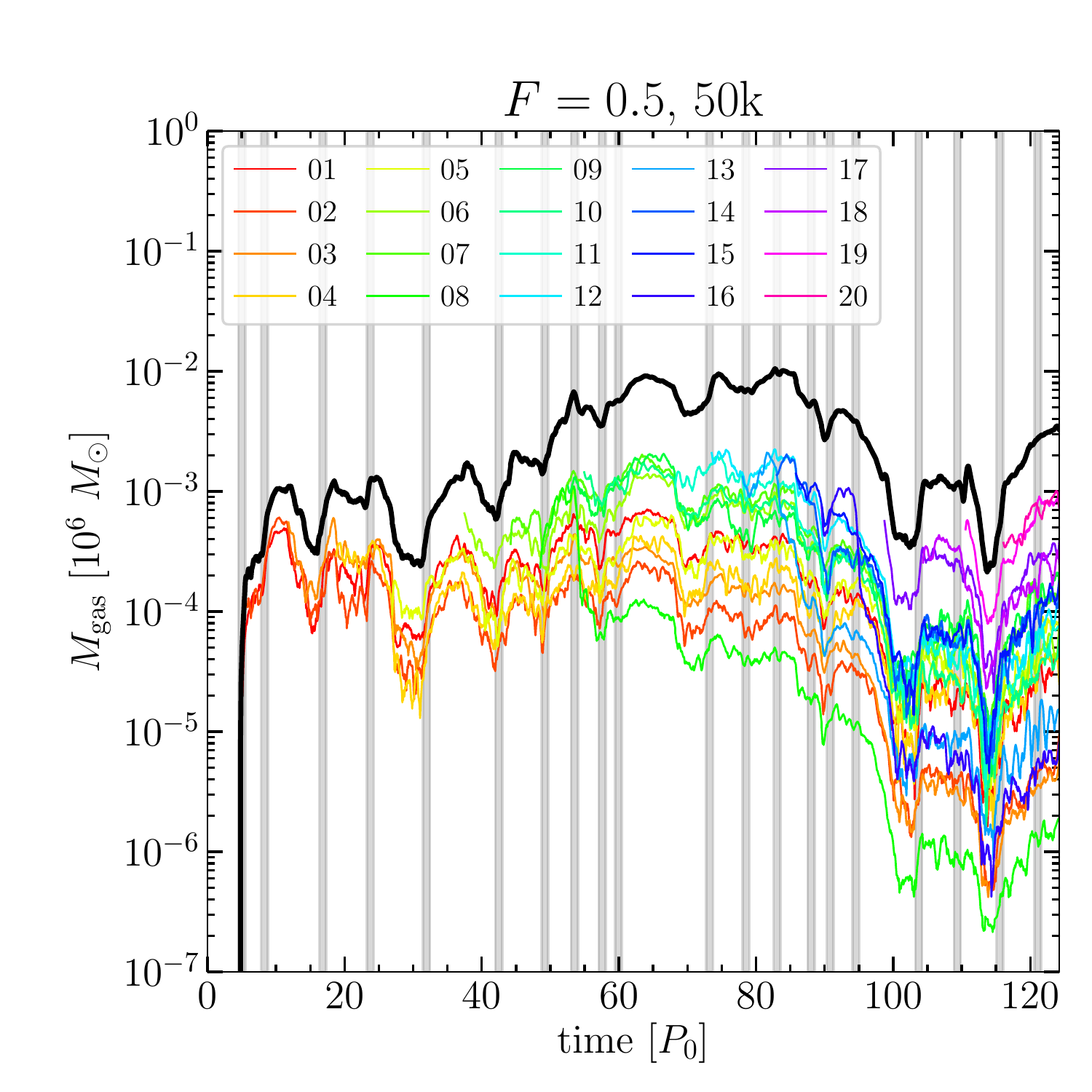}
      & \includegraphics[width=0.3\textwidth]{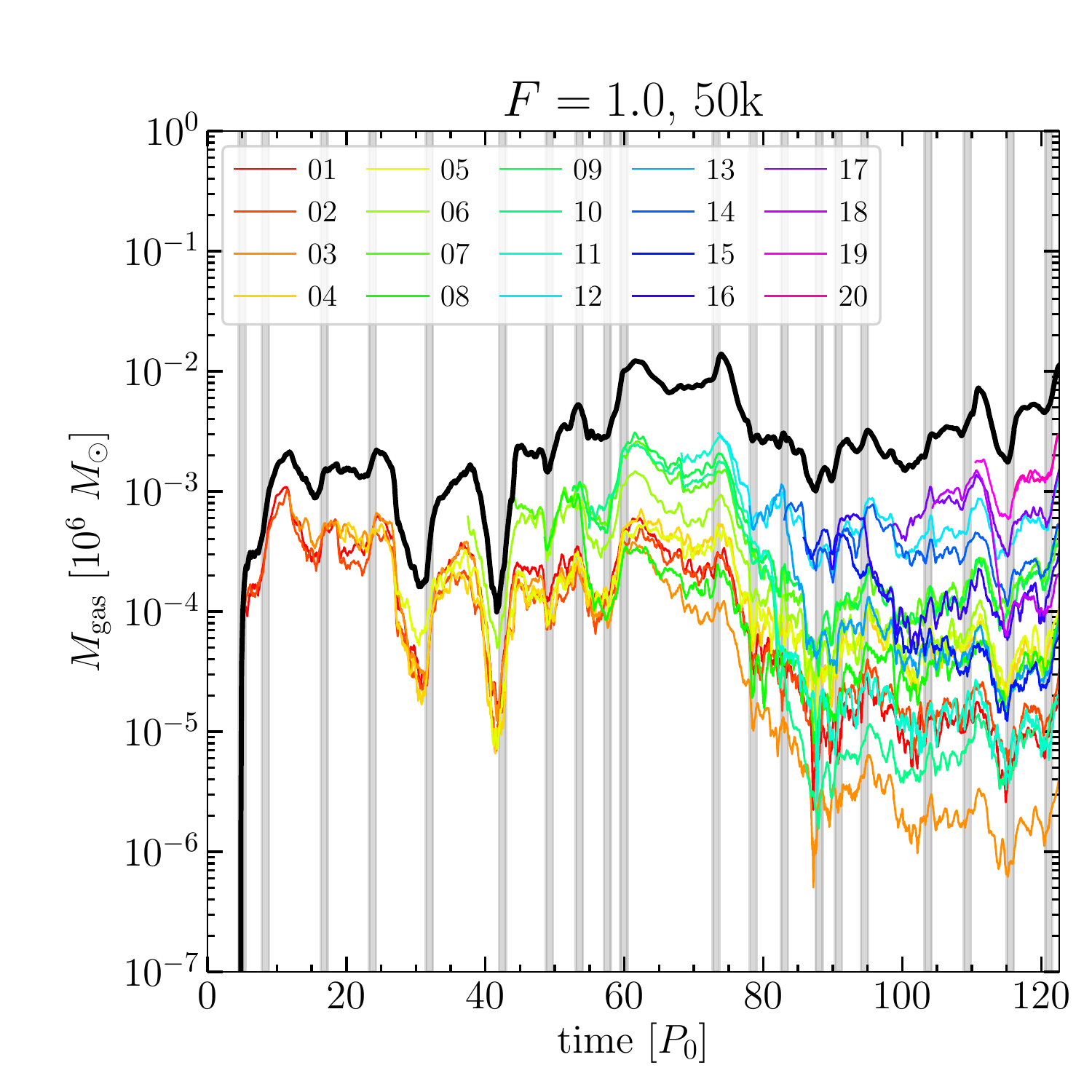}
    \end{tabular}
    \caption{Mass evolution of the circumbinary structure for the
        models $F = 0.0$ (left), $F = 0.5$ (centre) and $F = 1.0$ (right). The first
        row represents {\runa} while the bottom row represents {\runb}. The different
        coloured lines are the mass contribution from each infalling cloud while the
        black line is the sum of the contribution from all clouds}
    \label{fig:mass_disc}
\end{figure*}

We first look at the evolution of the (angle averaged) gas density profile as
a function of distance to the {\mbhb} centre of mass.
This is shown Figure~\ref{fig:vol_density} as runs advance and more clouds are
added to the system.  {As a general trend, we see that in all cases the gas
density outside the binary orbit tends to follow a $\rho\propto r^{-2}$
distribution, even though the profile is necessarily time dependent and there
is a large scatter. In the long run, this has to be expected; by throwing gas
at the binary from all directions, the resulting envelope will be almost
at rest with respect to the binary centre of mass. One can then estimate
the Bondi radius as $R_B=2GM/c_s^2$. For our fiducial system, the gas
temperature is $T=100$ K, resulting in $c_s\approx 2$ km s$^{-1}$. This means
that the Bondi radius is several tens of parsecs i.e. much bigger than the
domain of the simulation. We are hence injecting within the Bondi
radius of the binary gas at a roughly constant rate, which is therefore
expected to settle into an $r^{-2}$ density profile.}

The ``humps'' -- observed for example in the middle top-panel on the 22th cloud
line around $5$pc, or bottom-right panel at $0.5$pc and $1$pc for the 10th and
15th cloud lines, respectively -- correspond to new clouds infalling into the
system, and are not properties generated by bound material that is falling
back to the binary after cloud disruption. Each new infalling cloud will
generate a ``hump'' that will move from right to left until is disrupted by the
{\mbhb}. Conversely, the sharp lines accumulating at $R>1$pc are due to high
density clumps formed during the phase of cloud compression in the interaction
with the {\mbhb} and then ejected at large distances.

The figure also highlights few interesting features specific to each of the
$F-$distributions. On the \F{0.0} panels (left column) we can observe a clear
dip in the density profile around 0.2-0.3pc, comparable to the binary orbital
separation, building up over time as more clouds get into the system. This is
a clear indication that the action of the {\mbhb} is carving a cavity in the
gas distribution. In fact in the \F{0.0} case, clouds are mostly co-rotating
and we expect a lot of gas will re-arrange in a co-rotating circumbinary
structure. Lindblad resonances are then expected to carve a hole in the gas
distribution of a size of $\approx2a$ \citep{1994ApJ...421..651A}. The large
overdensities at the {\mbh} location are instead indicative of prominent
{\mds}, that are also expected to form in the co-rotating case
\citep{GoicovicEtAl2016}. Note, moreover, how the density just outside the
cavity in {\runb} builds up with time much more prominently than in {\runa},
which is a sign that a more massive circumbinary disc is being built in the
former case (as we will see below, cf Figure~\ref{fig:mass_disc} and
Figure~\ref{fig:circumbinaryAB_splash}). Conversely, clouds are mostly
counter-rotating in the \F{1.0} case (right panels), and Lindblad resonances do
not operate. No steady dip is in fact observed in this case, however prominent
transient overdensities can be seen forming around 0.2pc, which are indicative
of the formation of compact dense rings that get disrupted in the interaction
with new incoming clouds (again, see example in
Figure~\ref{fig:circumbinaryAB_splash}). The isotropic nature of the \F{0.5}
runs can be also appreciated (central panels), which shows a relatively smooth
and steady shape. In the following, we will examine in detail the evolution of
these gas structures, paying particular attention to the formation of {\mds},
circumbinary discs and rings.

\begin{figure*}
    \centering
    \begin{tabular}{c}
        \includegraphics[width=0.9\textwidth]{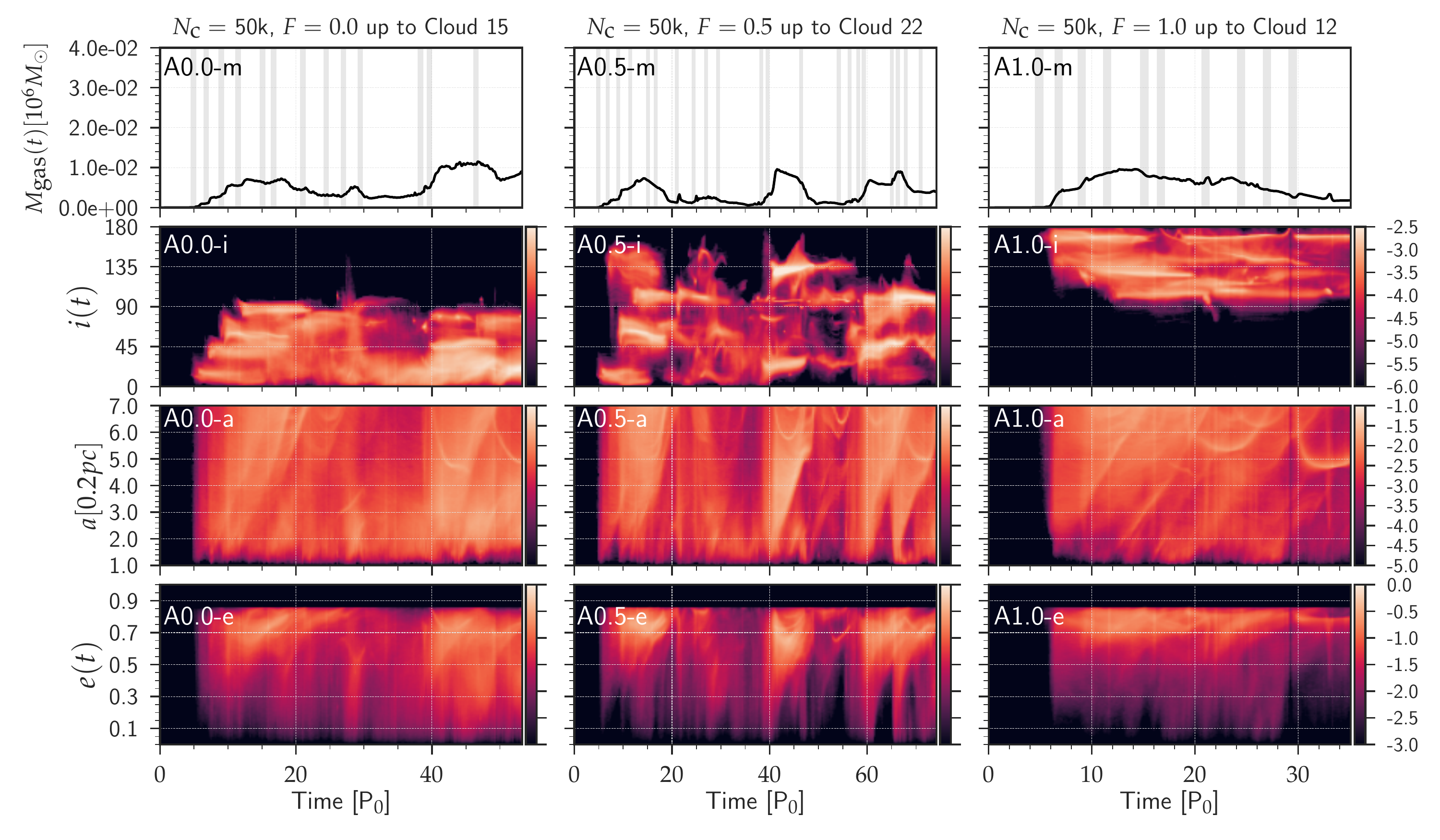}\\
        \includegraphics[width=0.9\textwidth]{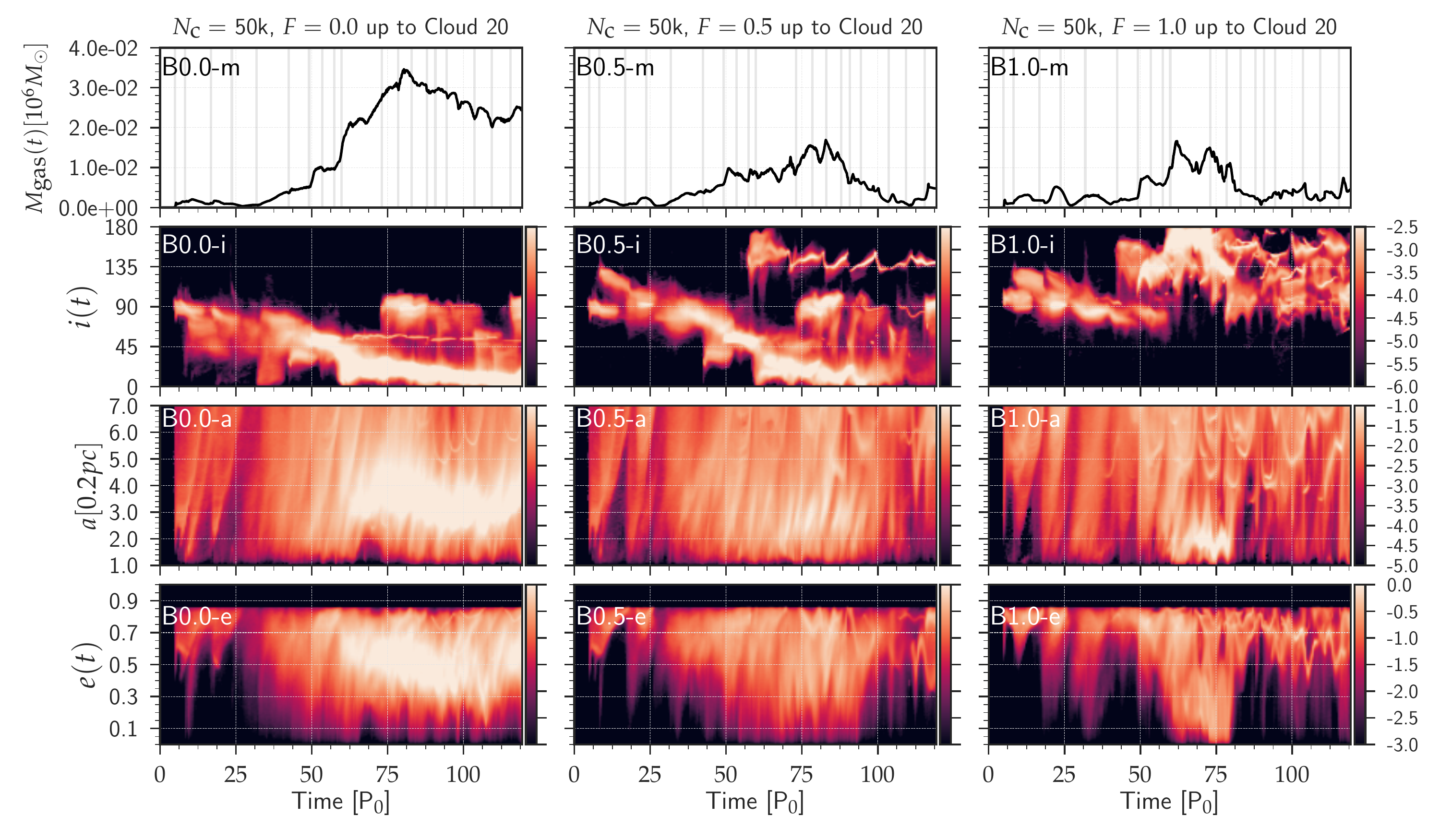}
    \end{tabular}
    \caption{Time evolution of the circumbinary structure in each of the
    simulations. The three columns, left to right, are for the \F{0.0}, \F{0.5} and
    \F{1.0} scenarios; the top $3\times4$ grid of plots is for {\runa} and the
    bottom for {\runb}. In each grid, the top row shows the time evolution of the
    total mass of the structure, $M_{\rm gas}$. All the other rows show in
    log-scale the time evolution of selected particle distributions, normalised so
    that the integral of any vertical slice of any panel gives the total mass
    $M_{\rm gas}$ at that specific time. With such normalisation, the second row
    represents $dM_{\rm gas}/di$, the third row $dM_{\rm gas}/da$ ($a$ is displayed
    in units of initial binary separation) and the forth row $dM_{\rm gas}/de$. The
    colour gradient is displayed in log-scale, as indicated by the bars at the far
    right of each row. Grey lines on the $M_{\textrm{gas}}$ panels, indicate the
    time of arrival of infalling clouds.
    }
    \label{fig:circumbinaryAB}
\end{figure*}

\subsection{Circumbinary structures}
\label{sec:circumbinary}

Following the infall of each cloud, some of the non-accreted material will
remain bound to the binary, forming structures around it, as well as around
each {\mbh}. Previous work by \cite{GoicovicEtAl2016} studied the impact of the
orbital configuration of the infalling cloud on the formation of such
structures, showing a rich phenomenology depending on the initial orbital
inclination relative to the {\mbhb} plane and impact parameter. In particular
they found that the binary is generally unable to change significantly the
orientation of the gas, which produces discs that follow the initial cloud's
inclination.

\begin{figure*}
    \centering
    \begin{tabular}{ccc}
        \includegraphics[width=0.3\textwidth]{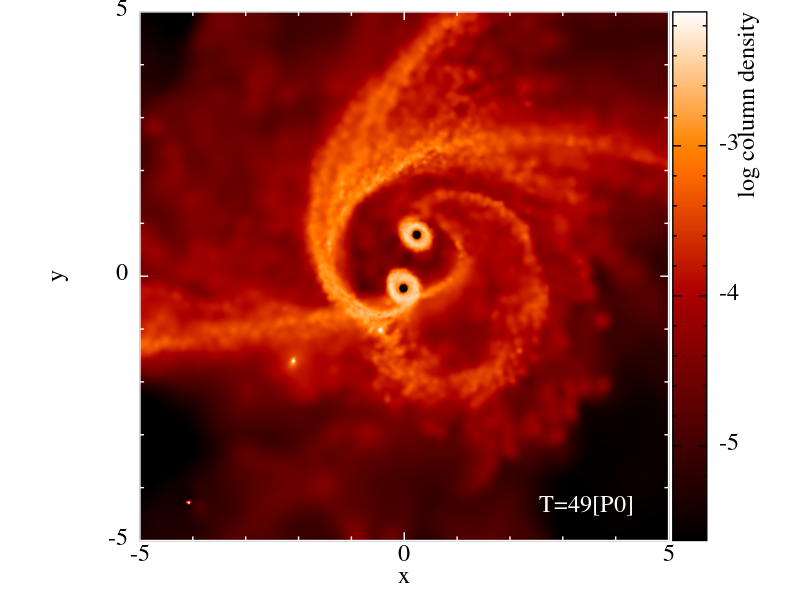}
        &\includegraphics[width=0.3\textwidth]{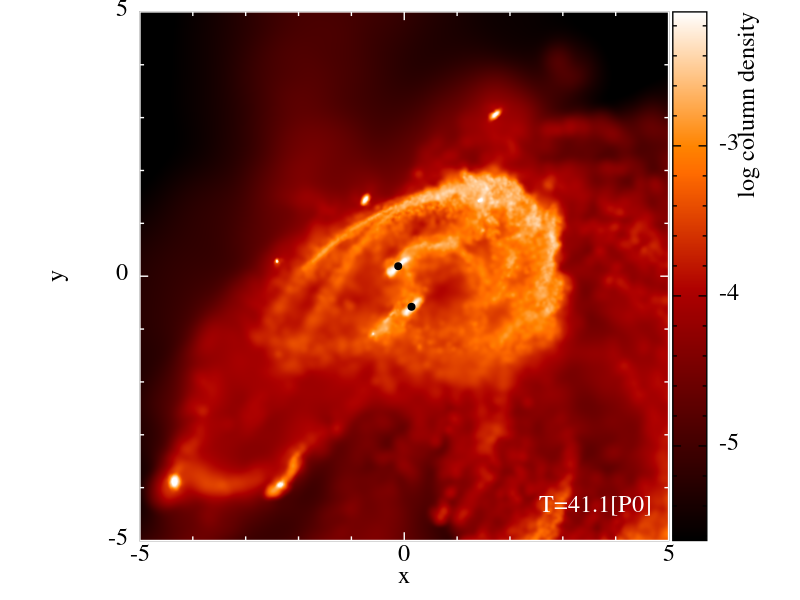}
        &\includegraphics[width=0.3\textwidth]{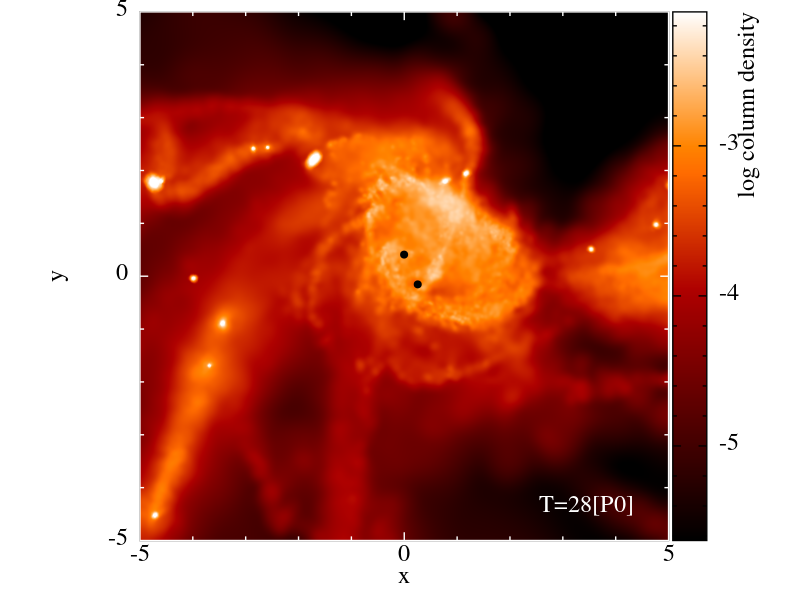}\\
        \includegraphics[width=0.3\textwidth]{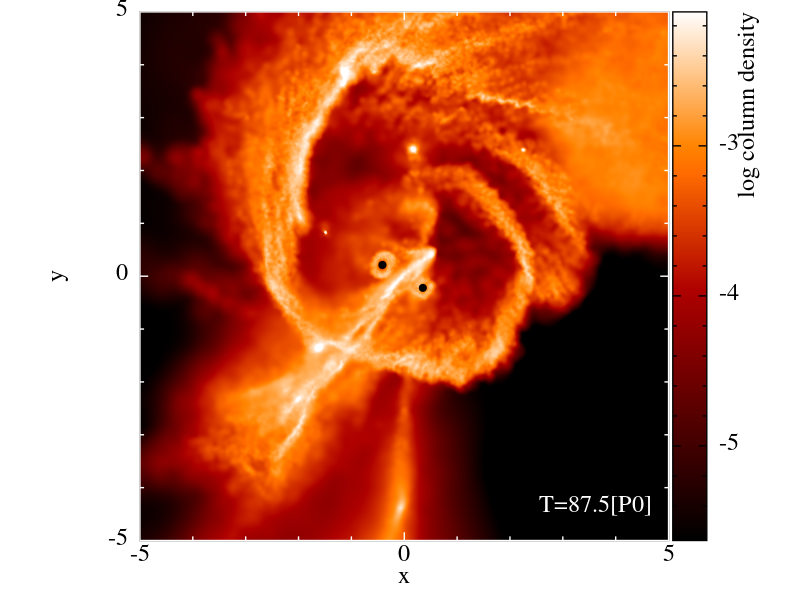}
        &\includegraphics[width=0.3\textwidth]{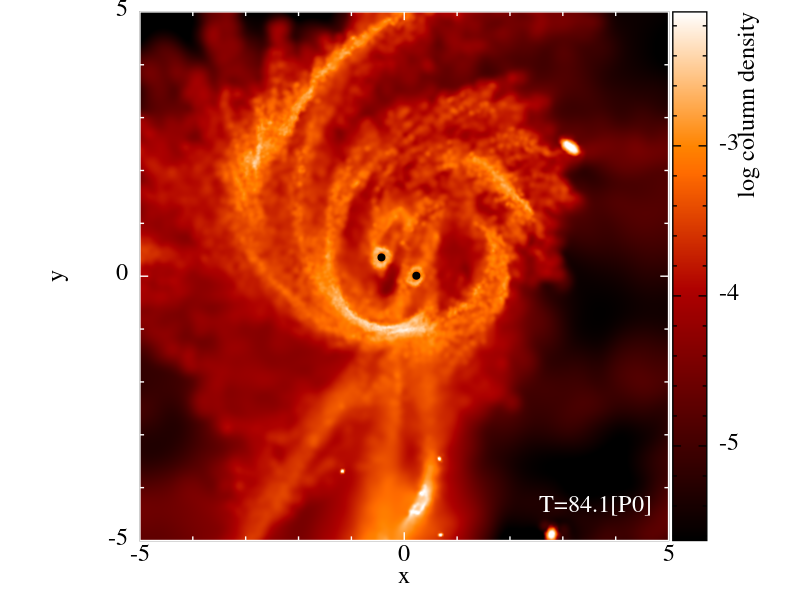}
        &\includegraphics[width=0.3\textwidth]{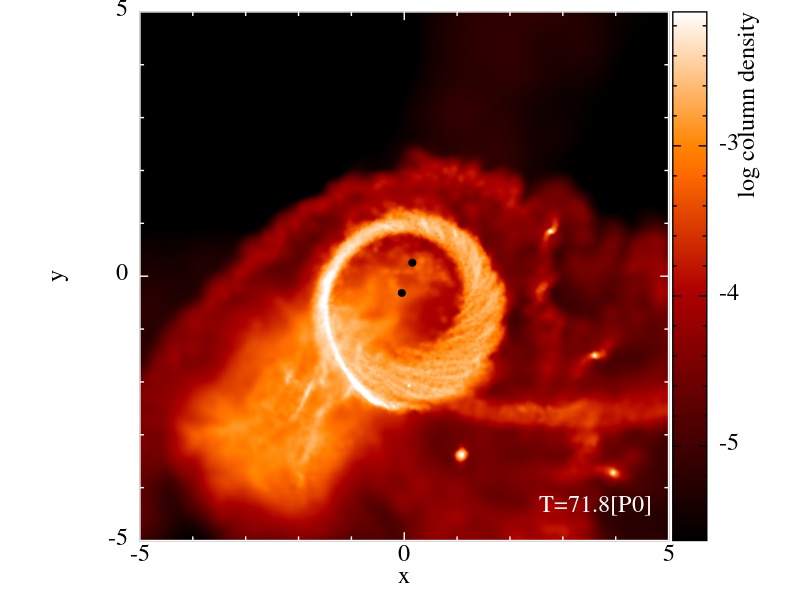}\\
    \end{tabular}
    \caption{Snapshots of {\runa} (top row) and {\runb} (bottom row)
    at selected time (as indicated in each panel), highlighting features described
    in figure~\ref{fig:circumbinaryAB}). In all panels, the binary is viewed
    face-on in the $x-y$ plane, and the gas column density (integrated along the
    $z-$axis) is shown in log-scale as indicated on the right of each panel.}
    \label{fig:circumbinaryAB_splash}
\end{figure*}

For the simulations presented in this paper, the incoherence of the accretion
events produces a variety of gaseous streams that continue interacting with the
binary and between each other, making a clear identification of circumbinary
disc structures much harder. To study any type of stable structure around the
{\mbhb} is essential to devise a set of conditions defining whether an SPH
particle belongs or not to that structure. In practice we define circumbinary
structures by  considering all particles that:

\begin{itemize}
    \item are bound to the {\mbhb} (to avoid including unbound streams of gas flung
          away in the interaction with the {\mbhb});
    \item are inside a critical radius $r<7a_0$.
          The specific choice is arbitrary, but is motivated by the typical size of
          circumbinary discs found in \cite{GoicovicEtAl2016}.
\end{itemize}

For each SPH particles identified by this first cut, we can take 3-D position and
velocity, replace the {\mbhb} with a point mass centred in the centre of mass of the
system, and compute a Keplerian orbit around this point. This calculation is an
approximation of the true orbit of the particle, since it ignores the binary
nature of the central massive object, as well as the external potential.
The advantage of this approach
is that it allows to define orbital elements such as semimajor axis $a$,
eccentricity $e$ and inclination $i$ for each particle. We therefore impose the
further condition that the particle orbit
\begin{itemize}
    \item has $a(1-e)>1$, to make sure that it does not intersect the {\mbhb};
    \item has $e<0.9$. Although this is also a somewhat arbitrary choice, it allows
          to exclude particles belonging to new infalling clouds
          (which are on almost radial orbits).
\end{itemize}

Figure~\ref{fig:mass_disc} shows amount of gas that meets the aforementioned
conditions as a function of time, together with the breakdown of the
contribution of each individual cloud. Despite stochasticity and differences
between the runs, we identify some general trends. As new clouds get into the
system, their contribution to the circumbinary structure initially rumps up to
$M\approx 10^{-3} M_{\rm bin}$ (i.e. 10\% of the initial mass cloud), since
material settles on orbits with $a<7a_0$ and circularises to $e<0.9$.  This
process is evident for the initial clouds in {\runa} which have fairly large
periapsis passage and tend to not interact with each other (cf the top panel of
Figure~\ref{fig:dist_pericentre}).  Up to $T\approx 15P_0$, the circumbinary
structure mass tends to steadily grow in time to about 1\% of the {\mbhb} mass.
Eventually, incoming new clouds on intersecting orbits prompts accretion of
pre-existing circumbinary gas, preventing the structure to significantly
further grow in mass. This is why the total mass in these structures never
grows to much more than few\% of the {\mbhb} mass. This process is more evident
right from the start in {\runb}. In this case, the first clouds have rather
small periapsis and they strongly interact with each other upon arrival onto
the {\mbhb}. The net effect is that the circumbinary structure does not grow
much beyond $M\approx 10^{-3} M_{\rm bin}$ until a number of clouds with larger
periapsis contribute a substantial mass budget from $T\approx 40 P_0$ onwards
(cf the bottom panel of Figure~\ref{fig:dist_pericentre}).

It is clear that the contribution of each individual cloud to the circumbinary
structure tend to decrease in time (although with large fluctuation). With
``older'' clouds contributing less to the mass budget.  This causes the large
rainbow-like spread towards the end of the runs, whereby the contribution of
each individual cloud ranges between 10\% to about 0.01\% of their initial
mass. It is interesting to note that the spread is much larger in the \F{0.5}
and \F{1.0} cases, in which clouds experience more violent interactions among
themselves and with the binary. In the \F{0.0}, conversely, most of the
circumbinary material is stored into a disc held-up by Lindblad resonances. The
income of new clouds in mostly co-rotating orbit tends to add new material to
the disc without a substantial disruption of the pre-existing conditions. For
example, the first incoming cloud (red curve) still contributes about 5\% of
its mass to the circumbinary disc by the end of the simulations in both {\runa}
and {\runb}.

Although Figure~\ref{fig:mass_disc} quantifies the amount of material that
forms a circumbinary structure, it does not provide much information about the
nature of that structure. For example, the gas might be configured in a disc or
in a cloud or in multiple rings and it would not make a difference. For each
SPH particle belonging to the structure, we reconstructed the orbit and we
computed $a$, $e$ and $i$. We can now construct the distributions of these
quantities, and follow their evolution in time.

This is shown in Figure~\ref{fig:circumbinaryAB} for all our runs, together
with a replication of the total circumbinary gas mass shown in figure
~\ref{fig:mass_disc} (top rows). Each panel is built as follows. We take
a uniform grid in the desired quantity and, at each simulation snapshot, we
construct an histogram by adding particles to the bins and normalising so that
the integral over the bins gives the total mass in the structure. Histograms at
subsequent steps are then concatenated along the $x$-axis and smoothed to
produce the 2-D density maps displayed in the figure (in logarithmic scale).
Figure~\ref{fig:circumbinaryAB} clearly show the connection between the
$F-$distribution and the geometry of the circumbinary structures for both
{\runa} and {\runb}. We now describe the diverse phenomenology of each
$F-$distribution, referring to representative examples of each individual case
shown in Figure~\ref{fig:circumbinaryAB_splash}.

For the co-rotating case (\F{0.0}, left column), after an initial transient
phase, most of the particles have less than $50^{o}$ respect to the binary, and
tend to distribute in a co-rotating, extended circumbinary disc, displaying a
variety of eccentricities. In {\runa}, the disc does not build up as a coherent
structure, as hinted by the low total mass (panel A0.0-m) and large range of
inclinations (panel A0.0-i). The presence of large amount of gas at high
inclinations, prevents resonances from being efficient, and a well defined
cavity cannot be seen in panel A0.0-a, where we see gas been distributed in the
whole range $1<a<7$, although a concentration of gas at $a<5$ appears from
$T\approx40 P_0$ onwards. Conversely, in {\runb}, a prominent thin disc builds
up coherently and progressively aligns with the {\mbhb}. This is demonstrated
by the mass build-up up to about $M_{\rm gas}=0.03M$ (panel B0.0-m) and by the
narrow range of inclinations decreasing with time (panel B0.0-i).  The density
contrast in panel  B0.0-a highlights that the bulk of the disc lies in the
range $2<a<4$, with the decline at $a<2$ indicative of the resonance-sustained
cavity. Gas within the disc remains in fairly eccentric orbits spread around
$e=0.5$ (panel B0.0-e).

The left column of Figure~\ref{fig:circumbinaryAB_splash} displays
representative snapshots of these two simulations. The top panel is taken when
the 13th cloud of {\runa} \F{0.0}
interacts with the system, at $T\approx 49P_0$. An extended
circumbinary structure, almost in the binary orbital plane ($i\approx20^{o}$)
is clearly visible, but it is about to be partially disrupted be the 13th cloud
streaming-in from the left (which will cause the subsequent drop in $M_{\rm
gas}$ observed in panel A0.0-m). The bottom panel shows the status of {\runb}
\F{0.0} at the peak of $M_{\rm gas}$ at $T\approx 95P_0$. Comparisons between
the two highlights the prominence of the circumbinary disc forming in {\runb},
featuring a massive spiral that extends to $a\approx5$,

Not surprisingly, in the counter-rotating case (\F{1.0}, right column), the
gas distribution around the binary has high inclinations. Also in this case,
the two runs display quite different behaviours. In {\runa}, we see the
formation of a persistent structure. The mass gets to $M_{\rm gas} \approx
10^{-2}M$ at the peak, declining to $M_{\rm gas} \approx 10^{-3}M$ by the end
of the simulation (panel A1.0-m), is spread across the whole $a$ range (panel
A1.0-a), and displays a wide range of inclinations (panel A1.0-i). The gas
essentially configures into a tenuous and low mass counter-rotating cloud. This
is clearly shown in the top right panel of
Figure~\ref{fig:circumbinaryAB_splash} taken at $T=28 P_{0}$; no clear
disc-like structure is present, and the binary is surrounded by a relatively
compact cloudy envelope with streams extending to $a>5$.

The situation is strikingly different in {\runb}. Here we see that the violent
interaction with the binary causes most of the gas to be either accreted or
expelled and no steady circumbinary structure is formed. A prominent eccentric
transient ring forms around $T=60P_0$, being disrupted about 20 periods later
(note that in our fiducial system the ring would last for about 0.2 Myr), as
clear from panel B1.0-m. The structure is almost perfectly counter-aligned to
the binary (panel B1.0-i), is confined within $a<2.5$ (panel B1.0-a) and has an
average eccentricity $e \approx 0.3$ (panel B1.0-e). The ring is clearly
visualised in the bottom right panel of Figure~\ref{fig:circumbinaryAB_splash}.

\begin{figure*}
    \begin{tabular}{c}
        \includegraphics[width=0.9\textwidth]{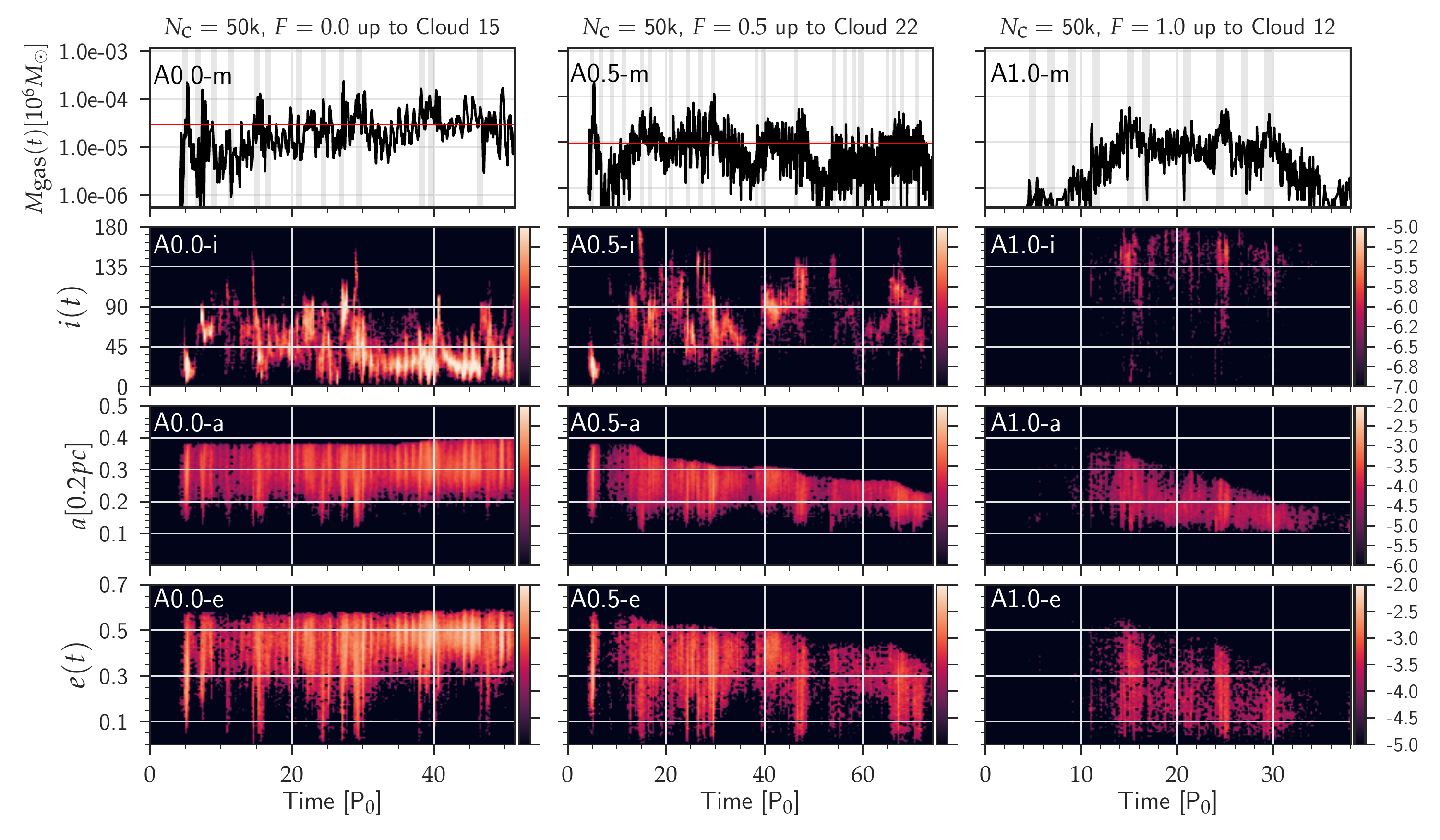}\\
        \includegraphics[width=0.9\textwidth]{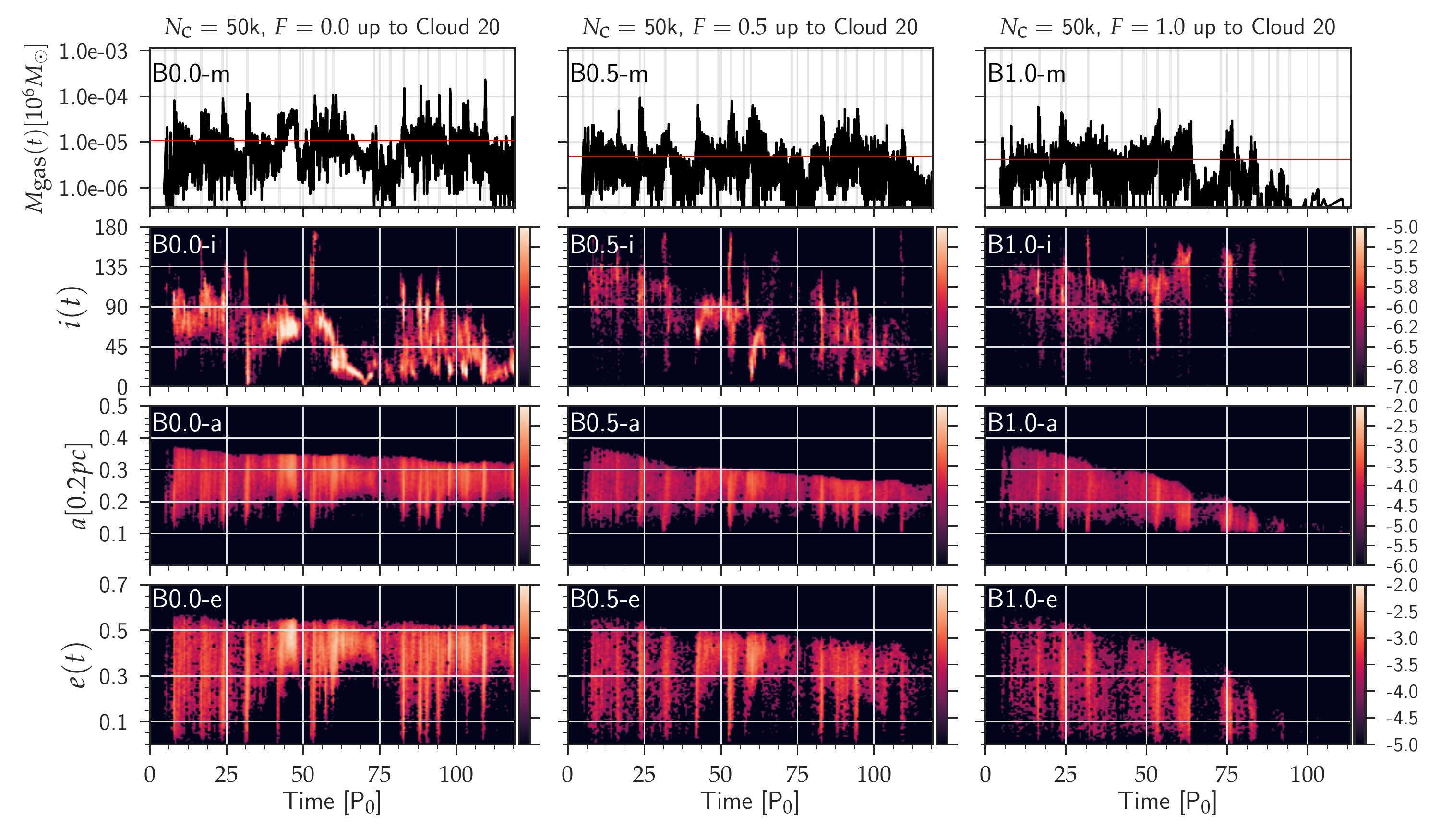}
    \end{tabular}
    \caption{Same as Figure~\ref{fig:circumbinaryAB} but for the {\md} of {\mbh}1 in
    the \F{0.0}, \F{0.5} and \F{1.0} cases (left to right), for
    {\runa} (top $3\times4$ grid) and  {\runb} (bottom grid). Each panel is
    labelled as in Figure~\ref{fig:circumbinaryAB}, note the log-scale in the top
    rows showing the total mass in the {\mds}, $M_{\rm gas}$.}
    \label{fig:RunABMD}
\end{figure*}

\begin{figure}
    \centering
    \resizebox{0.8\hsize}{!}
        {\includegraphics[scale=1,clip]{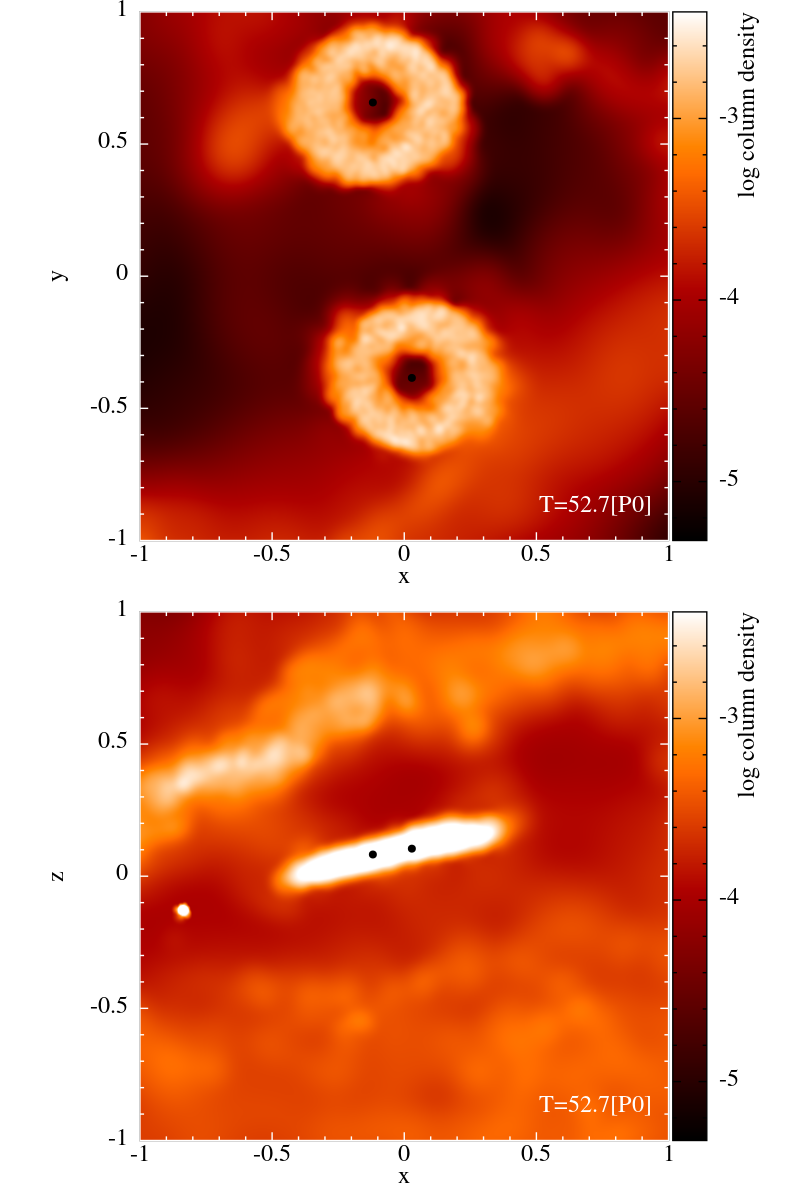}}
    \caption{Snapshot of the last stage of {\runa} \F{0.0} (shown in
        Figure~\ref{fig:RunABMD}, first column of the top grid). The top and bottom
        panels show the $x-y$ and $x-z$ views respectively.}
    \label{fig:RunAMDF0p0}
\end{figure}

The \F{0.5} simulations, shown in the central column of
Figure~\ref{fig:circumbinaryAB}, highlight the impact of the specific cloud
distribution on the formation of circumbinary structures. {\runa} features
a number of transient, incoherent structures extending at all radii (panel
A0.5-a) and inclinations (panel A0.5-i). The top central panel of
Figure~\ref{fig:circumbinaryAB_splash} shows the system at $T=41 P_{0}$. Cloud
12 and 13 are bringing fresh gas into the system, causing a temporary
enhancement of the mass in the circumbinary structure, that looks like
an incoherent, extended envelope. Conversely, {\runb}
shows a behaviour similar to the \F{0.0} case; at $T\approx60 P_0$, a
co-rotating circumbinary structure forms, progressively aligning with the
{\mbhb} as it gains mass in subsequent interactions (panels B0.5-m and B0.5-i).
The structure persists for about 30 binary orbits ($\approx$0.3 Myr,
when scaled to our fiducial system), before being completely disrupted
by the arrival of new clouds on small pericentre orbits
(clouds 13-to-16, cf Figure~\ref{fig:dist_pericentre}). A snapshot
of the system at $T=89 P_{0}$, shown in the bottom central panel of
Figure~\ref{fig:circumbinaryAB_splash}, highlights the similarity with the
\F{0.0} run; in this case, the circumbinary spiral is less massive and slightly
more compact.

Interestingly, in both {\runa} and {\runb},
there is the tendency to form more co-rotating than counter-rotating structures.
This can be clearly seen in panels A0.5-i and B0.5-i of
Fig.~\ref{fig:circumbinaryAB},  where the particles tend
to cluster below 90 degrees in inclination even though half of the events
come from the southern hemisphere.
This occurs simply because resonances with the {\mbhb} orbital motion
efficiently transfer angular momentum to co-rotating gas, which can
therefore settle into more extended and stable structures. An effect that is
absent for the counter-rotating material.

\subsection{Mini-discs}
\label{sec:minidiscs}
\begin{table}
    \centering
    \resizebox{\hsize}{!}{
        \begin{tabular}{llrr}
        \hline
        Simulation & Distribution & $M_{MD1} [10^{6}M_{\odot}]$ & $M_{MD2}
        [10^{6}M_{\odot}]$ \\ \hline \multirow{3}{*}{\texttt{RunA}} & \F{0.0}
        & $3.58\times 10^{-5}$ & $3.39\times 10^{-5}$\\ & \F{0.5} & $6.64\times
        10^{-6}$ & $6.82\times 10^{-6}$\\ & \F{1.0} & $1.17\times 10^{-6}$
        & $9.93\times 10^{-7}$\\ \hline

        \multirow{3}{*}{\texttt{RunB}} & \F{0.0} & $1.08\times 10^{-5}$
        & $1.07\times 10^{-5}$\\ & \F{0.5} & $4.84\times 10^{-6}$ & $5.15\times
        10^{-6}$\\ & \F{1.0} & $4.15\times 10^{-6}$ & $4.32\times 10^{-6}$\\
        \hline

        \hline
        \end{tabular}}
    \caption{Average mass of the {\mds} surrounding the {\mbh}s in all our runs.}
    \label{tab:minidiscs}
\end{table}

It is also interesting to study the dynamics of gaseous structures {\it inside}
the {\mbhb} corotation radius, which are generally directly responsible for the
binary feeding and the associated high energy electromagnetic radiation. In the
standard picture of a steady circumbinary disc, the forcing imposed by the
binary quadrupolar potential induces gas streams that feed prominent {\mds}
around each {\mbh} \citep{2013MNRAS.436.2997D}. Conversely, in the
counter-rotating case, the absence of Lindblad resonances allows gas at the
edge of a putative circumbinary disc to impact directly onto the binary. Since
the {\mbh} and the gas have opposite directions, $|\vec{v}_{\rm
BH}-\vec{v}_g|\approx 2v_{\rm BH}$; this means that the capture cross section
of the gas is very small, and only gas {\it inside} the sink radius becomes
bound to the {\mbh} being promptly accreted. In practice, {\mds} extending
beyond $r_{\rm sink}$ are unlikely to form in this case.

Similarly to Section~\ref{sec:circumbinary}, we now select all particles inside
either of the {\mbh} Roche Lobes, defined as circles around each MBH
with radius
\begin{equation}
  R_{\textrm{RL}} = \frac{0.49\ q^{2/3}}{0.6\ q^{2/3} + \log(1+q^{1/3})}a
\end{equation}
\citep{Eggleton1983}, where $q$ is the binary mass ratio,
which we assume to be unity throughout
all of our simulations.
We then use 3-D positions and velocities to compute their Keplerian orbit
around the closest {\mbh}. We define the {\md} around each {\mbh} as the
collection of gas particles on orbits that:
\begin{itemize}
\item are bound to that {\mbh}; \item have a pericentre larger than the
  sink radius, i.e.: $a(1-e) > r_{\rm sink} = 0.1$ (equivalent to
  $0.02\textrm{pc}$ for our fiducial system);
\item have an apocentre smaller than the Roche Lobe size, i.e.:
  $a(1+e) < R_{\textrm{RL}}$.
\end{itemize}

The evolution of the {\md} around {\mbh}1 in each run is summarised in
Figure~\ref{fig:RunABMD}. There we show the time evolution of the {\md} total
mass (top row), and the distribution of inclination (second row), semimajor
axis (third raw) and eccentricity (bottom row) of the particles belonging to
it, constructed as detailed in \S\ref{sec:circumbinary}.
The first thing we notice is that, due to the complex dynamics triggered by the
infall of incoherent gas clouds, {\mds} are far from being well defined, stable
structures in our simulations. However, some general features of relaxed
`steady state simulations' are preserved. In particular, we see that the
prominence and persistence of {\mds} is a strong function of $F$. This is also
supported by numbers reported in Table~\ref{tab:minidiscs}, demonstrating that
the average mass in the {\mds} is much higher in the \F{0.0} runs. Note that
this is particularly true for {\runa}: here the frequent supply of fresh clouds
has the effect of feeding the {\mds} in the \F{0.0} case, whereas increases the
chance of disrupting them quickly in the less coherent \F{0.5} or in the
counter-rotating \F{1.0} cases. In {\runb}, clouds are supplied at a lower
rate, leaving more time for the gas in the {\mds} to be accreted. This mostly
affects the \F{0.0} run, in which we see a smaller amount of mass accumulating
in the {\mds} on average.

In general Figure~\ref{fig:RunABMD} show that {\mds} are more massive and
persistent in the \F{0.0} runs (left column). Note that both in {\runa} and
{\runb} there is a significant scatter in the particle inclination
distribution. This is because particles partially preserve memory of the
inclination of their parent cloud and tend to form {\mds} aligned with their
incoming orbital angular momentum. In the long run, however, the {\mbhb}
potential torques the disks causing a partial alignment with its orbital
angular momentum (effect visible both in panels A0.0-i and B0.0-i). An example
of such persistent {\mds} forming in the co-rotating case is shown in
Figure~\ref{fig:RunAMDF0p0}, displaying the last snapshot of {\runa} \F{0.0}.

As expected, {\mds} becomes much more intermittent as $F$ increases. This is
essentially because {\md} formation is a more natural outcome in the
interaction with co-rotating clouds. In fact, we can see that in the \F{0.5}
case (central column) the formation of structures with $i<90^o$ is strongly
preferred ( panels A0.5-i and B0.5-i). In the extreme \F{1.0} case, {\mds} are
very intermittent and, contrary to the circumbinary structures, often show
a significant fraction of co-rotating material (i.e. with $i<90^o$, see panel
B1.0-i). The shrinking of the binary as the simulations advance, is clearly
noticeable in the $a$ and $e$ panels of all the \F{0.5} and \F{1.0}
simulations. In fact, as the two {\mbh}s get closer to each other, their Roche
Lobes contract and the size of the {\mds} that they can accommodate shrink
accordingly.

Finally, we note that the study of {\mds} is a delicate matter in this kind of
simulations, due to the relatively large sink radius and to the inherent small
size of these structures which makes their resolution difficult. In fact, even
the most prominent {\mds} in the \F{0.0} simulations have an average mass of
about $10^{-5}$, which means that in the {\lowa} simulations they are resolved
with about 50 particles. We therefore checked convergence of our results by
comparing {\mds} at {\lowa} and {\higha} resolutions finding essentially no
difference neither in the intermittent behaviour nor in the average masses,
indicating that our results are robust. Likewise, we tested that shrinking the
sink radius by a factor of two does not appreciably affect the {\mds}
evolution.

\subsection{`Forked' simulations: stopping the supply of clouds}

\begin{figure*}
    \resizebox{\hsize}{!}
        {\includegraphics[scale=1,clip]{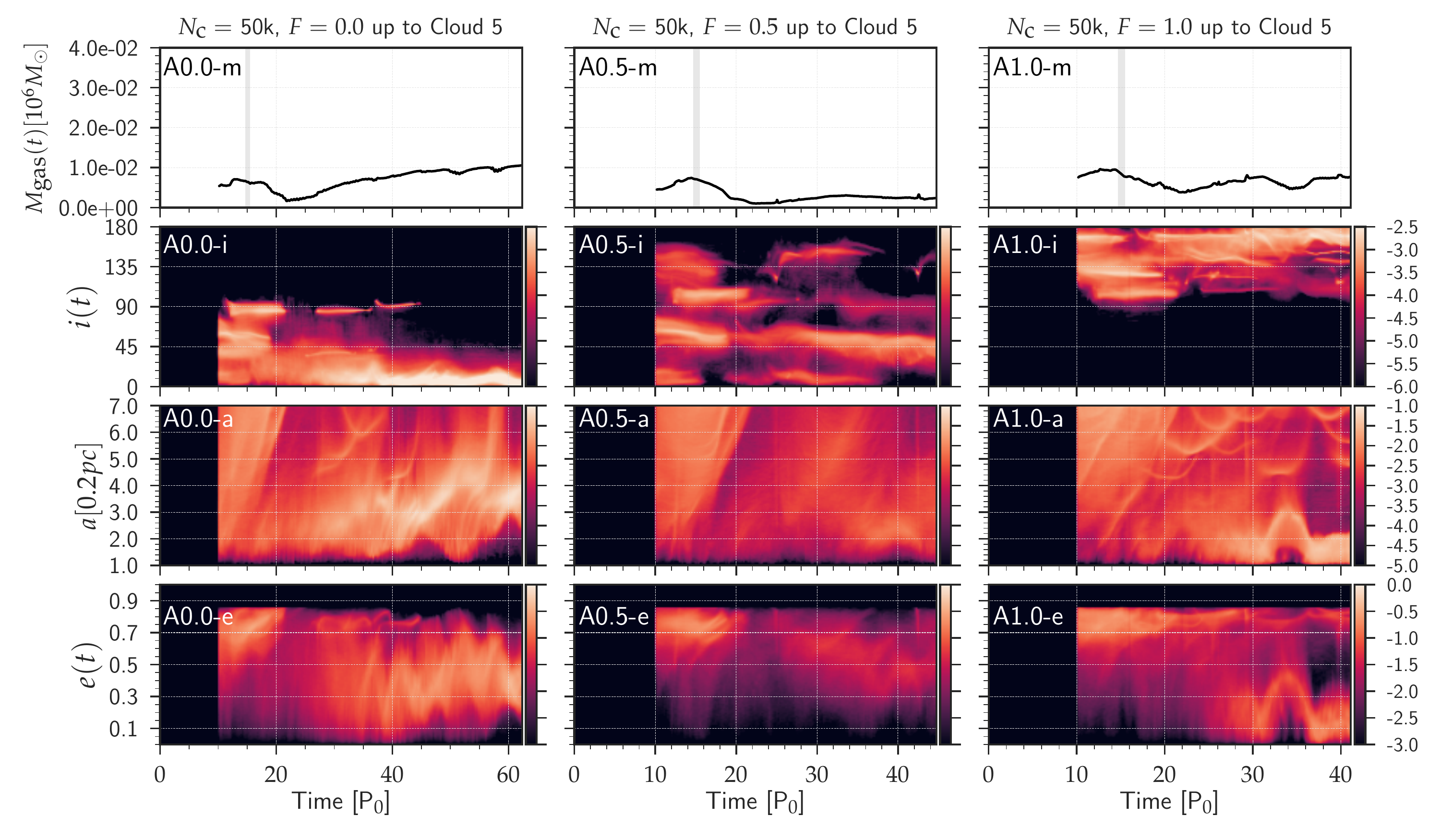}}
    \caption{Same as Figure~\ref{fig:circumbinaryAB} but for the set of `forked'
        simulations. The \F{0.0}, \F{0.5} and \F{1.0} distributions are shown from
        left to right. The vertical grey stripes in the first row mark the time of
        arrival of the 5th cloud. No further cloud is included into the system.
    }
    \label{fig:fork05}
\end{figure*}

\begin{figure*}
    \resizebox{\hsize}{!}
        {\includegraphics[scale=1,clip]{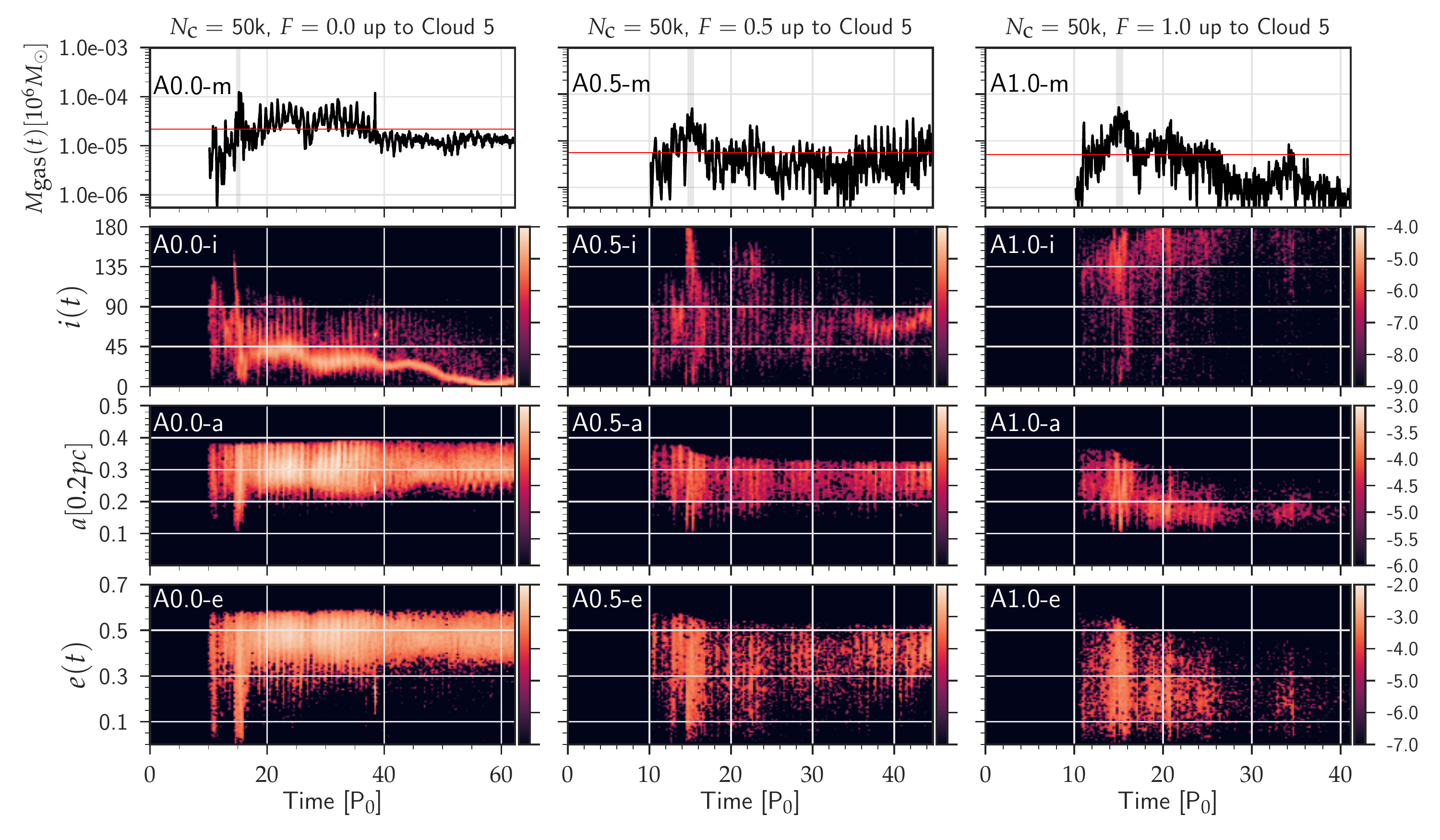}}
    \caption{Same as Figure~\ref{fig:RunABMD} but for the  ``forked''
    simulations. Note the log-scale in the top row.}
    \label{fig:fork05MD}
\end{figure*}

\begin{figure*}
    \begin{tabular}{ccc}
        \includegraphics[width=0.3\textwidth]{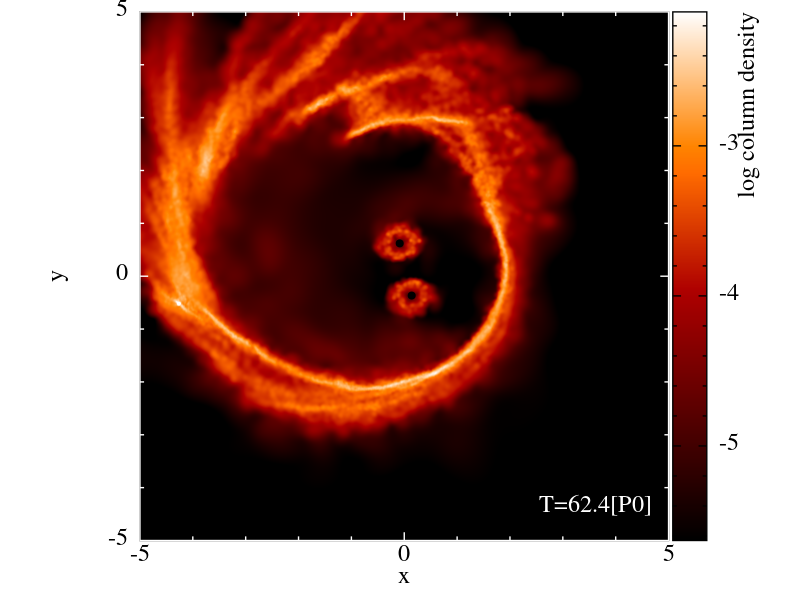}
        &\includegraphics[width=0.3\textwidth]{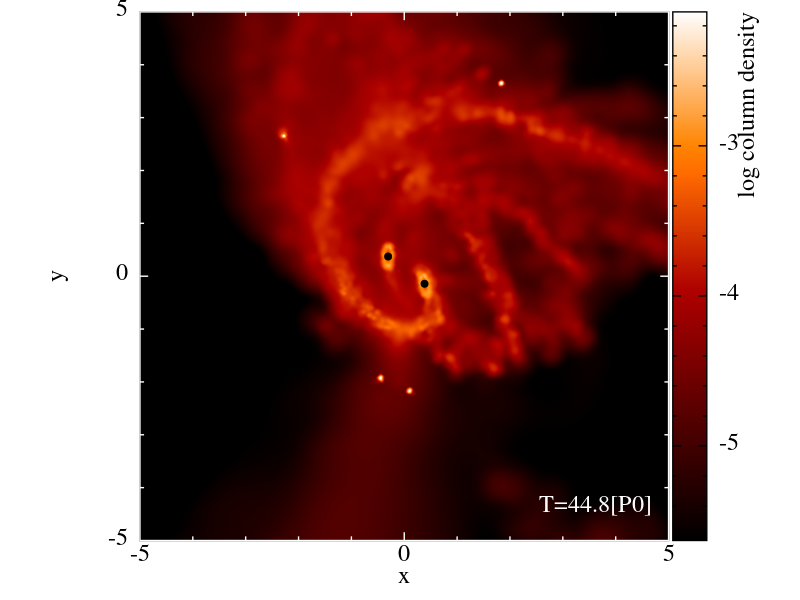}
        &\includegraphics[width=0.3\textwidth]{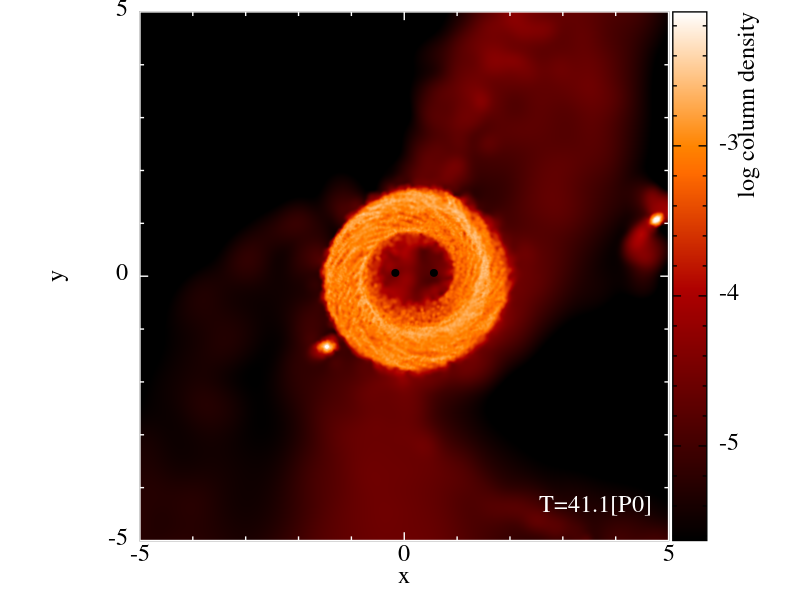}\\
        \includegraphics[width=0.3\textwidth]{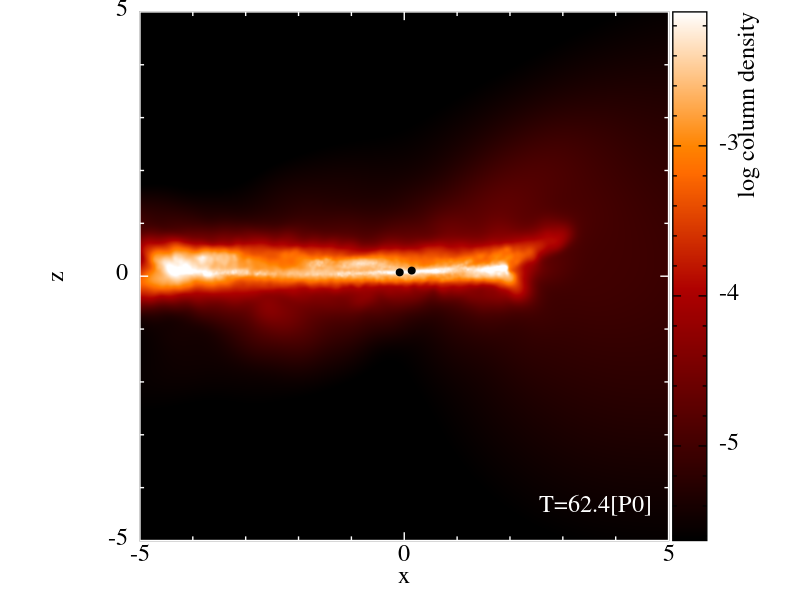}
        &\includegraphics[width=0.3\textwidth]{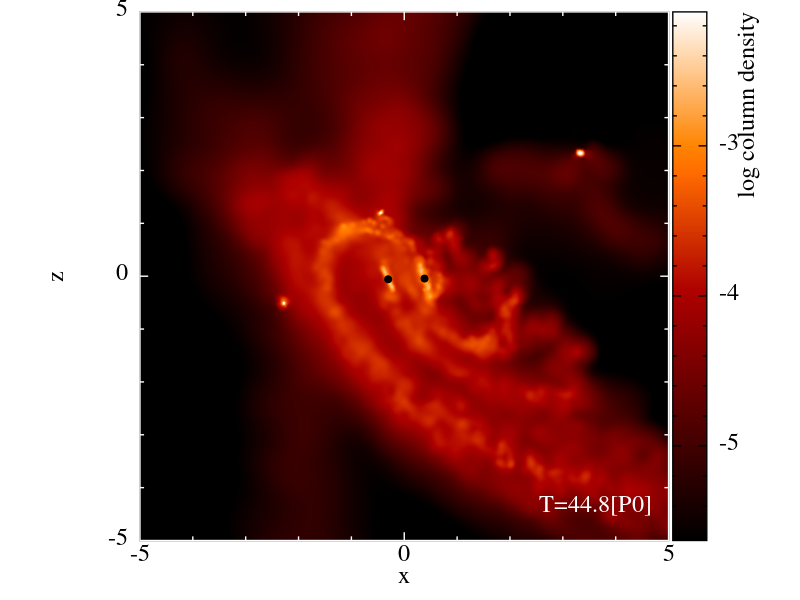}
        &\includegraphics[width=0.3\textwidth]{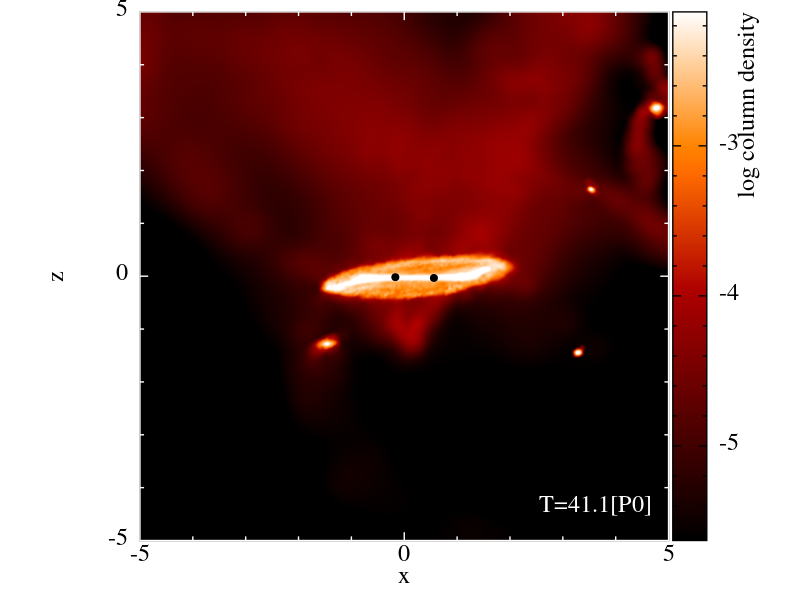}\\
    \end{tabular}
    \caption{Visualisation of the last snapshot of the ``forked''
    simulations. \F{0.0}, \F{0.5} and \F{1.0} runs are shown left to right. The top
    and bottom row display $x-y$ and $x-z$ views of the system.}
    \label{fig:fork05splash}
\end{figure*}

The variability of a system being constantly affected by multiple chaotic
accretion events has clearly emerged in the previous sections, where we showed
the continuous formation and disruption of structures as new clouds were added
to the system. An interesting question is what happens when the supply of new
clouds ceases and the system is allowed to relax. To answer this question, we
took two snapshots from all the {\runa} simulations at the
moment when the 5th and the 10th clouds were added to the system and
`forked' them, thus starting two parallel series of runs in which no
further clouds were added, allowing the system to evolve unperturbed.
We show in this section results from simulations forked after the
5th clouds, the outcome of the ones forked after the 10th cloud are
qualitatively very similar.

To illustrate the evolution of the system, we perform the same analysis
described in \S\ref{sec:circumbinary} and build 2-D density plots to study the
main properties of the particles forming both the circumbinary disc, and the
{\mds} around each {\mbh}. Those are shown in Figures ~\ref{fig:fork05} and
~\ref{fig:fork05MD} respectively. We now discuss the main long term features of
the \F{0.0}, \F{1.0} and \F{0.5} run separately.

In the left column of Figure~\ref{fig:fork05} we can see that in the \F{0.0}
run, after the infall of the 5th cloud, the remaining material from the other
disrupted clouds falls back forming a prominent circumbinary disc around the
{\mbhb} (panel 0.0-m). The disc is relatively thin, with an inclination of
about $10^{o}$ with respect to the binary orbital plane (panel 0.0-i), and
it most of the mass is concentrated at a distance $2\leq a \leq 5$ from
the binary centre of mass, with a broad range of eccentricities $0.1<e<0.7$ (panels
0.0-a and 0.0-e). The left column of Figure~\ref{fig:fork05MD} allow us to
confirm the long term stability of the {\mds}, which is almost
aligned with the {\mbhb} orbital plane.

The circumbinary disc in the \F{1.0} simulation (right column) displays a very
similar behaviour. Disrupted cloud material falling back on orbits with
comparable inclinations, interact with each other leading to the formation of
a relatively massive circumbinary disc (panel 1.0-m). From panel 1.0-i, we note
that, due to the high initial inclination and to the fact that $L_{d}\ll L_{B}$,
the disc tends to counter-align with the {\mbhb}
\citep{2005MNRAS.363...49K}. By the end of
the simulation, an almost perfectly counter-rotating circumbinary disc has
formed, with $i\approx 170^{o}$. The circumbinary disc is much better defined
than in the \F{0.0} case; particles are mostly confined confined within $a\leq2.5$
(panel 1.0-a) and have lower eccentricities $e<0.3$ (panel 1.0-e). The right
column of Figure~\ref{fig:fork05MD} shows that {\mds} are much lighter,
intermittent, and tend to be consumed with time. Note the increase in the {\md}
mass around $T=35 P_0$ (panel 1.0-m), in correspondence of the decrease in the
circumbinary disc mass. This is due to a stream of gas partially disrupting the
circumbinary structure and feeding gas to the central {\mbhb}.

The \F{0.5} simulation does not seem to converge towards a specific relaxed
state. The very diverse orbits of infalling clouds prevent the stream to
efficiently interact, dissipate angular momentum and circularise into any
specific circumbinary structure. In fact, the mass in a putative `circumbinary
disc' does not grow in time (panel 0.5-m) and streams at different inclinations
are clearly recognisable until the end of the simulation  (panel 0.5-i).
Figure~\ref{fig:fork05MD} shows that {\mds} are not very prominent, and also
tend to be consumed with time.

A summary of the overall structures formed by the end of these ``forked''
simulations is displayed on Figure~\ref{fig:fork05splash}. The $x-y$ and $x-z$
views highlight the thin circumbinary discs forming in both the \F{0.0} and
\F{1.0} runs, along with the well defined {\mds} in the former case.
Conversely, in the \F{0.5} case, no coherent structure is recognisable, and at
least four distinct streams originating from different clouds are clearly
visible. The aftermath of an epoch of incoherent accretion events is therefore
strongly dependent on the properties of the infalling clouds. In general,
ordered circumbinary structures persist only if the sum of all infalling
material has a substantial net angular momentum (either co- or counter-aligned
with the {\mbhb}).

\section{Discussion and future work}
\label{sec:disc}

In this work we have used detailed SPH simulations performed with a modified
version of the code {\gadget}-3 to study the interaction with a circular, equal
mass {\mbhb} with a series of infalling clouds in very eccentric orbits. We
performed six main runs, considering two distributions of cloud pericentre
distances and arrival times (defining {\runa} and {\runb}), and three
distribution of angular momenta with different degree of anisotropy that we
labelled \F{0.0} (co-rotating clouds), \F{0.5} (isotropically distributed
clouds) and \F{1.0} (counter-rotating clouds). The goal is to study the
dynamics of the {\mbhb}-gas interaction when the binary is supplied with gas in
incoherent discrete `pockets', which might be a typical situation in the
turbulent environment of high redshift, gas-rich galaxies
\citep[e.g.][]{2017ApJ...836..216P} and merger remnants
\citep[e.g.][]{2014A&A...562A...1P}
, relevant to the
early build-up of {\mbh}s \citep[see][ for a review]{2010A&ARv..18..279V} and
to future low frequency gravitational wave observations with LISA
\citep{2017arXiv170200786A}. The main focus of this paper was on the formation
and evolution of bound gaseous structures, in a companion paper ({\paperbhb})
we will turn our attention on the evolution of the {\mbhb}.

Our main findings are summarised in the following points.

\emph{Post interaction gas distribution}. In general, the density profile
distribution of the gas post-interaction with the {\mbhb} follows a $\rho_{\rm
gas}\propto r^{-2}$ distribution. This is expected since the speed of sound of
the gas is just few km s$^{-1}$, implying a large Bondi radius of the {\mbhb},
extending beyond the bulk of the extended gas.  Compression of the clouds at
pericentre also causes the formation of several, extremely dense
self-gravitating clumps that can be favourable sites for \emph{in-situ} star
formation. In this work, we stopped gravitational collapse of the clumps by
imposing an adiabatic behaviour of gas region above a critical gas density.

\emph{Circumbinary structures}. Many theoretical and numerical studies of
{\mbhb} evolution in gaseous environments rely on the presence of a relatively
stable, extended circumbinary disc that can efficiently extract energy and
angular momentum from the binary
\citep[e.g.][]{2008ApJ...672...83M,CuadraEtAl09,2011MNRAS.415.3033R,2012MNRAS.427.2680K,2017MNRAS.469.4258T}.
However, the route to the formation of such stable, extended disc has hardly
been investigated. Our simulations show that, when the gas is fed to the binary
in incoherent pockets, it is hard to form a massive circumbinary structure.
Adding more clouds to the system is a two-edged blade: on the one hand, new
clouds supply fresh material that can add-up to the mass budget of the disc; on
the other hand, clouds intersecting at high inclinations with a pre-existent
disc can cause its partial disruption. Even in the co-rotating simulations
(\F{0.0}), after the addition of 20 clouds, injecting in the system a gas mass
equal to 20\% of the {\mbhb} mass, the circumbinary disc mass does not exceed a
mere 1\% of the binary mass ($\approx 10^4$ solar masses for our fiducial
system). Extrapolating from \cite{GoicovicEtAl2016}, the naive expectation
would be that about 50\% of the injected mass ends up in a circumbinary
structure ($\approx 10^5$ solar masses for our fiducial system).  More
isotropic and counter-rotating cloud distributions (\F{0.5} and \F{1.0}) result
in even lighter and more transient circumbinary discs.

\emph{Counter-rotating rings}. The interaction with counter-rotating gas clouds
result in the formation of compact counter-rotating rings. Due to the absence
of Lindblad resonances for co-rotating orbits, the inner edge of the ring has
about the size of the {\mbhb} orbit and is not sustained by its forcing
quadrupolar potential, resulting in an extremely unstable structure. The net
consequence is that interaction with clouds on small impact parameters easily
disrupt those rings, triggering copious accretion on the {\mbhb}. Therefore,
building-up extended counter-rotating structures is virtually impossible in our
scenario, since it would require a much more gentle supply of material with
larger angular momentum, within a relatively small range of inclinations.

\emph{Mini-discs formation}. As expected from previous studies
\citep{GoicovicEtAl2016}, a co-rotating cloud distribution (\F{0.0}) results in
the formation of prominent {\mds}. Although those {\mds} can also be
significantly affected by the infall of new clouds, they are usually able to
maintain certain stability, and they re-build quite efficiently following
disruption from particularly aggressive cloud interactions. There is a notable
tendency of the {\mds} to align after a few orbital periods (see,
Figure~\ref{fig:RunABMD}). {\mds} are much less prominent in the \F{0.5} and
\F{1.0} runs. Because of the higher relative velocity between the gas particles
and the {\mbh}s, the cross section of gas capture within the {\mbh} Roche Lobe
is often smaller than the sink radius. It is possible that {\mds} form also in
those cases, but on much smaller scales, that cannot be resolved by our current
set-up.

\emph{Post interaction relaxation.} When the supply of gas clouds ceases, the
system tends to relax into a configuration that depends on the overall angular
momentum distribution of the gas. This was investigated in our ``forked''
simulations, in which we limited the supply of gas to the 5th cloud. The
stopping of cloud infall avoid further disruption of the circumbinary
structures, that grow their mass in time, approaching a stable configuration.
After about $20 P_0$ from the infall of the last cloud, a prominent co-rotating
circumbinary disc forms in the \F{0.0} case, whereas a well defined compact
thin ring forms in the \F{1.0} case. The mass in these structures is about
$10\%$ of the total mass content of the supplied clouds. Post-infall relaxation
does not lead to any well defined structure in the isotropic case \F{0.0}, and
several incoherent streams are still present at the end of the simulation. We
speculate that in the long term, interaction within the stream will result in
a relatively homogeneous, tenuous gas envelope.

\emph{Accretion onto the MBHB}. A consequence of the violent interaction
between different clouds and the continuous disruption of circumbinary
structure is the triggering of the infall of enormous amount of gas onto the
{\mbh}s. This has been sees in larger scale simulations of {\mbh} fuelling
\citep{Hobbs11,2015MNRAS.453.1608C}, and in simulations of accretion onto
a {\mbhb} due to `disk tearing' \citep{2015MNRAS.449.1251D}; the interaction of
gas streams coming from different direction cause efficient cancellation of
their respective angular momenta, resulting in efficient infall onto the
binary. In the test runs presented here, we showed that the associated
accretion rate can be as high as $\approx 100 \dot{M}_{\rm Edd}$, when scaled
to our fiducial system. This has a strong effect on the dynamics of the binary,
which we investigate in detail in a companion paper (\paperbhb).

Most importantly, our results show that the effect of multiple cloud
interactions with a {\mbhb}, does not sum-up to the effect of the single clouds
taken individually. Cloud-cloud interactions have a fundamental role in shaping
the gaseous structures forming around the binary, promoting continuous
formation and disruption of circumbinary discs or ring and triggering episodes
of enhanced accretion onto the {\mbhb}. In general, we found that it is
difficult to grow extended, massive circumbinary structures. This means that
the evolution of the {\mbhb} in this incoherent-feeding scenario is driven
mostly by direct gas capture and accretion rather than resonant torques exerted
by a circumbinary disc, as we explore in \paperbhb. Our
simulations are the first to explore in detail this incoherent {\mbhb} feeding
scenario. We focused on some specific aspects of the system evolution, but
there is a number of different properties of the systems that can be further
investigated. For example, the distribution of dense clumps can be used to
simulate star cluster formation in the vicinity of the binary and to study
their further interaction with the binary. Likewise, the final state of the
system can be evolved for longer time to better assess the stability and fate
of the gaseous structures on longer timescales. Due to computational
constraints, we could follow ``forked'' simulations only for 20-25$P_0$,
corresponding to only 0.2 Myr for our fiducial system. Finally, we remark that
we implemented an extremely simplified hydrodynamic scheme, featuring an
effective isothermal/adiabatic equation of state and ignoring any feedback from
accretion onto the {\mbhb}. Eventually, our result should be tested against
enhanced simulations including realistic cooling prescription capturing disc
fragmentation, together with a scheme tracing accretion feedback on the
surrounding gas.

\section*{Acknowledgements}
The simulations were partially run between the \textit{minerva}
cluster at AEI, and the \textit{sandy-bridge} nodes at HITS.  CMF acknowledges
support from the Transregio 7 ``Gravitational Wave Astronomy'' financed by the
Deutsche Forschungsgemeinschaft DFG (German Research Foundation).  CMF
acknowledges support from the DFG Project  ``Supermassive black holes,
accretion discs, stellar dynamics and tidal disruptions'', awarded to PAS, and
the International Max-Planck Research School.  FG acknowledges support from the
CONICYT-PCHA Doctorado Nacional scholarship, the ERC-StG grant EXAGAL-308037,
and the Klaus Tschira Foundation.  AS is supported by the Royal Society.  PAS
acknowledges support from the Ram{\'o}n y Cajal Programme of the Ministry of
Economy, Industry and Competitiveness of Spain, as well as the COST Action
GWverse CA16104. This work has been partially supported by the CAS President's
International Fellowship Initiative.

\bibliographystyle{style/mnras}

\begin{thebibliography}{}
\makeatletter
\relax
\def\mn@urlcharsother{\let\do\@makeother \do\$\do\&\do\#\do\^\do\_\do\%\do\~}
\def\mn@doi{\begingroup\mn@urlcharsother \@ifnextchar [ {\mn@doi@}
  {\mn@doi@[]}}
\def\mn@doi@[#1]#2{\def\@tempa{#1}\ifx\@tempa\@empty \href
  {http://dx.doi.org/#2} {doi:#2}\else \href {http://dx.doi.org/#2} {#1}\fi
  \endgroup}
\def\mn@eprint#1#2{\mn@eprint@#1:#2::\@nil}
\def\mn@eprint@arXiv#1{\href {http://arxiv.org/abs/#1} {{\tt arXiv:#1}}}
\def\mn@eprint@dblp#1{\href {http://dblp.uni-trier.de/rec/bibtex/#1.xml}
  {dblp:#1}}
\def\mn@eprint@#1:#2:#3:#4\@nil{\def\@tempa {#1}\def\@tempb {#2}\def\@tempc
  {#3}\ifx \@tempc \@empty \let \@tempc \@tempb \let \@tempb \@tempa \fi \ifx
  \@tempb \@empty \def\@tempb {arXiv}\fi \@ifundefined
  {mn@eprint@\@tempb}{\@tempb:\@tempc}{\expandafter \expandafter \csname
  mn@eprint@\@tempb\endcsname \expandafter{\@tempc}}}

\bibitem[\protect\citeauthoryear{{Abramowicz}, {Czerny}, {Lasota}  \&
  {Szuszkiewicz}}{{Abramowicz} et~al.}{1988}]{1988ApJ...332..646A}
{Abramowicz} M.~A.,  {Czerny} B.,  {Lasota} J.~P.,   {Szuszkiewicz} E.,  1988,
  \mn@doi [\apj] {10.1086/166683}, \href
  {http://adsabs.harvard.edu/abs/1988ApJ...332..646A} {332, 646}

\bibitem[\protect\citeauthoryear{{Amaro-Seoane}, {Freitag}  \&
  {Spurzem}}{{Amaro-Seoane} et~al.}{2004}]{Amaro-SeoaneEtAl2004}
{Amaro-Seoane} P.,  {Freitag} M.,   {Spurzem} R.,  2004, \mn@doi [\mnras]
  {10.1111/j.1365-2966.2004.07956.x}, \href
  {http://adsabs.harvard.edu/abs/2004MNRAS.352..655A} {352, 655}

\bibitem[\protect\citeauthoryear{{Amaro-Seoane} et~al.,}{{Amaro-Seoane}
  et~al.}{2012}]{Amaro-SeoaneEtAl2012}
{Amaro-Seoane} P.,  et~al., 2012, \mn@doi [Classical and Quantum Gravity]
  {10.1088/0264-9381/29/12/124016}, \href
  {http://adsabs.harvard.edu/abs/2012CQGra..29l4016A} {29, 124016}

\bibitem[\protect\citeauthoryear{{Amaro-Seoane} et~al.,}{{Amaro-Seoane}
  et~al.}{2013a}]{Amaro-SeoaneEtAl2013}
{Amaro-Seoane} P.,  et~al., 2013a, GW Notes, Vol.~6, p.~4-110, \href
  {http://adsabs.harvard.edu/abs/2013GWN.....6....4A} {6, 4}

\bibitem[\protect\citeauthoryear{{Amaro-Seoane}, {Brem}  \&
  {Cuadra}}{{Amaro-Seoane} et~al.}{2013b}]{Pau2013}
{Amaro-Seoane} P.,  {Brem} P.,   {Cuadra} J.,  2013b, \mn@doi [\apj]
  {10.1088/0004-637X/764/1/14}, \href
  {http://adsabs.harvard.edu/abs/2013ApJ...764...14A} {764, 14}

\bibitem[\protect\citeauthoryear{{Amaro-Seoane}, {Maureira-Fredes}, {Dotti}  \&
  {Colpi}}{{Amaro-Seoane} et~al.}{2016}]{Amaro-SeoaneEtAl2016a}
{Amaro-Seoane} P.,  {Maureira-Fredes} C.,  {Dotti} M.,   {Colpi} M.,  2016,
  \mn@doi [\aap] {10.1051/0004-6361/201526172}, \href
  {http://adsabs.harvard.edu/abs/2016A%26A...591A.114A} {591, A114}

\bibitem[\protect\citeauthoryear{{Amaro-Seoane} et~al.,}{{Amaro-Seoane}
  et~al.}{2017a}]{Amaro-SeoaneEtAl2017}
{Amaro-Seoane} P.,  et~al., 2017a, preprint, \href
  {http://adsabs.harvard.edu/abs/2017arXiv170200786A} {} (\mn@eprint {arXiv}
  {1702.00786})

\bibitem[\protect\citeauthoryear{{Amaro-Seoane} et~al.,}{{Amaro-Seoane}
  et~al.}{2017b}]{2017arXiv170200786A}
{Amaro-Seoane} P.,  et~al., 2017b, preprint, \href
  {http://adsabs.harvard.edu/abs/2017arXiv170200786A} {} (\mn@eprint {arXiv}
  {1702.00786})

\bibitem[\protect\citeauthoryear{{Armitage} \& {Natarajan}}{{Armitage} \&
  {Natarajan}}{2005}]{ArmNat05}
{Armitage} P.~J.,  {Natarajan} P.,  2005, \mn@doi [\apj] {10.1086/497108},
  \href {http://adsabs.harvard.edu/abs/2005ApJ...634..921A} {634, 921}

\bibitem[\protect\citeauthoryear{{Artymowicz} \& {Lubow}}{{Artymowicz} \&
  {Lubow}}{1994}]{1994ApJ...421..651A}
{Artymowicz} P.,  {Lubow} S.~H.,  1994, \mn@doi [\apj] {10.1086/173679}, \href
  {http://adsabs.harvard.edu/abs/1994ApJ...421..651A} {421, 651}

\bibitem[\protect\citeauthoryear{{Barnes} \& {Hut}}{{Barnes} \&
  {Hut}}{1986}]{BarnesHut1986}
{Barnes} J.,  {Hut} P.,  1986, \mn@doi [\nat] {10.1038/324446a0}, \href
  {http://adsabs.harvard.edu/abs/1986Natur.324..446B} {324, 446}

\bibitem[\protect\citeauthoryear{{Bate}, {Bonnell}  \& {Price}}{{Bate}
  et~al.}{1995}]{Bate1995}
{Bate} M.~R.,  {Bonnell} I.~A.,   {Price} N.~M.,  1995, \mnras, \href
  {http://adsabs.harvard.edu/abs/1995mnras.277..362B} {277, 362}

\bibitem[\protect\citeauthoryear{{Begelman}, {Blandford}  \& {Rees}}{{Begelman}
  et~al.}{1980}]{BegelmanEtAl1980}
{Begelman} M.~C.,  {Blandford} R.~D.,   {Rees} M.~J.,  1980, \mn@doi [\nat]
  {10.1038/287307a0}, \href {http://adsabs.harvard.edu/abs/1980Natur.287..307B}
  {287, 307}

\bibitem[\protect\citeauthoryear{{Berczik}, {Merritt}, {Spurzem}  \&
  {Bischof}}{{Berczik} et~al.}{2006}]{2006ApJ...642L..21B}
{Berczik} P.,  {Merritt} D.,  {Spurzem} R.,   {Bischof} H.-P.,  2006, \mn@doi
  [\apjl] {10.1086/504426}, \href
  {http://adsabs.harvard.edu/abs/2006ApJ...642L..21B} {642, L21}

\bibitem[\protect\citeauthoryear{{Bonnell}}{{Bonnell}}{1994}]{Bonnell1994}
{Bonnell} I.~A.,  1994, \mn@doi [\mnras] {10.1093/mnras/269.3.837}, \href
  {http://adsabs.harvard.edu/abs/1994MNRAS.269..837B} {269}

\bibitem[\protect\citeauthoryear{{Bonnell} \& {Rice}}{{Bonnell} \&
  {Rice}}{2008}]{BonnellRice2008}
{Bonnell} I.~A.,  {Rice} W.~K.~M.,  2008, \mn@doi [Science]
  {10.1126/science.1160653}, \href
  {http://adsabs.harvard.edu/abs/2008Sci...321.1060B} {321, 1060}

\bibitem[\protect\citeauthoryear{{Carmona-Loaiza}, {Colpi}, {Dotti}  \&
  {Valdarnini}}{{Carmona-Loaiza} et~al.}{2015}]{2015MNRAS.453.1608C}
{Carmona-Loaiza} J.~M.,  {Colpi} M.,  {Dotti} M.,   {Valdarnini} R.,  2015,
  \mn@doi [\mnras] {10.1093/mnras/stv1749}, \href
  {http://adsabs.harvard.edu/abs/2015MNRAS.453.1608C} {453, 1608}

\bibitem[\protect\citeauthoryear{{Colpi} \& {Dotti}}{{Colpi} \&
  {Dotti}}{2011}]{ColpiDotti2011}
{Colpi} M.,  {Dotti} M.,  2011, \mn@doi [Advanced Science Letters]
  {10.1166/asl.2011.1205}, \href
  {http://adsabs.harvard.edu/abs/2011ASL.....4..181C} {4, 181}

\bibitem[\protect\citeauthoryear{{Cuadra}, {Armitage}, {Alexander}  \&
  {Begelman}}{{Cuadra} et~al.}{2009}]{CuadraEtAl09}
{Cuadra} J.,  {Armitage} P.~J.,  {Alexander} R.~D.,   {Begelman} M.~C.,  2009,
  \mn@doi [\mnras] {10.1111/j.1365-2966.2008.14147.x}, \href
  {http://adsabs.harvard.edu/abs/2009MNRAS.393.1423C} {393, 1423}

\bibitem[\protect\citeauthoryear{{D'Orazio}, {Haiman}  \&
  {MacFadyen}}{{D'Orazio} et~al.}{2013a}]{DOrazio2013}
{D'Orazio} D.~J.,  {Haiman} Z.,   {MacFadyen} A.,  2013a, \mn@doi [\mnras]
  {10.1093/mnras/stt1787}, \href
  {http://ads.astro.puc.cl/abs/2013mnras.436.2997D} {436, 2997}

\bibitem[\protect\citeauthoryear{{D'Orazio}, {Haiman}  \&
  {MacFadyen}}{{D'Orazio} et~al.}{2013b}]{2013MNRAS.436.2997D}
{D'Orazio} D.~J.,  {Haiman} Z.,   {MacFadyen} A.,  2013b, \mn@doi [\mnras]
  {10.1093/mnras/stt1787}, \href
  {http://adsabs.harvard.edu/abs/2013MNRAS.436.2997D} {436, 2997}

\bibitem[\protect\citeauthoryear{{Dabringhausen}, {Hilker}  \&
  {Kroupa}}{{Dabringhausen} et~al.}{2008}]{Dabringhausen2008}
{Dabringhausen} J.,  {Hilker} M.,   {Kroupa} P.,  2008, \mn@doi [\mnras]
  {10.1111/j.1365-2966.2008.13065.x}, \href
  {http://adsabs.harvard.edu/abs/2008MNRAS.386..864D} {386, 864}

\bibitem[\protect\citeauthoryear{{Dotti}, {Colpi}  \& {Haardt}}{{Dotti}
  et~al.}{2006}]{DottiEtAl2006}
{Dotti} M.,  {Colpi} M.,   {Haardt} F.,  2006, \mn@doi [\mnras]
  {10.1111/j.1365-2966.2005.09956.x}, \href
  {http://adsabs.harvard.edu/abs/2006MNRAS.367..103D} {367, 103}

\bibitem[\protect\citeauthoryear{{Dotti}, {Volonteri}, {Perego}, {Colpi},
  {Ruszkowski}  \& {Haardt}}{{Dotti} et~al.}{2010}]{DottiEtAl2010}
{Dotti} M.,  {Volonteri} M.,  {Perego} A.,  {Colpi} M.,  {Ruszkowski} M.,
  {Haardt} F.,  2010, \mn@doi [\mnras] {10.1111/j.1365-2966.2009.15922.x},
  \href {http://adsabs.harvard.edu/abs/2010MNRAS.402..682D} {402, 682}

\bibitem[\protect\citeauthoryear{{Dotti}, {Colpi}, {Pallini}, {Perego}  \&
  {Volonteri}}{{Dotti} et~al.}{2013}]{DottiEtAl2013}
{Dotti} M.,  {Colpi} M.,  {Pallini} S.,  {Perego} A.,   {Volonteri} M.,  2013,
  \mn@doi [\apj] {10.1088/0004-637X/762/2/68}, \href
  {http://adsabs.harvard.edu/abs/2013ApJ...762...68D} {762, 68}

\bibitem[\protect\citeauthoryear{{Do{\u g}an}, {Nixon}, {King}  \&
  {Price}}{{Do{\u g}an} et~al.}{2015}]{2015MNRAS.449.1251D}
{Do{\u g}an} S.,  {Nixon} C.,  {King} A.,   {Price} D.~J.,  2015, \mn@doi
  [\mnras] {10.1093/mnras/stv347}, \href
  {http://adsabs.harvard.edu/abs/2015MNRAS.449.1251D} {449, 1251}

\bibitem[\protect\citeauthoryear{{Dunhill}, {Alexander}, {Nixon}  \&
  {King}}{{Dunhill} et~al.}{2014}]{Dunhill2014}
{Dunhill} A.~C.,  {Alexander} R.~D.,  {Nixon} C.~J.,   {King} A.~R.,  2014,
  \mn@doi [\mnras] {10.1093/mnras/stu1914}, \href
  {http://adsabs.harvard.edu/abs/2014mnras.445.2285D} {445, 2285}

\bibitem[\protect\citeauthoryear{{Eggleton}}{{Eggleton}}{1983}]{Eggleton1983}
{Eggleton} P.~P.,  1983, \mn@doi [\apj] {10.1086/160960}, \href
  {http://adsabs.harvard.edu/abs/1983ApJ...268..368E} {268, 368}

\bibitem[\protect\citeauthoryear{{Escala}, {Larson}, {Coppi}  \&
  {Mardones}}{{Escala} et~al.}{2004}]{EscalaEtAl04}
{Escala} A.,  {Larson} R.~B.,  {Coppi} P.~S.,   {Mardones} D.,  2004, \mn@doi
  [\apj] {10.1086/386278}, \href
  {http://adsabs.harvard.edu/abs/2004ApJ...607..765E} {607, 765}

\bibitem[\protect\citeauthoryear{{Escala}, {Larson}, {Coppi}  \&
  {Mardones}}{{Escala} et~al.}{2005a}]{EscalaEtAl05}
{Escala} A.,  {Larson} R.~B.,  {Coppi} P.~S.,   {Mardones} D.,  2005a, \mn@doi
  [\apj] {10.1086/431747}, \href
  {http://adsabs.harvard.edu/abs/2005ApJ...630..152E} {630, 152}

\bibitem[\protect\citeauthoryear{{Escala}, {Larson}, {Coppi}  \&
  {Mardones}}{{Escala} et~al.}{2005b}]{EscalaEtAl2005}
{Escala} A.,  {Larson} R.~B.,  {Coppi} P.~S.,   {Mardones} D.,  2005b, \mn@doi
  [\apj] {10.1086/431747}, \href
  {http://adsabs.harvard.edu/abs/2005ApJ...630..152E} {630, 152}

\bibitem[\protect\citeauthoryear{{Ferrarese} \& {Merritt}}{{Ferrarese} \&
  {Merritt}}{2000}]{FerrareseMerrit2000}
{Ferrarese} L.,  {Merritt} D.,  2000, \mn@doi [\apjl] {10.1086/312838}, \href
  {http://adsabs.harvard.edu/abs/2000ApJ...539L...9F} {539, L9}

\bibitem[\protect\citeauthoryear{{Frank} \& {Rees}}{{Frank} \&
  {Rees}}{1976}]{FR76}
{Frank} J.,  {Rees} M.~J.,  1976, \mn@doi [\mnras] {10.1093/mnras/176.3.633},
  \href {http://adsabs.harvard.edu/abs/1976MNRAS.176..633F} {176, 633}

\bibitem[\protect\citeauthoryear{{Gallego-Cano}, {Sch{\"o}del}, {Dong},
  {Nogueras-Lara}, {Gallego-Calvente}, {Amaro-Seoane}  \&
  {Baumgardt}}{{Gallego-Cano} et~al.}{2017}]{Gallego-CanoEtAl2017}
{Gallego-Cano} E.,  {Sch{\"o}del} R.,  {Dong} H.,  {Nogueras-Lara} F.,
  {Gallego-Calvente} A.~T.,  {Amaro-Seoane} P.,   {Baumgardt} H.,  2017,
  preprint, \href {http://adsabs.harvard.edu/abs/2017arXiv170103816G} {}
  (\mn@eprint {arXiv} {1701.03816})

\bibitem[\protect\citeauthoryear{{Gaspari}, {Ruszkowski}  \& {Oh}}{{Gaspari}
  et~al.}{2013}]{Gaspari2013}
{Gaspari} M.,  {Ruszkowski} M.,   {Oh} S.~P.,  2013, \mn@doi [\mnras]
  {10.1093/mnras/stt692}, \href
  {http://adsabs.harvard.edu/abs/2013MNRAS.432.3401G} {432, 3401}

\bibitem[\protect\citeauthoryear{{Gaspari}, {Brighenti}  \& {Temi}}{{Gaspari}
  et~al.}{2015}]{Gaspari2015}
{Gaspari} M.,  {Brighenti} F.,   {Temi} P.,  2015, \mn@doi [\aap]
  {10.1051/0004-6361/201526151}, \href
  {http://adsabs.harvard.edu/abs/2015A%26A...579A..62G} {579, A62}

\bibitem[\protect\citeauthoryear{{Gaspari}, {Temi}  \& {Brighenti}}{{Gaspari}
  et~al.}{2017}]{Gaspari2017b}
{Gaspari} M.,  {Temi} P.,   {Brighenti} F.,  2017, \mn@doi [\mnras]
  {10.1093/mnras/stw3108}, \href
  {http://adsabs.harvard.edu/abs/2017MNRAS.466..677G} {466, 677}

\bibitem[\protect\citeauthoryear{{Gebhardt} et~al.,}{{Gebhardt}
  et~al.}{2000}]{GebhardtEtAl2000}
{Gebhardt} K.,  et~al., 2000, \mn@doi [\aj] {10.1086/301240}, \href
  {http://adsabs.harvard.edu/abs/2000AJ....119.1157G} {119, 1157}

\bibitem[\protect\citeauthoryear{{Genzel}, {Eisenhauer}  \&
  {Gillessen}}{{Genzel} et~al.}{2010}]{GenzelEtAl10}
{Genzel} R.,  {Eisenhauer} F.,   {Gillessen} S.,  2010, \mn@doi [Reviews of
  Modern Physics] {10.1103/RevModPhys.82.3121}, \href
  {http://adsabs.harvard.edu/abs/2010RvMP...82.3121G} {82, 3121}

\bibitem[\protect\citeauthoryear{{Goicovic}, {Cuadra}, {Sesana}, {Stasyszyn},
  {Amaro-Seoane}  \& {Tanaka}}{{Goicovic} et~al.}{2016}]{GoicovicEtAl2016}
{Goicovic} F.~G.,  {Cuadra} J.,  {Sesana} A.,  {Stasyszyn} F.,  {Amaro-Seoane}
  P.,   {Tanaka} T.~L.,  2016, \mn@doi [\mnras] {10.1093/mnras/stv2470}, \href
  {http://adsabs.harvard.edu/abs/2016MNRAS.455.1989G} {455, 1989}

\bibitem[\protect\citeauthoryear{{Goicovic}, {Sesana}, {Cuadra}  \&
  {Stasyszyn}}{{Goicovic} et~al.}{2017}]{Goicovic2017}
{Goicovic} F.~G.,  {Sesana} A.,  {Cuadra} J.,   {Stasyszyn} F.,  2017, \mn@doi
  [\mnras] {10.1093/mnras/stx1996}, \href
  {http://adsabs.harvard.edu/abs/2017MNRAS.472..514G} {472, 514}

\bibitem[\protect\citeauthoryear{{Goicovic}, {Maureira-Fredes}, {Sesana},
  {Amaro-Seoane}  \& {Cuadra}}{{Goicovic} et~al.}{2018}]{Goicovic2018}
{Goicovic} F.~G.,  {Maureira-Fredes} C.,  {Sesana} A.,  {Amaro-Seoane} P.,
  {Cuadra} J.,  2018, preprint, \href
  {http://adsabs.harvard.edu/abs/2018arXiv180104937G} {} (\mn@eprint {arXiv}
  {1801.04937})

\bibitem[\protect\citeauthoryear{{Haiman}, {Kocsis}  \& {Menou}}{{Haiman}
  et~al.}{2009}]{Haiman2009}
{Haiman} Z.,  {Kocsis} B.,   {Menou} K.,  2009, \mn@doi [\apj]
  {10.1088/0004-637X/700/2/1952}, \href
  {http://adsabs.harvard.edu/abs/2009\apj...700.1952H} {700, 1952}

\bibitem[\protect\citeauthoryear{{H{\"a}ring} \& {Rix}}{{H{\"a}ring} \&
  {Rix}}{2004}]{HaeringRix2004}
{H{\"a}ring} N.,  {Rix} H.-W.,  2004, \mn@doi [\apjl] {10.1086/383567}, \href
  {http://adsabs.harvard.edu/abs/2004ApJ...604L..89H} {604, L89}

\bibitem[\protect\citeauthoryear{{Hernquist}}{{Hernquist}}{1990}]{Hernquist1990}
{Hernquist} L.,  1990, \mn@doi [\apj] {10.1086/168845}, \href
  {http://adsabs.harvard.edu/abs/1990ApJ...356..359H} {356, 359}

\bibitem[\protect\citeauthoryear{{Hobbs}, {Nayakshin}, {Power}  \&
  {King}}{{Hobbs} et~al.}{2011}]{Hobbs11}
{Hobbs} A.,  {Nayakshin} S.,  {Power} C.,   {King} A.,  2011, \mn@doi [\mnras]
  {10.1111/j.1365-2966.2011.18333.x}, \href
  {http://adsabs.harvard.edu/abs/2011MNRAS.413.2633H} {413, 2633}

\bibitem[\protect\citeauthoryear{{Khan}, {Just}  \& {Merritt}}{{Khan}
  et~al.}{2011}]{2011ApJ...732...89K}
{Khan} F.~M.,  {Just} A.,   {Merritt} D.,  2011, \mn@doi [\apj]
  {10.1088/0004-637X/732/2/89}, \href
  {http://adsabs.harvard.edu/abs/2011ApJ...732...89K} {732, 89}

\bibitem[\protect\citeauthoryear{{King} \& {Pringle}}{{King} \&
  {Pringle}}{2006}]{KingPringle2006}
{King} A.~R.,  {Pringle} J.~E.,  2006, \mn@doi [\mnras]
  {10.1111/j.1745-3933.2006.00249.x}, \href
  {http://adsabs.harvard.edu/abs/2006MNRAS.373L..90K} {373, L90}

\bibitem[\protect\citeauthoryear{{King}, {Lubow}, {Ogilvie}  \&
  {Pringle}}{{King} et~al.}{2005a}]{KingEtAl2005}
{King} A.~R.,  {Lubow} S.~H.,  {Ogilvie} G.~I.,   {Pringle} J.~E.,  2005a,
  \mn@doi [\mnras] {10.1111/j.1365-2966.2005.09378.x}, \href
  {http://adsabs.harvard.edu/abs/2005MNRAS.363...49K} {363, 49}

\bibitem[\protect\citeauthoryear{{King}, {Lubow}, {Ogilvie}  \&
  {Pringle}}{{King} et~al.}{2005b}]{2005MNRAS.363...49K}
{King} A.~R.,  {Lubow} S.~H.,  {Ogilvie} G.~I.,   {Pringle} J.~E.,  2005b,
  \mn@doi [\mnras] {10.1111/j.1365-2966.2005.09378.x}, \href
  {http://adsabs.harvard.edu/abs/2005MNRAS.363...49K} {363, 49}

\bibitem[\protect\citeauthoryear{{Kocsis}, {Haiman}  \& {Loeb}}{{Kocsis}
  et~al.}{2012a}]{Kocsis2012}
{Kocsis} B.,  {Haiman} Z.,   {Loeb} A.,  2012a, \mn@doi [\mnras]
  {10.1111/j.1365-2966.2012.22118.x}, \href
  {http://adsabs.harvard.edu/abs/2012\mnras.427.2680K} {427, 2680}

\bibitem[\protect\citeauthoryear{{Kocsis}, {Haiman}  \& {Loeb}}{{Kocsis}
  et~al.}{2012b}]{2012MNRAS.427.2680K}
{Kocsis} B.,  {Haiman} Z.,   {Loeb} A.,  2012b, \mn@doi [\mnras]
  {10.1111/j.1365-2966.2012.22118.x}, \href
  {http://adsabs.harvard.edu/abs/2012MNRAS.427.2680K} {427, 2680}

\bibitem[\protect\citeauthoryear{{Kormendy}}{{Kormendy}}{2003}]{Kormendy03}
{Kormendy} J.,  2003, in {Ho} L.,  ed., ``Coevolution of Black Holes and
  Galaxies'', Carnegie Observatories, Pasadena.

\bibitem[\protect\citeauthoryear{{Kormendy} \& {Ho}}{{Kormendy} \&
  {Ho}}{2013}]{KormendyHo2013}
{Kormendy} J.,  {Ho} L.~C.,  2013, \mn@doi [\araa]
  {10.1146/annurev-astro-082708-101811}, \href
  {http://adsabs.harvard.edu/abs/2013ARA%26A..51..511K} {51, 511}

\bibitem[\protect\citeauthoryear{{Larson}}{{Larson}}{1981}]{Larson1981}
{Larson} R.~B.,  1981, \mn@doi [\mnras] {10.1093/mnras/194.4.809}, \href
  {http://adsabs.harvard.edu/abs/1981MNRAS.194..809L} {194, 809}

\bibitem[\protect\citeauthoryear{{MacFadyen} \&
  {Milosavljevi{\'c}}}{{MacFadyen} \&
  {Milosavljevi{\'c}}}{2008}]{2008ApJ...672...83M}
{MacFadyen} A.~I.,  {Milosavljevi{\'c}} M.,  2008, \mn@doi [\apj]
  {10.1086/523869}, \href {http://adsabs.harvard.edu/abs/2008ApJ...672...83M}
  {672, 83}

\bibitem[\protect\citeauthoryear{{Maccagni}, {Morganti}, {Oosterloo}, {Oonk}
  \& {Emonts}}{{Maccagni} et~al.}{2018}]{2018arXiv180103514M}
{Maccagni} F.~M.,  {Morganti} R.,  {Oosterloo} T.~A.,  {Oonk} J.~B.~R.,
  {Emonts} B.~H.~C.,  2018, preprint, \href
  {http://adsabs.harvard.edu/abs/2018arXiv180103514M} {} (\mn@eprint {arXiv}
  {1801.03514})

\bibitem[\protect\citeauthoryear{{Magorrian} et~al.,}{{Magorrian}
  et~al.}{1998}]{MagorrianEtAl1998}
{Magorrian} J.,  et~al., 1998, \mn@doi [\aj] {10.1086/300353}, \href
  {http://adsabs.harvard.edu/abs/1998AJ....115.2285M} {115, 2285}

\bibitem[\protect\citeauthoryear{{Merritt} \& {Milosavljevi{\'c}}}{{Merritt} \&
  {Milosavljevi{\'c}}}{2005}]{MM05}
{Merritt} D.,  {Milosavljevi{\'c}} M.,  2005, Living Reviews in Relativity,
  \href
  {http://adsabs.harvard.edu/cgi-bin/nph-bib_query?bibcode=2005LRR.....8....8M&db_key=AST}
  {8, 8}

\bibitem[\protect\citeauthoryear{{Milosavljevi{\' c}} \&
  {Merritt}}{{Milosavljevi{\' c}} \& {Merritt}}{2003}]{MM03}
{Milosavljevi{\' c}} M.,  {Merritt} D.,  2003, \apj, \href
  {http://adsabs.harvard.edu/cgi-bin/nph-bib_query?bibcode=2003ApJ...596..860M&amp;db_key=AST}
  {596, 860}

\bibitem[\protect\citeauthoryear{{Milosavljevi{\'c}} \&
  {Merritt}}{{Milosavljevi{\'c}} \& {Merritt}}{2001}]{MilosavljevicMerrit2001}
{Milosavljevi{\'c}} M.,  {Merritt} D.,  2001, \mn@doi [\apj] {10.1086/323830},
  \href {http://adsabs.harvard.edu/abs/2001ApJ...563...34M} {563, 34}

\bibitem[\protect\citeauthoryear{{Naab} \& {Burkert}}{{Naab} \&
  {Burkert}}{2001}]{NaabBurkert2001}
{Naab} T.,  {Burkert} A.,  2001, in {Knapen} J.~H.,  {Beckman} J.~E.,
  {Shlosman} I.,   {Mahoney} T.~J.,  eds,  Astronomical Society of the Pacific
  Conference Series Vol. 249, The Central Kiloparsec of Starbursts and AGN: The
  La Palma Connection. p.~735 (\mn@eprint {} {astro-ph/0110374})

\bibitem[\protect\citeauthoryear{{Nixon} \& {Lubow}}{{Nixon} \&
  {Lubow}}{2015}]{Nixon2015}
{Nixon} C.,  {Lubow} S.~H.,  2015, \mn@doi [\mnras] {10.1093/mnras/stv166},
  \href {http://adsabs.harvard.edu/abs/2015MNRAS.448.3472N} {448, 3472}

\bibitem[\protect\citeauthoryear{{Ohsuga}, {Mori}, {Nakamoto}  \&
  {Mineshige}}{{Ohsuga} et~al.}{2005}]{2005ApJ...628..368O}
{Ohsuga} K.,  {Mori} M.,  {Nakamoto} T.,   {Mineshige} S.,  2005, \mn@doi
  [\apj] {10.1086/430728}, \href
  {http://adsabs.harvard.edu/abs/2005ApJ...628..368O} {628, 368}

\bibitem[\protect\citeauthoryear{{Perret}, {Renaud}, {Epinat}, {Amram},
  {Bournaud}, {Contini}, {Teyssier}  \& {Lambert}}{{Perret}
  et~al.}{2014}]{2014A&A...562A...1P}
{Perret} V.,  {Renaud} F.,  {Epinat} B.,  {Amram} P.,  {Bournaud} F.,
  {Contini} T.,  {Teyssier} R.,   {Lambert} J.-C.,  2014, \mn@doi [\aap]
  {10.1051/0004-6361/201322395}, \href
  {http://adsabs.harvard.edu/abs/2014A%26A...562A...1P} {562, A1}

\bibitem[\protect\citeauthoryear{{Preto}, {Berentzen}, {Berczik}  \&
  {Spurzem}}{{Preto} et~al.}{2011}]{2011ApJ...732L..26P}
{Preto} M.,  {Berentzen} I.,  {Berczik} P.,   {Spurzem} R.,  2011, \mn@doi
  [\apjl] {10.1088/2041-8205/732/2/L26}, \href
  {http://adsabs.harvard.edu/abs/2011ApJ...732L..26P} {732, L26}

\bibitem[\protect\citeauthoryear{{Prieto}, {Escala}, {Volonteri}  \&
  {Dubois}}{{Prieto} et~al.}{2017}]{2017ApJ...836..216P}
{Prieto} J.,  {Escala} A.,  {Volonteri} M.,   {Dubois} Y.,  2017, \mn@doi
  [\apj] {10.3847/1538-4357/aa5be5}, \href
  {http://adsabs.harvard.edu/abs/2017ApJ...836..216P} {836, 216}

\bibitem[\protect\citeauthoryear{{Quinlan}}{{Quinlan}}{1996}]{1996NewA....1...35Q}
{Quinlan} G.~D.,  1996, \mn@doi [\na] {10.1016/S1384-1076(96)00003-6}, \href
  {http://adsabs.harvard.edu/abs/1996NewA....1...35Q} {1, 35}

\bibitem[\protect\citeauthoryear{{Roedig} \& {Sesana}}{{Roedig} \&
  {Sesana}}{2014}]{Roedig2014}
{Roedig} C.,  {Sesana} A.,  2014, \mn@doi [\mnras] {10.1093/mnras/stu194},
  \href {http://adsabs.harvard.edu/abs/2014MNRAS.439.3476R} {439, 3476}

\bibitem[\protect\citeauthoryear{{Roedig}, {Dotti}, {Sesana}, {Cuadra}  \&
  {Colpi}}{{Roedig} et~al.}{2011}]{2011MNRAS.415.3033R}
{Roedig} C.,  {Dotti} M.,  {Sesana} A.,  {Cuadra} J.,   {Colpi} M.,  2011,
  \mn@doi [\mnras] {10.1111/j.1365-2966.2011.18927.x}, \href
  {http://adsabs.harvard.edu/abs/2011MNRAS.415.3033R} {415, 3033}

\bibitem[\protect\citeauthoryear{{Sanders} \& {Mirabel}}{{Sanders} \&
  {Mirabel}}{1996}]{SandersMirabel1996}
{Sanders} D.~B.,  {Mirabel} I.~F.,  1996, \mn@doi [\araa]
  {10.1146/annurev.astro.34.1.749}, \href
  {http://adsabs.harvard.edu/abs/1996ARA%26A..34..749S} {34, 749}

\bibitem[\protect\citeauthoryear{{Sch{\"o}del}, {Ott}, {Genzel}, {Eckart},
  {Mouawad}  \& {Alexander}}{{Sch{\"o}del} et~al.}{2003}]{SchoedelEtAl03}
{Sch{\"o}del} R.,  {Ott} T.,  {Genzel} R.,  {Eckart} A.,  {Mouawad} N.,
  {Alexander} T.,  2003, \mn@doi [\apj] {10.1086/378122}, \href
  {http://adsabs.harvard.edu/abs/2003ApJ...596.1015S} {596, 1015}

\bibitem[\protect\citeauthoryear{{Sch{\"o}del}, {Feldmeier}, {Kunneriath},
  {Stolovy}, {Neumayer}, {Amaro-Seoane}  \& {Nishiyama}}{{Sch{\"o}del}
  et~al.}{2014}]{SchoedelEtAl2014b}
{Sch{\"o}del} R.,  {Feldmeier} A.,  {Kunneriath} D.,  {Stolovy} S.,  {Neumayer}
  N.,  {Amaro-Seoane} P.,   {Nishiyama} S.,  2014, \mn@doi [\aap]
  {10.1051/0004-6361/201423481}, \href
  {http://adsabs.harvard.edu/abs/2014A%26A...566A..47S} {566, A47}

\bibitem[\protect\citeauthoryear{{Sch{\"o}del}, {Gallego-Cano}, {Dong},
  {Nogueras-Lara}, {Gallego-Calvente}, {Amaro-Seoane}  \&
  {Baumgardt}}{{Sch{\"o}del} et~al.}{2017}]{SchoedelEtAl2017}
{Sch{\"o}del} R.,  {Gallego-Cano} E.,  {Dong} H.,  {Nogueras-Lara} F.,
  {Gallego-Calvente} A.~T.,  {Amaro-Seoane} P.,   {Baumgardt} H.,  2017,
  preprint, \href {http://adsabs.harvard.edu/abs/2017arXiv170103817S} {}
  (\mn@eprint {arXiv} {1701.03817})

\bibitem[\protect\citeauthoryear{{Sesana} \& {Khan}}{{Sesana} \&
  {Khan}}{2015}]{2015MNRAS.454L..66S}
{Sesana} A.,  {Khan} F.~M.,  2015, \mn@doi [\mnras] {10.1093/mnrasl/slv131},
  \href {http://adsabs.harvard.edu/abs/2015MNRAS.454L..66S} {454, L66}

\bibitem[\protect\citeauthoryear{{Sesana}, {Haardt}  \& {Madau}}{{Sesana}
  et~al.}{2007}]{2007ApJ...660..546S}
{Sesana} A.,  {Haardt} F.,   {Madau} P.,  2007, \mn@doi [\apj]
  {10.1086/513016}, \href {http://adsabs.harvard.edu/abs/2007ApJ...660..546S}
  {660, 546}

\bibitem[\protect\citeauthoryear{{Sesana}, {Barausse}, {Dotti}  \&
  {Rossi}}{{Sesana} et~al.}{2014}]{Sesana2014}
{Sesana} A.,  {Barausse} E.,  {Dotti} M.,   {Rossi} E.~M.,  2014, \mn@doi
  [\apj] {10.1088/0004-637X/794/2/104}, \href
  {http://adsabs.harvard.edu/abs/2014ApJ...794..104S} {794, 104}

\bibitem[\protect\citeauthoryear{{Springel} \& {Hernquist}}{{Springel} \&
  {Hernquist}}{2002}]{SpringelHernquist2002}
{Springel} V.,  {Hernquist} L.,  2002, \mn@doi [\mnras]
  {10.1046/j.1365-8711.2002.05445.x}, \href
  {http://adsabs.harvard.edu/abs/2002MNRAS.333..649S} {333, 649}

\bibitem[\protect\citeauthoryear{{Tang}, {MacFadyen}  \& {Haiman}}{{Tang}
  et~al.}{2017a}]{Tang2017}
{Tang} Y.,  {MacFadyen} A.,   {Haiman} Z.,  2017a, \mn@doi [\mnras]
  {10.1093/mnras/stx1130}, \href
  {http://adsabs.harvard.edu/abs/2017MNRAS.469.4258T} {469, 4258}

\bibitem[\protect\citeauthoryear{{Tang}, {MacFadyen}  \& {Haiman}}{{Tang}
  et~al.}{2017b}]{2017MNRAS.469.4258T}
{Tang} Y.,  {MacFadyen} A.,   {Haiman} Z.,  2017b, \mn@doi [\mnras]
  {10.1093/mnras/stx1130}, \href
  {http://adsabs.harvard.edu/abs/2017MNRAS.469.4258T} {469, 4258}

\bibitem[\protect\citeauthoryear{{Tombesi}, {Sambruna}, {Reeves}, {Braito},
  {Ballo}, {Gofford}, {Cappi}  \& {Mushotzky}}{{Tombesi}
  et~al.}{2010}]{2010ApJ...719..700T}
{Tombesi} F.,  {Sambruna} R.~M.,  {Reeves} J.~N.,  {Braito} V.,  {Ballo} L.,
  {Gofford} J.,  {Cappi} M.,   {Mushotzky} R.~F.,  2010, \mn@doi [\apj]
  {10.1088/0004-637X/719/1/700}, \href
  {http://adsabs.harvard.edu/abs/2010ApJ...719..700T} {719, 700}

\bibitem[\protect\citeauthoryear{{Tombesi}, {Mel{\'e}ndez}, {Veilleux},
  {Reeves}, {Gonz{\'a}lez-Alfonso}  \& {Reynolds}}{{Tombesi}
  et~al.}{2015}]{2015Natur.519..436T}
{Tombesi} F.,  {Mel{\'e}ndez} M.,  {Veilleux} S.,  {Reeves} J.~N.,
  {Gonz{\'a}lez-Alfonso} E.,   {Reynolds} C.~S.,  2015, \mn@doi [\nat]
  {10.1038/nature14261}, \href
  {http://adsabs.harvard.edu/abs/2015Natur.519..436T} {519, 436}

\bibitem[\protect\citeauthoryear{{Tremblay} et~al.,}{{Tremblay}
  et~al.}{2016}]{TremblayEtAl2016}
{Tremblay} G.~R.,  et~al., 2016, \mn@doi [\nat] {10.1038/nature17969}, \href
  {http://adsabs.harvard.edu/abs/2016Natur.534..218T} {534, 218}

\bibitem[\protect\citeauthoryear{{Vasiliev}, {Antonini}  \&
  {Merritt}}{{Vasiliev} et~al.}{2015}]{2015ApJ...810...49V}
{Vasiliev} E.,  {Antonini} F.,   {Merritt} D.,  2015, \mn@doi [\apj]
  {10.1088/0004-637X/810/1/49}, \href
  {http://adsabs.harvard.edu/abs/2015ApJ...810...49V} {810, 49}

\bibitem[\protect\citeauthoryear{{Volonteri}}{{Volonteri}}{2010}]{2010A&ARv..18..279V}
{Volonteri} M.,  2010, \mn@doi [\aapr] {10.1007/s00159-010-0029-x}, \href
  {http://adsabs.harvard.edu/abs/2010A%26ARv..18..279V} {18, 279}

\bibitem[\protect\citeauthoryear{{Volonteri}, {Haardt}  \& {Madau}}{{Volonteri}
  et~al.}{2003}]{VolonteriEtAl03}
{Volonteri} M.,  {Haardt} F.,   {Madau} P.,  2003, \mn@doi [\apj]
  {10.1086/344675}, \href
  {http://adsabs.harvard.edu/cgi-bin/nph-bib_query?bibcode=2003ApJ...582..559V&db_key=AST}
  {582, 559}

\makeatother
\end{thebibliography}

\appendix

\section{Generating the clouds' initial conditions from the angular momentum vector}
\label{Ap:ics}

The initial angular momentum vector of each cloud orbit is determined by
sampling the orientation and the pericentre distance. However, the orbit of
each cloud is not fully determined, as a Keplerian trajectory is defined by a
total of 6 parameters.
Having ($\theta_\ell$, $\phi_\ell$, $r_p$), together with the initial distance
and speed of the cloud, leaves one degree of freedom, which basically means
that we can choose the initial position to be oriented in \textit{any}
direction as long as it lies on the plane defined by the unitary vector
\begin{equation}
\vu*{\ell} = (\cos\phi_\ell\sin\theta_\ell, \sin\phi_\ell\sin\theta_\ell, \cos\theta_\ell),
\end{equation}
where $\theta_\ell$ and $\phi_\ell$ are the polar and azimuthal angles of the
angular momentum direction, respectively.

The initial position and velocity vectors are generated by first obtaining
an arbitrary unitary vector $\vu*{e}$ lying on the aforementioned plane by
taking the cross product with the $x$-axis
\begin{equation}
\vu*{e} = \vu*{\ell}\cross\vu*{x}.
\end{equation}
Then we rotate this vector by a random angle $\beta$, drawn uniformly between
$0$ and $2\pi$, using the expression
\begin{equation}
\vb{r}^\prime = \vu*{e}\cos\beta + (\vu*{\ell}\cross\vu*{e})\sin\beta.
\label{eq:rot_r}
\end{equation}
This ensures that there is no preferential direction of incoming gas.

On the other hand, the velocity vector also lies on the plane defined by
the angular momentum ($\vu*{\ell}$), but is rotated by an angle
$\theta_{\rm vel}$ with respect to the position vector, and therefore it can
be obtained using a similar expression
\begin{equation}
\vb{v}^\prime = \vu*{e}\cos(\beta-\theta_{\rm vel}) + (\vu*{\ell}\cross\vu*{e})\sin(\beta-\theta_{\rm vel}),
\end{equation}
where $\beta$ has the same value as used in eq.~\eqref{eq:rot_r}. The angle
$\theta_{\rm vel}$ is directly related to the pericentre distance as follows
\begin{equation}
\theta_{\rm vel} = \arcsin\left( \frac{v_{\rm p}}{v_{\rm ini}}\frac{r_{\rm p}}{d_{\rm ini}}\right),
\end{equation}
where $d_{\rm ini}=15a_0$ is the initial distance to the binary centre of mass,
$v_{\rm ini}=0.25\sqrt{GM_0/a_0}$ is the initial velocity and $v_{\rm p}$ is the
velocity at periapsis.

Finally, we normalise these vectors to the initial distance and velocity,
\begin{eqnarray}
\label{eq:r_ics}
\vb{r} = \frac{d_{\rm ini}}{\norm{\vb{r}^\prime}}\vb{r}^\prime,\\
\vb{v} = -\frac{v_{\rm ini}}{\norm{\vb{v}^\prime}}\vb{v}^\prime,
\label{eq:v_ics}
\end{eqnarray}
which yields the orbit defined by the sampled angular momentum vector.
The initial position and velocity vectors of each cloud are displayed in
Table~\ref{tab:RunABrvInfo}, listed here so that the results can be
reproduced by future studies.

\begin{table} \centering \resizebox{\hsize}{!}{
\begin{tabular}{|l|r|rrr|rrr|rrr|rrr|} \hline & & \multicolumn{6}{c|}{{\runa}}
& \multicolumn{6}{c|}{{\runb}} \\ \hline Distribution & Cloud
& \multicolumn{3}{c|}{$\vec{r}$ [$a_0$]} & \multicolumn{3}{c|}{$\vec{v}$
[$v_0$]} & \multicolumn{3}{c|}{$\vec{r}$ [$a_0$]}
& \multicolumn{3}{c|}{$\vec{v}$ [$v_0$]} \\ \hline
        \multirow{19}{*}{\F{0.0}} & 01 & 14.63  & 1.24   & 3.04   & -0.22
        & 0.09  & -0.04 & 1.68   & 10.45  & -10.63 & -0.05 & -0.23 & 0.09\\
        & 02 & -10.06 & -5.58  & -9.62  & 0.21  & -0.01 & 0.15  & -11.00
        & -7.22  & -7.21  & 0.22  & 0.10  & 0.05\\ & 03 & -12.28 & 6.96
        & -5.08  & 0.12  & -0.14 & 0.17  & -0.38  & 6.18   & -13.66 & -0.02
        & -0.17 & 0.18\\ & 04 & -4.13  & -10.12 & 10.28  & 0.06  & 0.09
        & -0.22 & -10.65 & -5.85  & -8.80  & 0.21  & 0.10  & 0.09\\ & 05
        & 12.21  & 7.61   & -4.25  & -0.22 & -0.08 & 0.07  & -13.12 & 7.25
        & 0.43   & 0.19  & -0.16 & 0.01\\ & 06 & 1.54   & -7.94  & 12.63
        & 0.05  & 0.21  & -0.12 & -12.88 & -0.82  & -7.65  & 0.22  & -0.08
        & 0.09\\ & 07 & 6.10   & 6.20   & -12.22 & -0.04 & 0.00  & 0.25
        & -4.80  & 11.49  & 8.36   & -0.03 & -0.15 & -0.20\\ & 08 & 11.08
        & 3.20   & 9.60   & -0.19 & 0.00  & -0.15 & 6.41   & 12.84  & -4.38
        & -0.11 & -0.22 & 0.06\\ & 09 & 10.63  & -10.37 & -2.10  & -0.16 & 0.19
        & -0.04 & -13.07 & -0.03  & -7.35  & 0.18  & -0.11 & 0.14\\ & 10
        & 13.52  & 4.21   & -4.93  & -0.22 & -0.02 & 0.12  & 12.04  & 8.12
        & -3.75  & -0.24 & -0.03 & 0.05\\ & 11 & 10.23  & 10.79  & 1.97
        & -0.21 & -0.09 & -0.07 & -5.79  & -9.17  & -10.36 & 0.13  & 0.20
        & 0.08\\ & 12 & -1.35  & 11.93  & -8.99  & -0.07 & -0.23 & 0.07
        & -0.85  & -14.98 & -0.06  & 0.01  & 0.22  & -0.11\\ & 13 & -2.67
        & -6.03  & -13.47 & 0.14  & 0.10  & 0.19  & 8.24   & -5.29  & 11.36
        & -0.11 & 0.11  & -0.20\\ & 14 & 8.18   & -1.46  & -12.49 & -0.09
        & 0.05  & 0.23  & -10.31 & -8.45  & 6.88   & 0.20  & 0.15  & -0.05\\
        & 15 & 4.68   & 5.60   & 13.11  & -0.09 & 0.01  & -0.22 & 6.82   & 3.50
        & -12.90 & -0.16 & -0.03 & 0.19\\ & 16 & ---    & ---    & ---    & ---
        & ---   & ---   & -9.59  & -10.59 & 4.56   & 0.17  & 0.18  & -0.04\\
        & 17 & ---    & ---    & ---    & ---   & ---   & ---   & -4.74
        & -12.65 & 6.52   & 0.15  & 0.20  & -0.04\\ & 18 & ---    & ---
        & ---    & ---   & ---   & ---   & -13.91 & 4.85   & 2.84   & 0.20
        & -0.13 & -0.09\\ & 19 & ---    & ---    & ---    & ---   & ---   & ---
        & 14.41  & -3.42  & 2.38   & -0.19 & 0.05  & -0.16\\ & 20 & ---
        & ---    & ---    & ---   & ---   & ---   & -12.34 & -8.50  & 0.73
        &  0.13 &  0.21 &  0.02\\

        \hline \multirow{22}{*}{\F{0.5}} & 01 & 14.63  & 1.24   & 3.04
        & -0.22 & 0.09  & -0.04 & 1.68   & 10.45  & -10.63 & -0.05 & -0.23
        & 0.09\\ & 02 & -10.06 & 10.81  & 2.62   & 0.21  & -0.09 & -0.10
        & -11.00 & 0.12   & -10.20 & 0.22  & 0.04  & 0.11\\ & 03 & -12.28
        & 6.96   & -5.08  & 0.12  & -0.14 & 0.17  & -0.38  & -8.35  & -12.46
        & -0.02 & 0.07  & 0.24\\ & 04 & -4.13  & -14.40 & -0.76  & 0.06  & 0.23
        & -0.08 & -10.65 & -1.93  & -10.39 & 0.21  & 0.06  & 0.12\\ & 05
        & 12.21  & -8.52  & -1.82  & -0.22 & 0.11  & 0.02  & -13.12 & -4.87
        & -5.39  & 0.19  & 0.12  & 0.10\\ & 06 & 1.54   & -7.94  & 12.63
        & 0.05  & 0.21  & -0.12 & -12.88 & -0.82  & -7.65  & 0.22  & -0.08
        & 0.09\\ & 07 & 6.10   & 11.37  & -7.64  & -0.04 & -0.12 & 0.22
        & -4.80  & 11.49  & 8.36   & -0.03 & -0.15 & -0.20\\ & 08 & 11.08
        & 3.20   & 9.60   & -0.19 & 0.00  & -0.15 & 6.41   & 12.84  & -4.38
        & -0.11 & -0.22 & 0.06\\ & 09 & 10.63  & -9.54  & 4.59   & -0.16 & 0.12
        & -0.14 & -13.07 & -5.28  & 5.12   & 0.18  & 0.18  & -0.02\\ & 10
        & 13.52  & 4.21   & -4.93  & -0.22 & -0.02 & 0.12  & 12.04  & 8.12
        & -3.75  & -0.24 & -0.03 & 0.05\\ & 11 & 10.23  & 10.79  & 1.97
        & -0.21 & -0.09 & -0.07 & -5.79  & -7.62  & -11.55 & 0.13  & 0.18
        & 0.11\\ & 12 & -1.35  & -13.21 & -6.97  & -0.07 & 0.16  & 0.19
        & -0.85  & -14.98 & -0.06  & 0.01  & 0.22  & -0.11\\ & 13 & -2.67
        & 5.47   & -13.71 & 0.14  & -0.06 & 0.21  & 8.24   & -5.29  & 11.36
        & -0.11 & 0.11  & -0.20\\ & 14 & 8.18   & 6.74   & -10.61 & -0.09
        & -0.10 & 0.21  & -10.31 & -5.12  & 9.61   & 0.20  & 0.11  & -0.10\\
        & 15 & 4.68   & 5.60   & 13.11  & -0.09 & 0.01  & -0.22 & 6.82   & 3.50
        & -12.90 & -0.16 & -0.03 & 0.19\\ & 16 & -1.93  & -13.95 & 5.17
        & 0.11  & 0.17  & -0.15 & -9.59  & -10.59 & 4.56   & 0.17  & 0.18
        & -0.04\\ & 17 & 12.22  & -6.54  & -5.73  & -0.22 & 0.12  & -0.02
        & -4.74  & -12.65 & 6.52   & 0.15  & 0.20  & -0.04\\ & 18 & 0.21
        & 7.90   & -12.75 & 0.03  & -0.02 & 0.25  & -13.91 & -4.08  & 3.86
        & 0.20  & 0.12  & -0.10\\ & 19 & 2.26   & -5.10  & 13.92  & -0.03
        & 0.04  & -0.24 & 14.41  & -3.28  & 2.56   & -0.19 & 0.04  & -0.16\\
        & 20 & 6.53   & -8.72  & -10.31 & -0.15 & 0.09  & 0.18  & -12.34 & 7.32
        & -4.38  & 0.13  & -0.20 & 0.07\\ & 21 & 13.38  & 0.58   & 6.75
        & -0.24 & -0.01 & -0.05 & ---    & ---    & ---    & ---   & ---
        & ---  \\ & 22 & -10.89 & -3.93  & -9.53  & 0.17  & -0.04 & 0.19  & ---
        & ---    & ---    & ---   & ---   & ---  \\ \hline
        \multirow{16}{*}{\F{1.0}} & 01 & 14.63  & -1.46  & -2.94 & -0.22
        & -0.08 & 0.06  & 1.68   & 1.55   & -14.83 & -0.05 & -0.12 & 0.21\\
        & 02 & -10.06 & 10.81  & 2.62  & 0.21  & -0.09 & -0.10 & -11.00 & 0.12
        & -10.20 & 0.22  & 0.04  & 0.11\\ & 03 & -12.28 & -2.25  & -8.31 & 0.12
        & 0.10  & 0.20  & -0.38  & -8.35  & -12.46 & -0.02 & 0.07  & 0.24\\
        & 04 & -4.13  & -14.40 & -0.76 & 0.06  & 0.23  & -0.08 & -10.65 & -1.93
        & -10.39 & 0.21  & 0.06  & 0.12\\ & 05 & 12.21  & -8.52  & -1.82
        & -0.22 & 0.11  & 0.02  & -13.12 & -4.87  & -5.39  & 0.19  & 0.12
        & 0.10\\ & 06 & 1.54   & 11.66  & 9.31  & 0.05  & -0.10 & -0.21
        & -12.88 & 6.29   & 4.43   & 0.22  & -0.01 & -0.12\\ & 07 & 6.10
        & 11.37  & -7.64 & -0.04 & -0.12 & 0.22  & -4.80  & -7.31  & 12.19
        & -0.03 & 0.18  & -0.17\\ & 08 & 11.08  & -6.62  & -7.64 & -0.19 & 0.06
        & 0.15  & 6.41   & 1.77   & -13.45 & -0.11 & -0.04 & 0.22\\ & 09
        & 10.63  & -9.54  & 4.59  & -0.16 & 0.12  & -0.14 & -13.07 & -5.28
        & 5.12   & 0.18  & 0.18  & -0.02\\ & 10 & 13.52  & 3.18   & 5.66
        & -0.22 & -0.09 & -0.06 & 12.04  & -8.91  & 0.81   & -0.24 & 0.05
        & -0.04\\ & 11 & 10.23  & -8.59  & -6.82 & -0.21 & 0.05  & 0.12
        & -5.79  & -7.62  & -11.55 & 0.13  & 0.18  & 0.11\\ & 12 & -1.35
        & -13.21 & -6.97 & -0.07 & 0.16  & 0.19  & -0.85  & -13.70 & -6.04
        & 0.01  & 0.25  & -0.01\\ & 13 & ---    & ---    & ---   & ---   & ---
        & ---   & 8.24   & 10.96  & 6.07   & -0.11 & -0.19 & -0.13\\ & 14 & ---
        & ---    & ---   & ---   & ---   & ---   & -10.31 & -5.12  & 9.61
        & 0.20  & 0.11  & -0.10\\ & 15 & ---    & ---    & ---   & ---   & ---
        & ---   & 6.82   & -7.47  & -11.08 & -0.16 & 0.12  & 0.15\\ & 16 & ---
        & ---    & ---   & ---   & ---   & ---   & -9.59  & -8.97  & 7.25
        & 0.17  & 0.16  & -0.09\\ & 17 & ---    & ---    & ---   & ---   & ---
        & ---   & -4.74  & 8.48   & 11.43  & 0.15  & -0.07 & -0.19\\ & 18 & ---
        & ---    & ---   & ---   & ---   & ---   & -13.91 & -4.08  & 3.86
        & 0.20  & 0.12  & -0.10\\ & 19 & ---    & ---    & ---   & ---   & ---
        & ---   & 14.41  & -3.28  & 2.56   & -0.19 & 0.04  & -0.16\\ & 20 & ---
        & ---    & ---   & ---   & ---   & ---   & -12.34 & -8.50  & 0.73
        & 0.13  & 0.21  & 0.02\\ \hline

    \end{tabular}}
    \caption{Initial 3-D position and velocity ($x,y,z$
    components) of the centre of mass of each cloud for {\runa} and {\runb}.}
    \label{tab:RunABrvInfo}
\end{table}

\section{Supplementary suite of simulations (\texttt{RunC})}
\label{appendix:runc}

\begin{figure*}
    \resizebox{\hsize}{!}
        {\includegraphics[scale=1,clip]{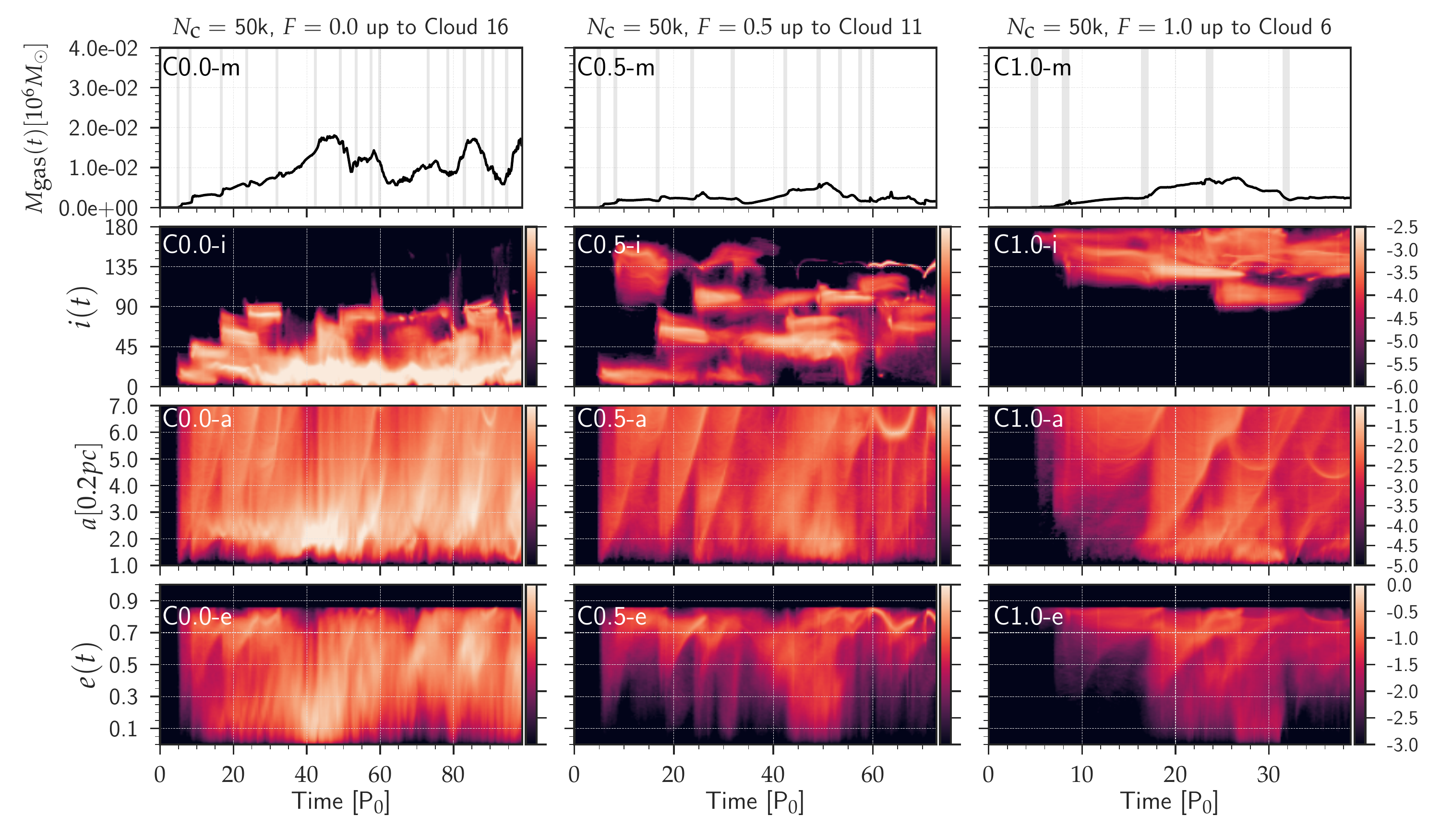}}
    \caption{Same as Figure~\ref{fig:circumbinaryAB} for {\runc}.}
    \label{fig:runCcircumbinary}
\end{figure*}

\begin{figure*}
    \resizebox{\hsize}{!}
    {\includegraphics[scale=1,clip]{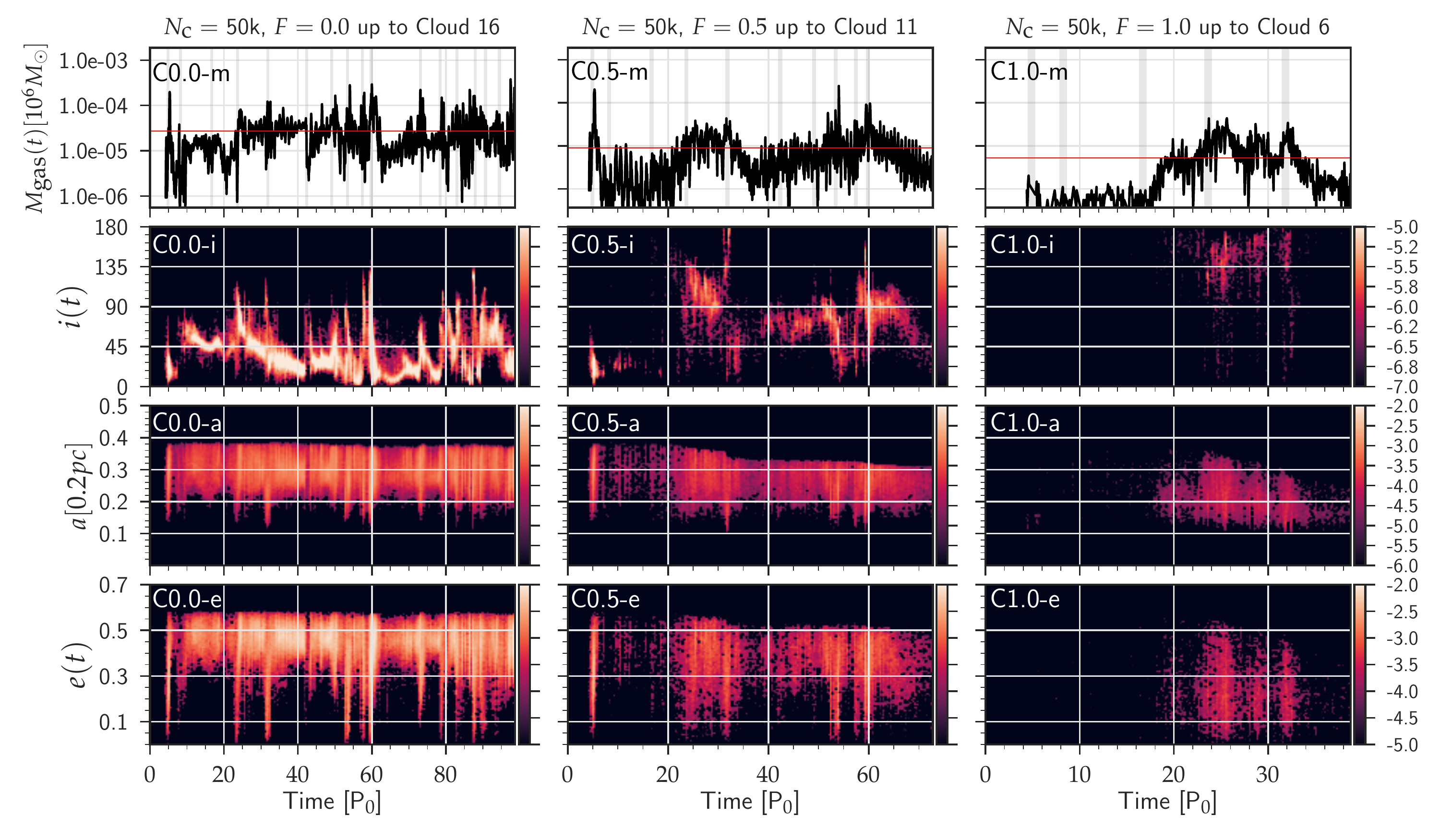}}
    \caption{Same as Figure~\ref{fig:RunABMD} for {\runc}.}
    \label{fig:runCminidiscs}
\end{figure*}

We show here some results for the extra suite of simulations ({\runc})
combining the cloud angular momentum distribution and pericentre distances of
{\runa} (see Figure~\ref{fig:dist_orientation} and \ref{fig:dist_pericentre})
with the cloud time of arrivals of {\runb} (see Table~\ref{tab:times}). Due to
limiting computing power, only 16 10 and 5 clouds arrived onto the binary in
run \F{0.0}, \F{0.5} and \F{1.0} respectively. And main properties of the gas
structures forming along the simulations are shown in
Figure~\ref{fig:runCcircumbinary} and \ref{fig:runCminidiscs}.

The overall behaviour of simulations is quite similar to what shown in
Figure~\ref{fig:circumbinaryAB} and \ref{fig:RunABMD} for {\runa}. Indicating,
not surprisingly, that the distribution of the cloud parameters, rather than
their times of arrival, leaves a strong imprint on the forming gas structures.
The \F{0.0} case (left column in both figures) initially builds up a prominent,
co-rotating circumbinary structure confined to $i<30^o$ and $a<3$
(Figure~\ref{fig:runCcircumbinary}), which is partially destroyed by clouds
incoming at $T\approx 50 P_0$. A similar behaviour is seen in the {\mds}
evolution (Figure~\ref{fig:runCminidiscs}). Note, however that the {\md}
inclination shows much larger fluctuations often reaching values larger than
$45^{o}$. Mass accumulation in circumbinary structures is much smaller in the
\F{0.5} and \F{1.0} cases (central and right columns, respectively), and {\mds}
are much lighter and intermittent (especially in the \F{1.0} case), in line
with the general finding of {\runa} and  {\runb}. Overall, this new suite of
runs confirm the robustness of the general features highlighted in the main
body of the paper.

\label{lastpage}

\bsp

\end{document}